\documentclass[aps,prx,longbibliography,twocolumn,superscriptaddress]{revtex4-2}

\usepackage{amsfonts}
\usepackage{amsmath}
\usepackage{graphicx}
\usepackage{dsfont}
\usepackage{diagbox}
\usepackage{array}
\usepackage{amssymb}
\usepackage{bbm}
\usepackage{float}
\usepackage{subfigure}
\usepackage{MnSymbol}
\usepackage{url}
\usepackage{times}
\usepackage{manfnt}
\usepackage{booktabs}
\usepackage{xcolor}
\usepackage{svg}
\definecolor{darkblue}{HTML}{004D6B}
\definecolor{darkred}{HTML}{8c1515}
\definecolor{darkgreen}{HTML}{006400}
\usepackage{hyperref}
\hypersetup{
    pdftitle={selfdual},
	colorlinks=true,
	urlcolor=darkred,
	citecolor=darkblue,
	linkcolor=darkred,
	breaklinks
}
\pagecolor{white}
\usepackage{ulem}
\usepackage{physics}
\usepackage{enumitem}
\usepackage{wasysym}
\usepackage{multirow}

\newcommand\redsout{\bgroup\markoverwith{\textcolor{red}{\rule[0.5ex]{2pt}{0.4pt}}}\ULon}

\begin{document}

\title{
Decoherence-induced self-dual criticality in topological states of matter
}

\author{Qingyuan Wang}
\affiliation{The Hong Kong University of Science and Technology (Guangzhou), Nansha, Guangzhou, 511400, Guangdong, China}

\author{Romain Vasseur}
\affiliation{Department of Theoretical Physics, University of Geneva, 24 quai Ernest-Ansermet, 1211 Gen\`eve, Switzerland}
%\affiliation{Department of Physics, University of Massachusetts, Amherst, MA 01003, USA}

\author{Simon Trebst}
\affiliation{Institute for Theoretical Physics, University of Cologne, Z\"ulpicher Straße 77, 50937 Cologne, Germany}

\author{Andreas W.W. Ludwig}
\affiliation{Department of Physics, University of California, Santa Barbara, California 93106, USA}

\author{Guo-Yi Zhu}
\email{guoyizhu@hkust-gz.edu.cn}
\affiliation{The Hong Kong University of Science and Technology (Guangzhou), Nansha, Guangzhou, 511400, Guangdong, China}

\date{\today}
\begin{abstract}
Quantum measurements performed on a subsystem of a quantum many-body state can generate entanglement for its remaining constituents. The whole system including the measurement record is described by a hybrid mixed state, which can exhibit exotic phase transitions and critical phenomena. We demonstrate that generic measurement-induced phase transitions (MIPTs) can be cast as decoherence-induced critical mixed states in one higher dimension, by constructing a projected entangled pair state (PEPS) prior to decoherence or measurement. 
In this context, a deeper conceptual understanding of such mixed-state criticality is called for, particularly with regard to algebraic symmetry as an advanced organizing principle for such entangled states of matter. Integrating these connections we investigate the role of self-dual symmetry -- a fundamental notion in theoretical physics -- in mixed states, showing that the decoherence of electric (e) and magnetic (m) vortices from the 2D bulk of the toric code, or equivalently, a 2D cluster state with symmetry- protected topological order, can leave a (1+1)D quantum critical mixed state protected by a weak Kramers-Wannier self-dual symmetry. The corresponding self-dual critical bulk is described by the $N\to 1$ limit of the 2D Non-linear Sigma Model in symmetry class D with target space SO(2N)/U(N) at $\Theta$-angle $\pi$, and represents a ``measurement-version’’ of the Cho-Fisher network model subjected to Born-rule randomness. Explicit breaking of self-duality, by incoherent noise amounting to fermion interactions or (non-interacting) coherent deformation, is shown to induce an RG crossover from this self-dual critical state to Nishimori criticality or to it from a novel type of Ising+ criticality, respectively, both related to the random-bond Ising model in different replica limits. Using an unbiased numerical approach combining tensor network, Monte Carlo, and Gaussian fermion simulations, we chart out a global phase diagram as diagnosed by coherent information and entanglement entropy measures. 
Our results point to a way towards a general understanding of mixed-state criticality in open quantum systems in terms of symmetry and topology, while also providing a concrete protocol amenable to simulation on near-term quantum devices.
\end{abstract}

\maketitle

%%%%%%%%%%%%%%%%%%%%%%%%%%%%%%%%%%%%%%%%%%%%%%%%%%
%%%%%%%%%%%%%%%%%%%%%%%%%%%%%%%%%%%%%%%%%%%%%%%%%%
%%%%%%%%%%%%%%%%           INTRODUCTION         %%%%%%%%%%%%%%%%%%%
%%%%%%%%%%%%%%%%%%%%%%%%%%%%%%%%%%%%%%%%%%%%%%%%%%
%%%%%%%%%%%%%%%%%%%%%%%%%%%%%%%%%%%%%%%%%%%%%%%%%%

\begin{figure*}[t]
   \centering
   \includegraphics[width=\textwidth]{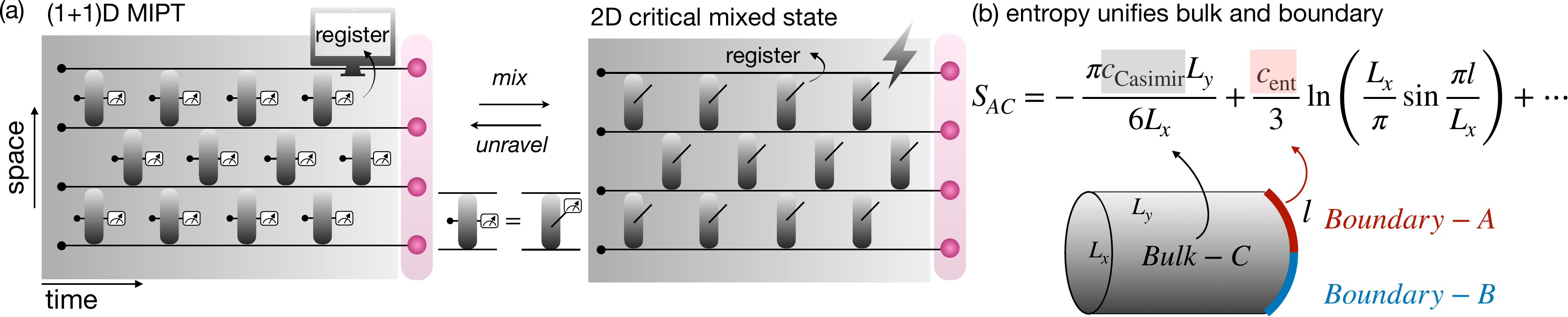}
   \caption{
   {\bf Equivalence between measurement induced phase transition (MIPT) and mixed state transition}. (a) A general scheme to map a generic 1-dimensional MIPT to a 2-dimensional mixed state, by means of a tensor network. Each gray box denotes a generic weak 2-body measurement gate, 
   by entangling the two qubits with a middle ancilla qubit and projectively measuring out the ancilla. 
   Our discussion applies to a generic MIPT that may or may not contain unitary gates; the latter would be considered part of the measurement gates in our notation.
   Such a measurement gate can be equivalently viewed as a rank-5 tensor, where the measurement outcome is the last,
   ``the physical''  index. The circuit can be embedded in a 2-dimensional tensor network
   state, specifically, a projected entangled pair state (PEPS), where the measurement record is materialized as the physical degrees of freedom of the PEPS. The map can be generalized to higher dimensions. 
   (b) The von Neumann entanglement entropy of the hybrid bulk-boundary system encodes the bulk and boundary conformal data of the underlying CFT.
   The PEPS can be written as a Rokhsar-Kivelson type state: $\sum_\mathbf{s} 
    \ket{\mathbf{s}}\otimes \ket{\psi(\mathbf{s})}$ for the un-normalized state $\ket{\psi(\mathbf{s})}$, where $\mathbf{s}$ denotes the set of measurement ourcomes from the space-time history of the circuit.
   }
   \label{fig:miptmixedstate}
   \end{figure*}

\section{Introduction}
\label{LabelSectionIntroduction}

Since the early days of quantum mechanics, Born's rule has been providing a link connecting quantum measurements to the probabilistic dynamics they induce. 
This paper addresses a class of continuous quantum phase transitions that occur in the presence of the randomness originating from quantum measurements. We characterize and quantify the critical behavior emerging in this class from the randomness governed by Born's rule.
A well-studied example of this class are noisy topological quantum memories~\cite{Preskill2002}, which are known  to be related to the random bond Ising model (RBIM) and to the emergence of an error threshold described by the famous Nishimori critical point (``Nishimori criticality'')~\cite{Nishimori1981}
as a ``quantum'' (Born-rule) randomness-induced generalization 
of the (ordinary) Ising crticial point (``Ising criticality'').
Such a link to Nishimori physics has recently been uncovered also in the context of monitored quantum circuit dynamics~\cite{NishimoriCat, JYLee},
many-body teleportation thresholds~\cite{teleportcode}, and mixed-state phase transitions in open quantum systems~\cite{Lee23decoher, Hsieh23mixedstatecriticality, You24weaksym, Wang24strtowksym}.  In a recent experiment performed on an IBM quantum processor~\cite{Chen24nishimori} 
a transition in the same,  Nishimori universality class was found to describe
 an entire phase boundary
interpolating between
a measurement-induced phase transition (MIPT) and
a noise-induced phase transition. 
Nevertheless, generic MIPTs~\cite{Li2018,Skinner2019,Ludwig2020, Altman2020weak, Pixley22MIPTCFT, Potter21review, Fisher2022reviewMIPT} and noise-induced mixed state transitions are often viewed as distinct phenomena in the literature. 
Here we use a spacetime duality to explicitly map general 1-dimensional MIPTs to corresponding 2-dimensional mixed states; this can be rigorously established in a tensor network representation. Moreover, we provide a variant of the von Neumann entanglement entropy of  mixed states which unifies the notions of general measurement-induced phase transitions and associated decoherence-induced mixed state criticality by simultaneously providing an overarching  diagnostic of distinct universal critical characteristics of both, the measurement record in the bulk and the entanglement of the quantum state at the boundary. 
We apply this quantity to demonstrate how statistical (``weak'') Kramers-Wannier self-dual symmetry~\cite{KW1941} can protect decoherence-induced criticality in mixed states. 
From a symmetry perspective, the most striking difference between clean Ising and disordered Nishimori criticality is their relation to Kramers-Wannier {\it self-duality}~\cite{KW1941}, which is an idiosyncratic symmetry for the clean Ising transition, while for Nishimori criticality it is inevitably and strongly broken~\cite{GRL2001, Read2000, Chalker02negative}. Self-duality has long been appreciated as one of the most elegant symmetries for a many-body system to exhibit, often serving as a guiding light
to deeper understanding, such as Wegner's seminal insight~\cite{Wegner71duality} connecting spin to gauge-invariant models and, subsequently, the interplay of gauge and matter shaped by t'Hooft \cite{tHooft1978}, and the Fradkin/Shenker phase diagram~\cite{FradkinShenker79}.
Today, self-duality is again providing a spotlight in the 
exploration of quantum circuits where its non-invertible algebraic character, its relation to anomalies and fractionalization are broadly discussed~\cite{Wen20categorysym, Rizi23selfdual, Ho23noninvert, Tachiwawa23cftnoninvertible, Verresen24measurespt, Seiberg24selfdual, Tachikawa24noninvertible, Ning24categorysym, Verstraete21mposym, You24noninvert}. 
This has led us to question whether in the context of {\it monitored} circuit dynamics there is a way to twist Born's rule into preserving self-duality,
what kind of criticality this might induce, and how decoherence or coherent deformations might then break it down, e.g., to Nishimori criticality.

The focus of the present paper is a generalization
 of the 2D Nishimori critical point to a critical point which exhibits a different universality class enforced by self-dual Kramers-Wannier symmetry. The Nishimori critical point is well known to describe the critical theory at the threshold in
  Ref.~\onlinecite{Preskill2002}, which can be formulated as a problem of Quantum Error Correction (QEC) of the 2D Toric Code, and we pursue a related formulation later in this paper. However, as described
in Ref.~\cite{GRL2001} which precedes \cite{Preskill2002}, the critical theory at the error correction threshold
in \cite{Preskill2002}
can be described as a problem  %Anderson localization 
of non-interacting  (Majorana) fermions in two spatial dimensions subjected to a particular type of quenched randomness, and this  allows for a convenient description of the field theory for the  transition.
This is because in general, problems of non-interacting   
fermions subjected to quenched randomness, including
Anderson localization problems which are subjected to generic uncorrelated disorder, but also a set of physically different problems where the quenched disorder is correlated and arises from  quantum mechanical measurement, are known to be classified using the  framework of ``10-fold way'' Cartan (Altland-Zirnbauer) classification of symmetry classes, which arises from the analysis of anti-unitary symmetries of quantum mechanics (reflected in the Wigner - von Neumann theorem).
The key property established in \cite{GRL2001} is that the Nishimori line, and thus its critical point, resides in what is known as 
Cartan's (Altland-Zirnbauer's)
symmetry class D. Since the universal properties of the critical point at the
QEC threshold are generally described by field theories, one is naturally interested in the field theory describing the Nishimori critical point. For every one of the ``10-fold way'' Cartan symmetry classes, corresponding field theories are well known to be classical two-dimensional
so-called Non-Linear-Sigma-Models (NLSMs). These are simply generalizations of the classical Heisenberg ferromagnet statistical mechanics model, where the order parameter resides on a so-called symmetric space that generalizes in a total of 10 possible ways, following foundational work by the mathematician E. Cartan, the well-known sphere swept out by the order parameter in the Heisenberg ferromagnet magnet. There are 10 such symmetric spaces corresponding to the 10 possible symmetry classes of Anderson localization problems. In symmetry class D of interest here,  there are two possible such symmetric spaces on which the order parameter of the corresponding NLSM field theory can reside. This is formulated in terms of replicas which typically arise in the formulation of systems with quenched randomness; in the Nishmori problem the randomness is {\it correlated} because it arises from the intrinsic randomness of the outcomes of quantum mechanical measurements described by the Born rule.
For this reason these are
in fact not Anderson localization problems
as the latter correspond to generic spatially uncorrelated randomness.
If we denote by $R$ the number of replicas, the two possible symmetric spaces in symmetry class D on which the order parameter of the `generalized Heisenberg magnet' resides are the coset~\footnote{Here $SO(K)$ and $U(K)$ denote the special orthogonal and unitary groups for any integer $K$.}
spaces (i):  $SO(2M+1)/U(M)$ or
(ii): $SO(2N)/U(N)$.
Since the physical problem describes systems with Born-rule measurements, we need to consider the replica limit $R \to 1$, where in the first case $R=2M+1$ is an odd integer, while in the second case $2R=2N$ is an even integer. (This follows, as discussed extensively in
\cite{Jian23measurefreefermion}, 
from the requirement that the partition function must be constant in that limit, as a consequence of the POVM condition on the Kraus operators of the underlying quantum measurement problem.~\footnote{In the mentioned limits of the number $R$ of replicas the degree of freedom of the NLSM, the `generalized Heisenberg model spin', is trivial and has no degree of freedom as the corresponding symmetric space on which it resides collapses into a single point, implying that the partition function is constant.})
Both field theories (i) and (ii) allow for a topological $\Theta$-term, where criticality is known to occur only
when the theta-angle $\Theta=\pi$ (for the same reasons as those familiar from the usual Heisenberg magnet). It was established in \cite{GRL2001} that (i) represents the critical field  theory for the {\it Nishimori} problem and that (ii) represents {\it a different} critical theory, both residing in symmetry class D. Most importantly for the present work,   Ref.~\onlinecite{GRL2001}  established 
 the key distinction between these two field theories, namely that (ii) is invariant under statistical Kramers-Wannier (KW)
 ``self-dual'' symmetry, while 
(i) lacks this (statistical)  symmetry.

The notion of KW symmetry naturally enters the description as the fermions appearing in the above-described problems arising upon Jordan-Wigner transformation from the underlying classical Ising spin degrees of freedom used by Nishimori~\cite{Nishimori1981} in his formulation of the $\pm J$ random bond Ising model (RBIM). This RBIM model clearly lacks (statistical) KW symmetry because the random $\pm J$ bonds are only those between two neighboring Ising {\it spin} degrees of freedom while there is no coupling of a corresponding type between neighboring Ising Kramers-Wannier {\it dual spins}. This lack of (statistical) KW duality of the RBIM model gets automatically transferred~\cite{GRL2001} to the fermion formulation and thus to the NLSM (i) mentioned above. In contrast, as already mentioned above, 
Ref.~\onlinecite{GRL2001} established that the NLSM
with target space (ii) listed above  possesses statistical KW symmetry.

These observations raise several fundamental questions that we address in this work: 
(i)~Is there a general framework that unifies measurement-induced phase transitions and decoherence-induced mixed-state criticality? 
(ii)~What is the connection between the 2D self-dual NLSM mentioned above, and monitored quantum circuits or decohered topological states? 
(iii)~What physical quantities can diagnose such transitions, and what are the identifying universal characteristics of the underlying critical theory?
(In the current ($d=1$)-dimensional situation, these are related to the underlying conformal symmetry). 
In the following section, we provide an overview of our main results. We first establish a general connection between MIPTs in $d$-dimensional position space, and $(d+1)$-dimensional mixed states via a Projected Entangled Pair State (PEPS) construction, and introduce a unified entanglement entropy measure that simultaneously captures universal characteristics of the bulk and of the
`temporal boundary' of the MIPT.
We then demonstrate an elegant prototype of this general connection: how weak Kramers-Wannier self-dual symmetry naturally emerges in the decohered toric code mixed state as well as the monitored Ising circuit, both described by self-dual critical mixed states, revealing their identifying universal characteristics (in the current dimension $d=1$ given in terms of corresponding conformal data).

\section{Overview of the main results}
\label{LabelSectionOverviewOfResults}

Before going into the details of our study, we provide an overview of our main findings.

\subsection{Unifying MIPT and mixed state transition}
\label{LabelSubsectionUnifyingMITPMixedState}

Here we first establish
a very general
connection between a generic MIPT in $d$-dimensional space, and a quantum state in $D=(d+1)$-dimensional space, which represents the space-time of the MIPT, subjected in the bulk to maximally dephasing or to measurements.
Fig.~\ref{fig:miptmixedstate}a shows a schematic picture of a ($d=1$)-dimensional monitored quantum circuit exhibiting an MIPT. There is an extensive number of measurement events in the spacetime. Note that the 1-dimensional quantum circuit itself is a 2-dimensional tensor network, specifically, a projected entangled pair state (PEPS)~\cite{Cirac04peps, Cirac06}. Then we remove all the measurement gadgets from the tensor network, which leaves a dangling leg for each measurement event, see Fig.~\ref{fig:miptmixedstate}a. In this way, the 2-dimensional tensor network is turned into a 2-dimensional tensor network {\it state}, specifically of the type of a projected entangled pair state (PEPS)~\cite{Cirac06}. 
This PEPS can be written as
\begin{equation}
\label{LabelEqPEPSMixedState}
|\Psi\rangle = \sum_\mathbf{s} 
\ket{\mathbf{s}}_{\rm bulk} \otimes \ket{\psi(\mathbf{s})}_{\rm boundary} \ ,
\end{equation}
where $\mathbf{s}$ labels the measurement record in the bulk 
(states $|\mathbf{s}\rangle$ are orthonormal), and 
$\ket{\psi(\mathbf{s})}_{\rm boundary}:=$
$\ket{\psi(\mathbf{s})}$ (for short)
is the conditional {\it un-normalized} state at the boundary, and 
$P(\mathbf{s})=\braket{\psi(\mathbf{s})}$ is the Born probability of observing $\mathbf{s}$. 
The resulting PEPS shares the properties of a Rokhsar-Kivelson state~\cite{Henley04RK,Fradkin04RK,DynamicsConformalQCP} in  that correlation functions are equal to those of the corresponding classical model for the statistical ensemble 
whose Gibbs distribution equals the probability distribution $P(\mathbf{s})$ for the measurement record $\mathbf{s}$ of the circuit.
When this higher dimensional PEPS is subjected to maximally dephasing noise or projective measurements on the physical legs of the PEPS in the bulk, the qubits in the spacetime collapse into fully decohered, i.e.
classical snapshots, which correspond to the measurement outcomes of the MIPT. 
Fixing a measurement record is generally
called ``unravelling''~\cite{Dalibard92unravelling}.
More generally, in this way one can embed a generic {\it monitored} quantum circuit in $d$-dimensional space in a quantum state in $d+1$-dimensional space subjected to a certain set of projective measurements.

When all the measurement outcomes are collected as a whole, the system is effectively described by a $(d+1)$-dimensional mixed quantum state, where the bulk is fully dephased while the boundary remains quantum, 
\begin{equation}
\mathcal{N}(\ketbra{\Psi}) = \sum_\mathbf{s} %P(\mathbf{s}) 
\ketbra{\mathbf{s}} \otimes \ketbra{\psi(\mathbf{s})} \ ,
\label{eq:PEPSmixedstate}
\end{equation}
see Fig.~\ref{fig:miptmixedstate}a. Here $\mathcal{N}(\cdot)$,
applied in the above equation to the pure state density matrix
$\ketbra{\Psi}$,
stands for the noisy channel that erases the off-diagonal elements from the density matrix of the qubits in the bulk in the (orthonormal) basis 
$\{|\mathbf{s}\rangle\}_{\mathbf{s}}$. For example,
in the case of a single qubit and the Pauli-$Z$ eigen-basis, this noisy channel reads
$\mathcal{N}(\cdot) = \left( (\cdot) + Z(\cdot)Z\right)/2$, for any $2 \times 2$ dimensional input density matrix $(\cdot)$. 
Note that the `bulk-decohered' density matrix of the total system composed of the bulk and the boundary qubits on the right hand side of Eq.~\eqref{eq:PEPSmixedstate} has `block-diagonal' structure where each block is specified by 
a measurement outcome ``$\mathbf{s}$'' of the set of bulk qubits. We also note in passing, as a somewhat unrelated comment,
that the noisy channel in
Eq.~\eqref{eq:PEPSmixedstate}
can be viewed as a projected ensemble~\cite{Ho23projectedensemble}:
$\{ P(\mathbf{s}), 
\ket{\psi(\mathbf{s})}/\sqrt{P(\mathbf{s}}\}$~\footnote{In the language of the projected ensemble, here we take a quantum state defined in the union of the bulk and boundary qubits, and projectively measure the bulk in the 
(orthonormal) basis 
$\{|\mathbf{s}\rangle\}_{\mathbf{s}}$,
which yields an ensemble of the quantum states supported on the boundary qubits}.

In the present paper we establish that the von Neumann entanglement entropy of the bulk-dephased $(d+1)$-dimensional PEPS, a mixed state, as depicted in Fig.~\ref{fig:miptmixedstate}b,
is a natural diagnostic that 
unifies the notions of measurement-induced phase transitions (MIPTs) and decoherence-induced mixed states. 
The MIPT typically focuses on the spacetime dynamics of a monitored quantum circuit, emphasizing the entanglement entropy of the quantum state at the final time, i.e.\ its `temporal boundary', while the latter often refers to a quantum state experiencing a noisy channel without necessarily considering the time dynamics. In our setting, we design the 2D system as a hybrid with a decohered classical bulk and an untouched quantum boundary, as a measurement-based quantum computation (MBQC) way of realizing the MIPT. In this way, the measurement outcomes are materialized and carried in the bulk, while the monitored evolved quantum state resides on the boundary. 
In this setting, the bulk undergoes a decoherence-induced mixed state transition, that is signaled by the Shannon entropy of its measurement record~\cite{Pixley22MIPTCFT}, yielding a universal finite-size correction of Casimir-energy type
at the critical point. The boundary undergoes an MIPT, whose quantum trajectory-averaged entanglement entropy obeys universal critical scaling at exactly the same critical point. 
In this work we show how to unify both universal quantities 
in a {\it single} entanglement entropy $S_{AC}$ of the 2D {\it mixed} state between (referring to the drawing below) the region $A$ of length $l$ on the boundary (standing for ``Alice'') and its boundary complement
$B$ (standing for ``Bob''),
see Fig.~\ref{fig:miptmixedstate}b, 
which encodes both the bulk and boundary conformal data. In particular, while the universal bulk finite-size scaling
correction reveals the {\it effective central charge} $c_{\rm Casimir}$ of the underlying non-unitary CFT, we caution that the 
{\it universal number}~\footnote{It characterizes the (typical) scaling dimension of a boundary operator (the ``boundary condition changing'' twist operator)~\cite{TensorNetworksPotterVasseur2019,Ludwig2020}}
$c_{\rm ent}$ extracted from the boundary entanglement entropy is not equivalent to this central charge, $c_{\rm Casimir}\neq c_{\rm ent}$. This feature sharply distinguishes the {\it monitored} non-unitary CFT  (random and thus not translationally invariant due to the intrinsic randomness of the measurement outcomes),
from a conventional `non-random', translationally invariant unitary CFT, where in the latter case the two universal quantities $c_{\rm Casimir}$
and $c_{\rm ent}$ are well known to be precisely identical, and are equal to the central charge $c$, which vanishes in the monitored (`random') case.
The so-defined variant $S_{AC}$  of the von Neumann entanglement
entropy is applicable to generic MIPTs as well as to
critical mixed states, and is discussed in more detail in Sec.~\ref{LabelSubsectionBoundaryEntanglementEntropy}, see in particular 
Eqs.~(\ref{eq:CalabreseCardy},
\ref{LabelEqSCFExplicit}), with the mixed state written in the form of
Eq.~(\ref{eq:mixedstate}).
These expressions apply to 
generic MIPTs, where ``$(em)$'' is replaced by the set of measurement outcomes ``$\mathbf{s}$''
from the entire space-time  history of the circuit, as in 
Eq.~(\ref{LabelEqPEPSMixedState}). 
On the other hand, in a generic mixed state ``$(em)$'' is replaced by the
the set of good quantum numbers ``$\mathbf{s}$'', with respect to which
the density matrix is block diagonalized. One example would be the 2D toric code subjected to bit-flip and phase-flip 
errors~\cite{Preskill2002} where the set ``$\mathbf{s}$''
of `good quantum numbers' corresponds to the (stabilizer) syndrome measurement outcomes, also see Ref.~\cite{Puetz25percolation} for related discussion. 

\begin{figure*}[t!]
   \centering
   \includegraphics[width=\textwidth]{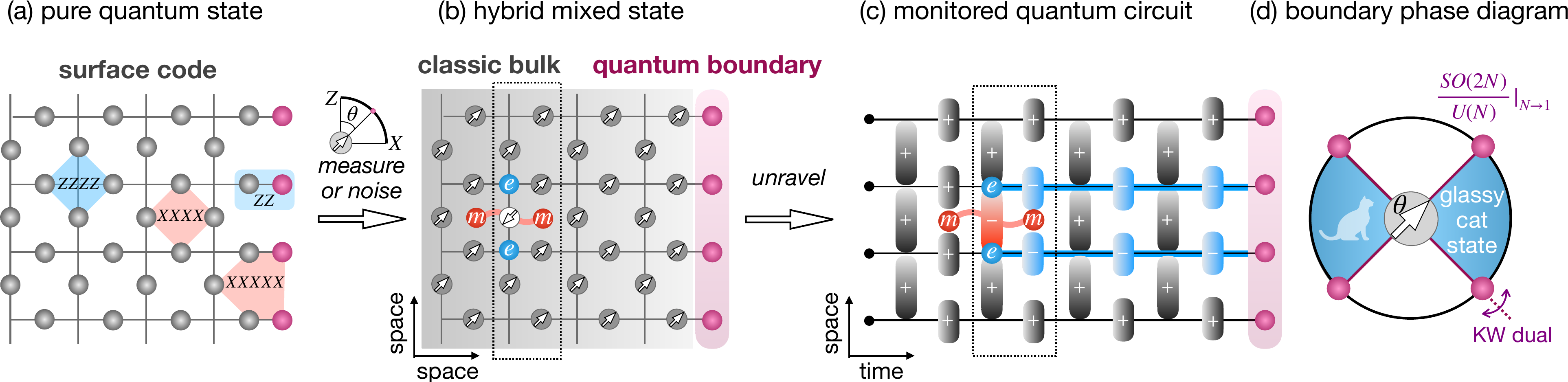}
   \caption{
      {\bf Bulk decoherence of surface code}. 
      (a) 
      A pure surface code state with a rough boundary on the right. Here we attach a chain of boundary qubits that are entangled with the rough boundary of the surface code, satisfying the stabilizers illustrated in the figure. Such boundary coupling is a natural consequence of measuring a cluster state~\cite{Raussendorf2001} (see Fig.~\ref{fig:classicquantumclusterroute} for details).
      (b) 
      The bulk qubits are collapsed (thus decohered)
      by measurements into classical snapshots of spin configurations. The site qubit at site $j$ collapses into the $X_j$ eigenstate
      $\ket{+}$, and is labeled by a right arrow in an orange disk, while $\ket{-}$ is labeled by
      a left arrow in a blue disk. 
      The qubit on the bond between sites $i$ and $j$ collapses to one of the eigenstates for rotated Pauli operator $\sigma_{ij}^\theta \equiv \cos(\theta)Z_{ij}+\sin(\theta)X_{ij}$,
      denoted by the tilted up arrow in gray circle, or down tilted arrow in white circle. We label the snapshot by $e$ particle (for a left arrow at site) and $m$-string (for a tilted down arrow at bond). The endpoint of an $m$-string is an $m$-particle. 
      The rightmost column of qubits remain the unmeasured quantum boundary, highlighted by purple disks.
      While the bulk becomes a classical mixed state, the boundary quantum state is a  pure quantum state conditioned upon the bulk snapshot. 
      (c)
      By fixing an arbitrary bulk snapshot, the boundary quantum state is effectively generated by a 1D deep monitored quantum circuit, where the measurement record corresponds to the bulk classical configuration. The derivation follows from the tensor network representation discussion in Eq.~\eqref{eq:tnderivation}. 
      (d) Self-dual phase diagram
         by tuning the bond measurement angle $\theta$, shown in a circle. When the bond measurement angle $\theta$ is tuned to the symmetric angle $\pm\pi/4$ or $\pm 3\pi/4$, the system exhibits the weak self-dual criticality, described by the $SO(2N)/U(N)$ Non-linear Sigma Model (NLSM) at topological $\Theta$-angle $\pi$ in the $N\to 1$ limit.
         Changing the measurement angle $\theta$ at the microscopic level has the same effect as changing the topological $\Theta$-angle in the long-wave-length theory. 
   }
   \label{fig:classicquantum}
   \end{figure*}

\subsection{Self-dual criticality}
\label{LabelSubsubsectionWeakSelfDualCriticality}

In this work, we use a formulation of the non-interacting  fermion systems described in the penultimate paragraph of 
Sect.~\ref{LabelSectionIntroduction},
in the language of the 2D Toric Code or, equivalently, in the language of a 2D cluster state to be described below. In particular, we
establish that specific types of   measurements and/or decoherence, when applied to the 2D  Toric Code or more generally the  2D cluster state,  lead to the critical point (ii) with statistical KW `self-dual' symmetry
discussed in Sect.~\ref{LabelSectionIntroduction}.
Specifically, the unmeasured and decohered 2D Toric Code  wavefunction possesses itself a  self-dual symmetry, corresponding to an exchange of the Pauli-$X$ and Pauli-$Z$ operators appearing in the vertex and plaquette stabilizers while simultaneously exchanging the latter two,
see the bulk of Fig.~\ref{fig:classicquantum}(a). 
Violations of the stabilizers at a vertex or a plaquette correspond to insertions of $e$ and $m$ particles (anyons), respectively. KW self-dual symmetry acts to exchange $e$ and $m$ particles. (Analogous physics appears in the 2D cluster state.)
Measurement (or decoherence) of solely $m$ particles at plaquettes is known to drive the transition at the noise threshold in  Ref.\cite{Preskill2002}
and at the measurement threshold in Ref.~\cite{teleportcode}. This maximally violates (statistical) KW self-dual symmetry because only $m$, and not $e$ particles are involved. In the present paper we show that measurement (or decoherence) of both, $e$ and $m$ particles with {\it equal strength} results in a novel {\it KW self-dual} criticality, which is precisely described by the field theory (ii) discussed in Sect.~\ref{LabelSectionIntroduction}.

In brief, the protocol is simply to projectively measure every bulk qubit of the toric code into the eigenstates of Pauli $\sigma^\theta = \sin(\theta)X+\cos(\theta)Z$, shown by the gray or white disks in  Fig.~\ref{fig:classicquantum}(b).
We use gray disk to denote positive measurement outcome: $\sigma^\theta =+1$, and white disk to label the opposite outcome: $\sigma^\theta =-1$. For a negative outcome, we equivalently label the outcome by associating a pair of $e$ or a pair of $m$ particles with its adjacent vertices and plaquettes, which correspond to inserting Pauli $Z$ and Pauli $X$ into the virtual legs of the tensor network~\footnote{in tensor network  language these are called `shadows' of anyons~\cite{haegeman2015shadows, Zhu19ToricCode}} 
-- compare Fig.~\ref{fig:classicquantum}. 
When all the outcomes are collected as a whole, the system is equivalent to a toric code subjected to a maximally dephasing noise channel 
\begin{equation}
   \mathcal{N}(\cdot) = \frac{1}{2}[(\cdot )+ \sigma^\theta(\cdot)\sigma^\theta] \ ,
\end{equation}
with tunable angle $\theta$. 
While Ref.~\cite{Grover24selfdual} tackled the model 
using an {\it annealed average} 2-replica
approximation which does not describe the measurement-based physics of the system at hand~\footnote{in which the Born-rule quenched random measurement outcomes  are treated as if they were in equilibrium
(i.e. ``annealed'') with the Toric Code degrees of freedom of the system, leading to a description of the critical point in terms of a completely conventional unitary CFT which occurs at a replica number dependent value of system parameters},
here we solve the problem in the required physical replica number $R \to 1$ limit describing the {\it quenched} random nature of the measurement outcomes~\cite{Altman2020weak,Ludwig2020}, and 
 thereby unveil the non-unitary CFT nature of the underlying critical state. 
As a probe of the quantum entanglement, we couple the rough surface of the toric code to a chain of boundary qubits (purple spheres in Fig.~\ref{fig:classicquantum}), which are left unmeasured and describe a pure quantum state conditioned upon the bulk snapshot. For an arbitrarily fixed bulk snapshot, called an
{\it unravelling}~\cite{Dalibard92unravelling},
the boundary state 
$\ket{\psi(\mathbf{s})}$ can be effectively generated by a (1+1)D monitored quantum circuit with measurement outcomes fixed by the bulk snapshot $\mathbf{s}$, a fact that
can be readily seen by viewing a 2D tensor network as a (1+1)D quantum circuit, see Fig.~\ref{fig:classicquantum}(c). 
Here the two-body gates are $e^{\pm \frac{\beta}{2} Z_j Z_{j+1}}$ and the one-body gates are $e^{\pm \frac{\beta'}{2} X_j}$, with the sign determined by the measurement outcome, and $\tanh\beta = \sin(\theta)\ ,\ \tanh\beta' = \cos(\theta)$ (derivation shown in Eq.~\eqref{eq:tnderivation}).
Namely, the $e$ particle and $m$ particle pulls strings in the original lattice and the dual lattice that effectively correspond to the  above mentioned changes of sign.
As a result, the boundary quantum of qubits undergoes a quantum phase transition from glassy GHZ-like~\cite{GHZ} cat~\footnote{
   A ``glassy (or: random) GHZ state'' is an equal weight superposition of an eigenstate of $\sigma_j^z$ at all lattice sites $j$  with eigenvalues of random signs $\pm 1$ and the same state with the signs of all eigenvalues reversed.
}
phase at $\theta=\pi/2$ to a product state at $\theta=0$. We find a sharp phase transition at the self-dual point $\theta=\pi/4$, which is described by the self-dual criticality of the circuit of Majorana fermions in symmetry class D, discussed in
Sect.~\ref{LabelSectionIntroduction}.

The relationship between these two formulations described above, one in terms of Majorana fermions and the other in terms of the Toric Code (or cluster state),  is precisely  provided in  a very convenient way by the completely general
equivalence of an (1+1)-dimensional MITP and a PEPS
quantum state in 2-dimensional space that we discussed  in Sect.~\ref{LabelSubsectionUnifyingMITPMixedState}
above. For systems of non-interacting fermions,
the MIPT is described by the transfer matrix of the 
2-dimensional statistical mechanics system representing the (1+1)-dimensional deep circuit, while following
Sect.~\ref{LabelSubsectionUnifyingMITPMixedState}
the quantum state in 2-dimensional space is a PEPS
which,  in the non-interacting fermion case,
can be written in the form of the well-established so-called Chalker-Coddington model (for a more detailed description/review, see below): In the language of the Chalker-Coddington model, the virtual legs at the bonds of the PEPS are the non-interacting fermion (or equivalently Ising model) degrees of freedom, while the physical legs are the $e$ and $m$ degrees of freedom mentioned above (describing $Z_2$ vortices in the Chalker-Coddington model). Measuring (or decohering) the $e$ and $m$ degrees of freedom at the physical legs of this PEPS drives the transitions described by the NLSM field theories discussed in Sect.~\ref{LabelSectionIntroduction} above.
From this perspective it is physically clear that when measurements (or decoherence) of the $e$ (at vertices) and $m$ (at plaquettes) degrees of freedom occurs with equal strength, the system (e.g., the PEPS) possesses statistical KW self-dual symmetry. If on the other hand %only 
measurements (or decoherence) of only, say, $m$ particles occurs, statistical KW self-dual symmetry is maximally violated. These two situations therefore correspond precisely to the critical field theories (ii) and (i) from Sect.~\ref{LabelSectionIntroduction}.

In addition to relating the descriptions of the critical points, 
described by the field theories (i) and (ii)
from  Sect.~\ref{LabelSectionIntroduction},
in terms of (1+1)-deep circuits  of non-interacting fermions via the transfer matrix and quantum wave functions in 2-dimensional space of the Toric Code type, a key focus of the present paper consists in quantitative descriptions of the two universality classes (i) and (ii) of transitions. To put this in context we note that both can be viewed as MIPTs describing an entanglement transition between two area law phases~\footnote{It is well known~\cite{Fidkowski2021howdynamicalquantum} that a volume law phase can generically not occur for non-interacting fermions, while it has been known for 25 years that a critical phase with logarithmic entanglement is forbidden in symmetry class D in replica limits $R\to 1$~\cite{GRL2001}.}. These (1+1)-dimensional MIPTs are known to possess infinite dimensional conformal symmetry~\cite{belavin1984infinite}, a fact which we exploit extensively to characterize the universality classes. In making this connection one needs to realize that the type of conformal field theories (CFTs) at hand are of significantly greater complexity than familiar CFTs of the Ising, Potts models, etc.. This is due to the fact that the intrinsic randomness of the outcomes of the quantum measurements implies that for a fixed set of measurement outcomes (i.e. in a single quantum trajectory) the system lacks translational invariance in the space-time of the deep circuit. There is thus no sense of conformal symmetry (which requires the presence of translational and rotational symmetry) in a fixed quantum trajectory. The only meaningful application of concepts of conformal symmetry can be to averaged observables. A formulation of all such averaged observables is provided by the replica formalism, and in particular for the two critical points (i) and (ii) by their formulation in terms of the critical NLSM field theories that we listed above.

\begin{table}[t]
\centering
\caption{{\bf Overview of the mixed-state criticality originating from Ising.} 
		Shown is a characterization of the three mixed-state critical theories of the phase diagram in Fig.~\ref{fig:1dcircuit}(c),
		and a comparison to the pure-state Ising universality class. 
		The asterisk indicates results with system size $L_x>1000$ taken from Ref.~\onlinecite{Puetz24}.
		Critical exponents: $c_{\rm ent}^{\rm vN}$ and $c_{\rm ent}^{(\infty)}$ 
        characterize the scaling dimensions of the boundary condition changing operators
        that govern the Born-averaged von Neumann and the $\infty$-Rényi entanglement entropies of the boundary 1+1D quantum states. $c_{\rm Casimir}$ is the effective central charge that governs the Shannon entropy density of the 2D bulk. $\Delta_m$ is the
        {\it typical} $m$-vortex scaling dimension that governs the change of the bulk entropy upon changing the boundary condition. 
	}
\begin{tabular}[t]{c | c  c  c | c}
\toprule
criticality 		& {\bf Nishimori	}		& {\bf  weak self-dual }		& {\bf Ising+ }				& Ising 	\\
\midrule
1D quantum 	& mixed 				& mixed 				& mixed 				& pure 	\\[2mm]
self-dual 		& \multirow{2}{*}{broken} 				& \multirow{2}{*}{weak} 				&  \multirow{2}{*}{broken} 				& \multirow{2}{*}{strong} 	\\
symmetry 		&  				&  				&   				&  	\\
\midrule
2D stat mech 	& disordered 			& disordered 			& disordered			& clean 	\\[2mm]
disorder  		& \multirow{2}{*}{$\mathcal{Z}(m)$} 		& \multirow{2}{*}{$\mathcal{Z}(em)^2$} 	& \multirow{2}{*}{$\mathcal{Z}(m)^2$} 	& \multirow{2}{*}{$0$} 	\\
probability		&					& & & \\[2mm]
$\nu$ 		& $1.52(2)^*$			& $1.72(8)$ 			& $1.00(5)$ 			& $1$ 	\\
\midrule 
conformal 		& \multirow{2}{*}{non-unitary} 	&\multirow{2}{*}{non-unitary} 			& \multirow{2}{*}{non-unitary} 		& \multirow{2}{*}{unitary} 	\\
symmetry 		&					& & & \\[2mm]
central charge 	& 0~\footnote{See Sec.~\ref{sec:classical_statmech} for a discussion of the definition of the free energy.}					& 0 				& 0 				& $1/2$ 	\\
$c_{\rm ent}^{\rm vN}$ 	& $0.410(1)^*$		& $0.795(1)$ 			& $0.310(2)$ 			& $1/2$	\\
$c_{\rm ent}^{(\infty)}$ 	& --- 				& $0.484(1)$ 			& --- 					& $1/4$	\\
$c_{\rm Casimir}$ 		& $0.464(4)^*$ 		& $0.447(1)$			& --- 					& $1/2$	\\
$\Delta_{m}$ 	& $0.341(1)^*$ 		& $0.156(1)$ 			& --- 					& $1/8$	\\
\bottomrule
\end{tabular}
\label{tab:criticalpoints}
\end{table}

\begin{figure*}[tb!] 
   \centering
   \includegraphics[width=\textwidth]{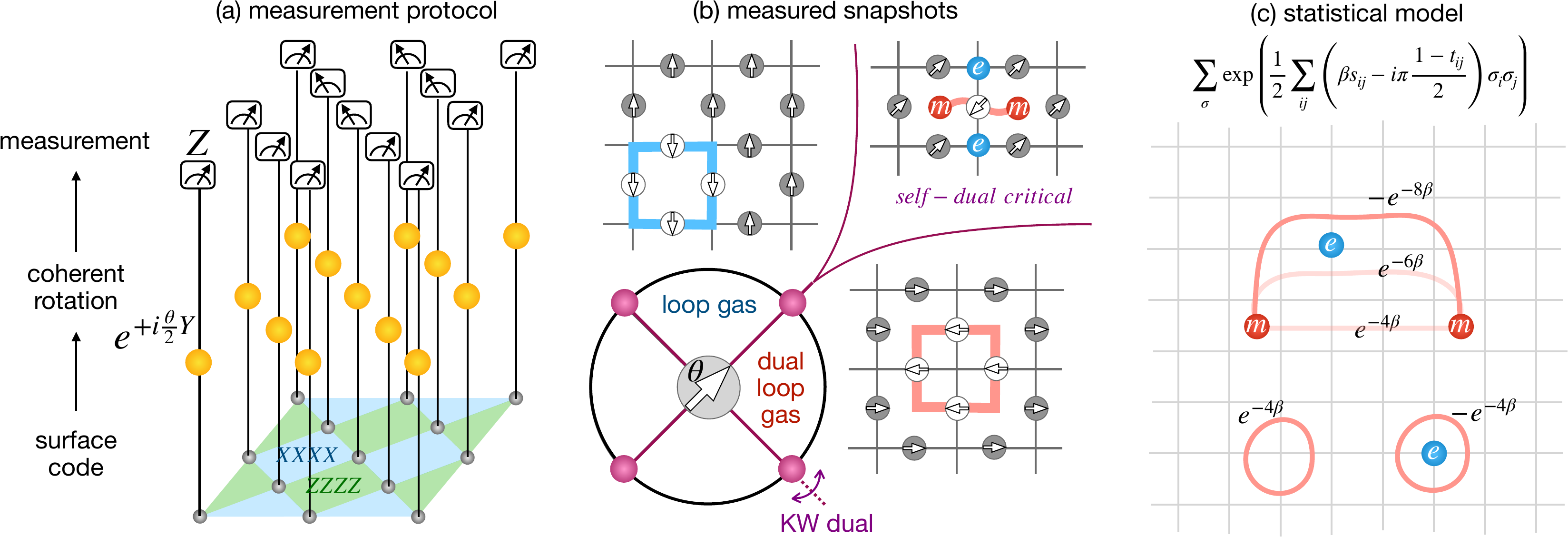} 
   \caption{{\bf Surface code with coherent rotation prior to measurement}. (a) The quantum circuit. The combination of coherent rotation and $Z$-measurement is effective measurement of qubit in $\cos\theta Z + \sin\theta X$ basis. (b) The disk phase diagram parametrized by the measurement angle $\theta$ are separated into four quadrants, separated by self-dual critical points, where the snapshots are schematically shown. 
   Measuring surface code in $Z$ basis results in a loop gas ensemble, where the blue shaded lines illustrate the loop defined by spin down. 
   Measuring surface code in $X$ basis results in a dual loop gas ensemble, where the red shaded lines illustrate the dual loop defined by spin left as eigenstate of Pauli $X$ with eigenvalue $-1$. 
   Measurement at the magic angle $\theta=\pi/4$ is self-dual, which not only excites the $e$ particle but also the $m$ particle, on equal footing. The Born's probability of the measurement outcomes is dictated by the statistical model in (c). (c) The statistical model is a random Ising model with both $e$ and $m$ particle disorder, in Eq.~\eqref{eq:statmech}. The schematic shows fluctuating strings (defined by $s_{ij}\sigma_i\sigma_j=-1$) that have to be closed or terminate at the $m$ particle. The Boltzmann weight of the string is given by the string tension $e^{-\beta}$ coupled to the length $l$ of the string. When the string passes through an $e$ particle, a $\pi$ phase is attached to the Boltzmann weight. 
  }
   \label{fig:surfacecode}
\end{figure*}

Alternatively, our protocol can also be viewed as dephasing a toric code after applying the coherent ``error"~\cite{teleportcode, bravyi2018correcting, Iverson2020, Beri23toriccodecoherenterror, Gullans23magic, Turkeshi24coherenterror, Beri24coherenterror, Behrends24Xcohernoise, Behrends24coheralldir, Moon24surfacecodecohererr, Ippoliti24designsfromerror, bao2024codecohererr, Ippolitti24coherenterror}, i.e. a uniform Pauli rotation $e^{i\theta \sum_{ij}Y_{ij}/2}$ at each link $\langle i,j \rangle$, see Fig.~\ref{fig:surfacecode}. Thus the final state can be viewed as $\mathcal{N}(e^{-i\theta \sum_{ij}Y_{ij}/2}\ket{\psi})$, where $\mathcal{N}$ refers to measurement or maximally dephasing noise that erases all off-diagonal elements from the density matrix in the $Z$ basis, and $|\psi\rangle$ refers to the perfect surface code state. 
As a purely classical mixed state, it can also be viewed as an ensemble of projected states: $\{\prod_j(1+s_j Z_j)e^{-i\theta \sum_{ij}Y_{ij}/2}\ket{\psi}\}$, whence the dephasing noise is interpreted as projective measurements of qubits at sites $j$, and each wave function norm determines the Born probability. 
The same type of coherent errors, but around the $Z$ or $X$ axis, was
studied in Ref.~\cite{bravyi2018correcting, Iverson2020, Beri23toriccodecoherenterror, Beri24coherenterror, Behrends24Xcohernoise, Behrends24coheralldir}, where the subsequent measurement is upon the star or plaquette stabilizer operators~\cite{Kitaev2003}: $\{\prod_\square(1\pm \prod_{j\in \square} X_j)e^{-i\theta \sum_{ij}Z_{ij}/2}\ket{\psi}\}$, instead of measuring Pauli $Z$ of each qubit. The distinction between
these works and ours thus lies in the type of measurement or channel that is applied subsequent to the action of the coherent error. Since the coherent errors are simply {\it local unitary} transformations, they alone do not cause the phase transition. Instead, it is the combination of the coherent errors and the subsequent channel or measurements that drives the transition at the threshold.

Furthermore, our consideration of coherent rotation by $Y$ preserves the electric-magnetic self-duality~\cite{Zhu19ToricCode, Grover24selfdual,Hsieh24miptcode,Tupitsyn2010,Nahum21selfdual, teleportcode} at $\theta=\pi/4$ is the criticality we unveil in this paper. 

Using extensive numerical simulations that combine elements from tensor network, Gaussian fermion, and Monte Carlo sampling, we estimate the critical exponents of this novel self-dual quantum critical point as well as the other two critical theories, relying primarily on the conformal scaling of Shannon entropy, entanglement entropy, and coherent information measures. In addition, we also show the single-particle and many-body Lyapunov exponents that form a spectrum akin to the energy spectrum in equilibrium. 
A summary of the three types of criticality is given above in Table~\ref{tab:criticalpoints}. 
Having established that these three critical theories fall into distinct universality classes, 
we explore the renormalization group (RG) flow between them by tracking the changes of critical exponent and central charge estimates as we break the self-dual symmetry upon moving away from the center line of our phase diagram.
Notably, we find clear numerical evidence that incoherent noise initiates an RG flow from self-dual to Nishimori criticality.
On the other hand, upon introducing a coherent wave function deformation that breaks
self-duality, we discover another critical theory (in the limit of infinitely strong deformation) that
can be described by a non-unitary conformal field theory (CFT) that can analytically 
be shown to retain
the unitary Ising CFT in a subset of its operators,
which is why we have dubbed it ``Ising+" criticality. 
In our phase diagram, we find that this critical point is an unstable fixed point with numerical evidence for an RG flow back to the weak self-dual criticality (Table~\ref{tab:criticalpoints}.)

The remainder of this manuscript illustrates this very general physics in detail for the example of  the specific system which forms the topic of this paper,
with Sec.~\ref{sec:model} introducing our mixed-state preparation protocol and its decoherence channels. 
Instead of starting directly from the toric code, we start from a 2D cluster state on the Lieb lattice, and project out all the qubits. It is known that measuring the 2D cluster state can result in the toric code, and thus this approach automatically involves the toric code as an intermediate state, but allows us to show the decoherence induced criticality in a bigger context.
We also briefly point out the underlying $Z_2$ gauge symmetry and its tensor network representation.
In Sec.~\ref{sec:representations} we then proceed to introduce four alternative model representations
in terms of a (1+1)D monitored circuit, a 2D classical stat mech model, followed by a (1+1)D Majorana circuit and a two-dimensional Chalker-Coddington network model,
along with a discussion of Kramers-Wannier self-duality in all four formulations.
The self-dual critical state is discussed in detail in Sec.~\ref{sec:self-dual} based on its coherent information, bulk Shannon entropy, and boundary entanglement entropy.
In Sec.~\ref{sec:phase_diagram} we then discuss the entire phase diagram of Fig.~\ref{fig:noisyphasediagram} in detail with a particular emphasis on the RG flows between the various critical theories. We close with a broader discussion and an outlook in Sec.~\ref{sec:discussion}. 

\setcounter{section}{1}

%%%%%%%%%%%%%%%%%%%%%%%%%%%%%%%%%%%%%%%%%%%%%%%%%%%%%%%%%%%%%%%%%%%%%%%%%%%%%%%%%%%%%%%
\section{Model}
\label{sec:model}
%%%%%%%%%%%%%%%%%%%%%%%%%%%%%%%%%%%%%%%%%%%%%%%%%%%%%%%%%%%%%%%%%%%%%%%%%%%%%%%%%%%%%%%

\begin{figure*}[t!]
   \centering
   \includegraphics[width=\textwidth]{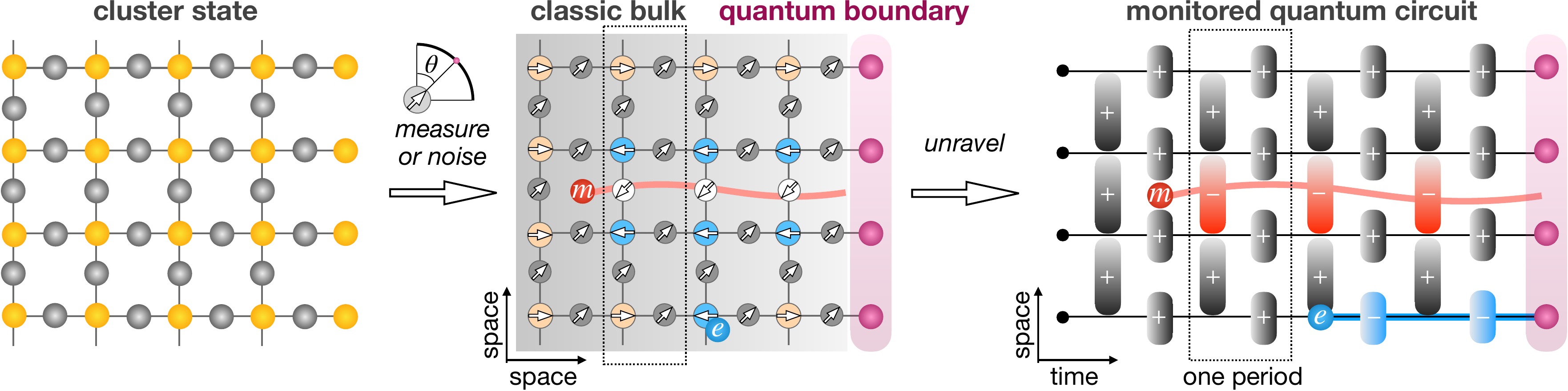}
   \caption{
      {\bf 2D protocol and effective 1D circuit}. 
      (a) A cluster state on the Lieb lattice is prepared by controlled-Z gates for every nearest neighbor pair between the orange and green qubits. It has a $Z_2^{(0)}\times Z_2^{(1)}$ SPT, where $Z_2^{(0)}$ symmetry applies to the site (orange) qubits and $Z_2^{(1)}$ symmetry applies to the bond (gray) qubits. 
      (b) 
      The bulk qubits are collapsed (thus decohered)
      by measurements into classical snapshots of spin configurations. The site qubit at site $j$ collapses into the $X_j$ eigenstate
      $\ket{+}$, and is labeled by a right arrow in an orange disk, while $\ket{-}$ is labeled by
      a left arrow in a blue disk. 
      The qubit on the bond between sites $i$ and $j$ collapses to one of the eigenstates for rotated Pauli operator $\sigma_{ij}^\theta \equiv \cos(\theta)Z_{ij}+\sin(\theta)X_{ij}$,
      denoted by the tilted up arrow in green circle, or down tilted arrow in red circle. We label the snapshot by $e$ particle (for a left arrow at site) and $m$-string (for a tilted down arrow at bond). The endpoint of an $m$-string is an $m$-particle. 
      The rightmost column of qubits remain the unmeasured quantum boundary, highlighted by purple disks.
      While the bulk becomes a classical mixed state, the boundary quantum state is a  pure quantum state conditioned upon the bulk snapshot. 
      (c)
      By fixing an arbitrary bulk snapshot, the boundary quantum state is effectively generated by a 1D deep monitored quantum circuit, where the measurement record corresponds to the bulk classical configuration. The derivation follows from the tensor network representation discussion in Sec.~\ref{sec:derivetns}. 
   }
   \label{fig:classicquantumclusterroute}
\end{figure*}

While we emphasize on the measurement and decoherence of toric code above, from now on we take the cluster state on a Lieb lattice as a simpler starting point. This is because measuring out a subset of the qubits of the cluster state results in a toric code~\cite{Raussendorf2005}. Thus they share the same decoherence induced criticality, which we will detail. Generally, this is because both of them can be interpreted as a $Z_2$ lattice gauge theory with matter field~\cite{Kitaev2003}. 

In particular, we show that self-duality can naturally appear (even at zero temperature) in the form of an average ``weak" symmetry~\cite{Prosen12weaksym, Jiang14weaksym, Gorshkov20weaksym} when decohering 
(i) a short-range entangled symmetry protected topological state (SPT~\cite{Chen2010}),
(ii) a long-range entangled toric code~\cite{Kitaev2003},
or (iii) a measurement prepared cat state~\cite{GHZ, NishimoriCat}, either by projective measurement or by maximal dephasing noise. 
Specifically, we demonstrate that the {\it bulk} decoherence of a cluster state~\cite{Raussendorf2001,Hsieh23mixedstatecriticality, Grover24separable, Hsieh24spacetimemarkov}, which on a Lieb lattice exhibits $Z_2^{(0)}\times Z_2^{(1)}$ (0-form and 1-form) SPT~\cite{Verresen22higgs, CZX11, LevinGu12, Raussendorf17compspt}, 
can lead to a mixed quantum critical state {\it on its boundary}, described by a self-dual variant of the random bond Ising model. 
The same boundary state can appear upon decohering a toric code or a mixed Greenberger-Horne-Zeilinger (GHZ~\cite{GHZ}) state (dubbed ``Nishimori's cat" state in Refs.~\onlinecite{NishimoriCat, Chen24nishimori}). 
This trio of states is, of course, not incidental as all three of them -- the cluster state on a Lieb lattice, the toric code on one of its sublattices, and the GHZ/cat state on the other sublattice -- can be unified via a $Z_2$ lattice gauge theory~\cite{Wegner71duality, Kogut79rmp, FradkinShenker79, stamp10toriccodefield, Verresen22higgs}, where the site and bond qubits in the underlying Lieb lattice are interpreted as matter and $Z_2$ gauge fields, respectively. 
When the matter field is decohered in a first step, one obtains a deconfined gauge theory as an intermediate state, which is the topological toric code~\cite{Kitaev2003}, see Fig.~\ref{fig:clusterroute}. 
If, on the other hand, the gauge field is first decohered, then one obtains a gauge-symmetrized $Z_2$ ferromagnetic order, i.e.\ the state dubbed Nishimori's cat~\cite{NishimoriCat}. 
Finally, if both the matter and the gauge degrees of freedom are decohered in the bulk, one might naively expect a trivial state. 
However, the emergence of self-dual symmetry prohibits triviality and protects criticality: 
a vestigial quantum critical state 
persists at the boundary --  this is the aforementioned critical boundary state that reveals itself upon close inspection of the mixed-state ensemble after bulk decoherence.
Fig.~\ref{fig:clusterroute} summarizes the conceptual idea of decohering the SPT order of a cluster state via the quantum circuit schematically illustrated in Fig.~\ref{fig:classicquantumclusterroute}. 
In panel (b), we show the 
quantum phase diagram where self-duality is preserved as a weak symmetry that transforms $\theta$ to $\pi/2-\theta$, such that the point $\theta=\pi/4$ is invariant under duality transform. 
Introducing both incoherent noise (parametrized by $p_s$) and a coherent deformation (parameterized by $p_\eta$) we have explored the phase diagram of panel (c), 
which exhibits a critical point whose universality and field theoretical description is distinct from both the Nishimori criticality induced by incoherent noise 
and yet another type of critical behavior, dubbed ``Ising+" induced by the coherent deformation -- both of which break the weak self-duality, present along the center line
of our phase diagram.

%%%%%%%%%%%%%%%%%%%%%%%%%%%%%%%%%%%%%%%%%%%%%%%%%%%%%%%%%%%%%%%%%%%%%%%%%%%%%%%%%%%%%%%
\subsection{Protocol: preparation and decoherence}\label{Subsec:protocol}
%%%%%%%%%%%%%%%%%%%%%%%%%%%%%%%%%%%%%%%%%%%%%%%%%%%%%%%%%%%%%%%%%%%%%%%%%%%%%%%%%%%%%%%
Our protocol comprises two principal steps: 
(i) the preparation of an entangled resource state -- the cluster state, which, 
in step (ii), is subject to decoherence by either projective measurements or $50\%$ Pauli noise in some orientable basis direction.
For the first step we prepare a short-ranged entangled cluster state~\cite{Raussendorf2001}: $\prod CZ\ket{+}^{\otimes N}$ using a finite-depth circuit on a (square) Lieb lattice as illustrated in Fig.~\ref{fig:classicquantum}, where we distinguish
between the bond qubits $Z_{ij}$ and the site qubits $Z_j$ 
with $i$ and $j$ denoting sites of the square lattice. 
Secondly, we dephase all the qubits in the bulk in a designated basis, by projectively measuring the site qubits in the $X$ direction and the bond qubits in a tilted direction
\[
	\sigma^\theta = \cos(\theta) Z + \sin(\theta) X \,.
\] 
The ``measurement angle" $\theta$ can also be understood as the angle of a coherent unitary rotation error around the $Y$-axis right before a measurement in the $Z$-basis~\cite{Verresen2021cat, NishimoriCat, JYLee}
(not to be confused with the topological $\Theta$-angle of the NLSM).  Schematically, an elementary building block of this circuit for a given bond then takes the following form,
\begin{equation*}
\includegraphics[width=.5\columnwidth]{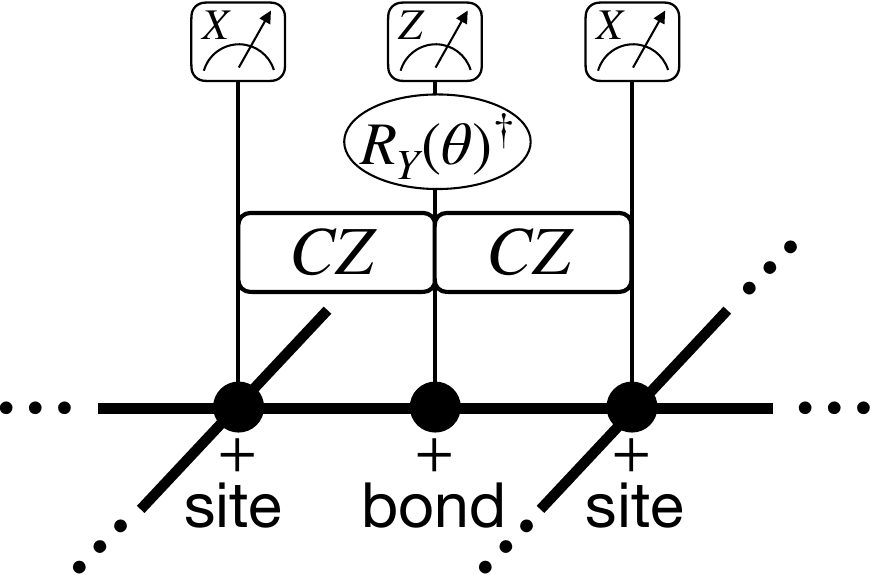} \ ,
\end{equation*}
for a bond of the square lattice with two qubits at the sites (denoted by $i$ and $j$) and one qubit on the center of the bond (denoted by $ij$). 
Here all qubits (at the sites as well as the bonds) are originally initialized in $X$-basis eigenstates $\ket{+}=X\ket{+}$, then subjected to two controlled-Z (CZ) gates
$CZ\times CZ={\rm exp}(-i \pi (1-Z_i)(1-Z_{ij})/4){\rm exp}(-i \pi (1-Z_j)(1-Z_{ij})/4)$ (as depicted in the unnumbered Figure above) that (maximally) entangle the nearest neighboring site- and bond-qubits. Before the final round of measurements
in the $X_j$ (at sites $j$) and $Z_{ij}$ (at bonds $(i,j)$) bases,
respectively, the bond qubit is rotated via  $R_Y(\theta)^\dag=\exp\left(i \theta Y/2\right)$ by the measurement angle $\theta$. 
The resulting state can be written as a random tensor network whose building block~\footnote{
   Here the Hadamard matrix $[1,1,;1,-1]$ on the bond captures the diagonal elements of the $CZ$ gate: ${\rm diag}(CZ)=[1,1,1,-1]$. } is
\begin{equation*}
\includegraphics[width=.5\columnwidth]{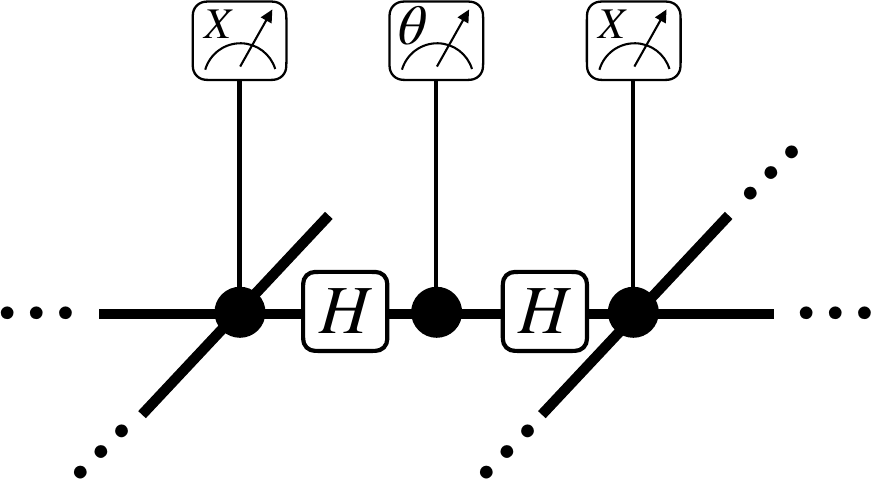} \ .
\end{equation*}
Depending on the measurement outcome, Pauli $X$ or $Z$ matrices are injected 
into the virtual leg of the tensor network, as summarized in the following figure
~\footnote{The rank-5 delta tensor with 5 legs can be viewed as a 5-body GHZ state, and projecting the physical leg into Pauli $X$ eigenstates results in the following two possible states: $\bra{+}(\ket{\uparrow\uparrow\uparrow\uparrow\uparrow}+\ket{\downarrow\downarrow\downarrow\downarrow\downarrow} )
= 
(\ket{\uparrow\uparrow\uparrow\uparrow}+\ket{\downarrow\downarrow\downarrow\downarrow} )$; 
and $\bra{-}(\ket{\uparrow\uparrow\uparrow\uparrow\uparrow}+\ket{\downarrow\downarrow\downarrow\downarrow\downarrow} )
= 
(\ket{\uparrow\uparrow\uparrow\uparrow}-\ket{\downarrow\downarrow\downarrow\downarrow} )$ if the measurement outcome is negative. 
Similarly, projecting a 3-body GHZ state into eigenstate of Pauli $\sigma^\theta$ results in a two-body entangling state conditioned upon $\theta$. 
},
\begin{equation}
   \includegraphics[width=\columnwidth]{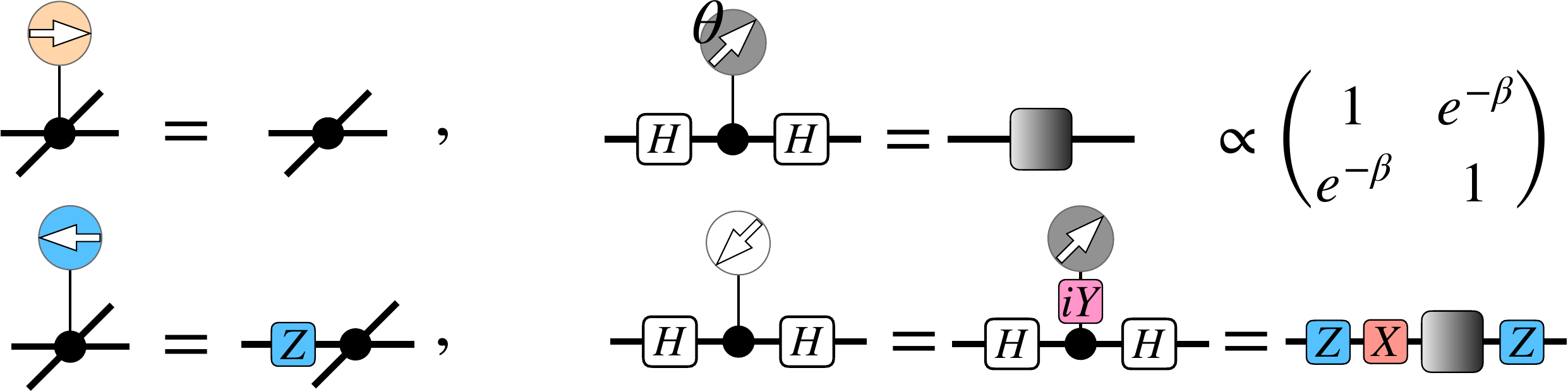} \ .
   \label{eq:tnderivation}
\end{equation}
Here the orange disk with right arrow denotes the projection to $\bra{+}$, and blue disk with left arrow to $\bra{-}$, and the green disk with northeast arrow denotes projection to the positive eigenstate of Pauli matrix $\sigma^\theta$, while the purple disk with southwest arrow to the negative eigenstate. Here $H=(X+Z)/\sqrt{2}$ denotes
the 2-by-2 Hadamard gate matrix acting on a bond.
The parameter $e^{\beta}= \tan(\frac{\theta}{2}+\frac{\pi}{4})$, (or equivalently $\tanh\beta = \sin\theta$), gives the element of 2-by-2 transfer matrix on a single bond.

Note that, from a symmetry perspective, there is a distinction of the site versus bond qubits -- the SPT order of the cluster state on the underlying Lieb lattice is protected by a 0-form $Z_2$ symmetry for the site qubits and a 1-form magnetic $Z_2^{(1)}$ symmetry for the bond qubits~\cite{Verresen22higgs}. 
Due to this inequivalence of the symmetry forms, the ordering of measuring the site and the bond qubits  
leads to two distinct intermediate states when decomposing it into two steps:
\begin{itemize}
\item[(i)]
If first the bond qubits are  measured and dephased~\cite{NishimoriCat, JYLee}, this leaves a 2D state which can be
tuned (by tuning the measurement angle $\theta$)
between a product state and a measurement induced random GHZ~\cite{GHZ} state, dubbed ``Nishimori cat" in Ref.~\cite{NishimoriCat}. When all the post-measurement states are collected as a mixed state, it exhibits strong-to-weak spontaneous symmetry breaking (SW-SSB) order of the $Z_2$ Ising symmetry~\cite{You24weaksym, Wang24strtowksym}~\footnote{Strictly speaking, a cat state does not break the symmetry on the microscopic level, but we still relate such cat state to an SSB order, because it has the defining long-range correlation.}.
This state can also be viewed as a randomized version of a Rokhsar-Kivelson state~\cite{Henley04RK, Fradkin04RK, Troyer10topocrit}, 
which can be cast as a 2D projected entangled pair state (PEPS)~\cite{Cirac06}. 
\item[(ii)]
 If the site qubits are measured first instead, this leaves a 2D toric code with $e$ particles at random but known locations~\cite{Raussendorf2005}, as eigenstates of the toric code Hamiltonian~\cite{Kitaev2003}. One can apply a Pauli string of $X$ on certain bond qubits to pair up and remove the $e$ particles~\footnote{Note that the star stabilizer that defines the $e$ particle is in the $Z$ basis following the convention of cluster state~\cite{Raussendorf2005}, compared with the convention in Ref.~\cite{Kitaev2003}.} for a clean toric code state. Here we donot need to do so, because we will measure the bond qubits in $\theta$ angle afterwards, which would also generate the $e$ and $m$ particles. The final critical state is equivalently labeled by configurations of $e$ and $m$ particles. 
\end{itemize}
Given these two routes through distinct intermediate states, the final state of our protocol can be equally interpreted as the vestigial boundary {\it quantum} state after the bulk decoherence of a cluster state with $Z_2^{(0)}\times Z_2^{(1)}$ SPT, a Nishimori cat state with SW-SSB, or a toric code with topological order.  

%%%%%%%%%%%%%%%%%%%%%%%%%%%%%%%%%%%%%%%%%%%%%%%%%%%%%%%%%%%%%%%%%%%%%%%%%%%%%%%%%%%%%%%
\subsection{$\mathbf{Z_2}$ gauge perspective for the intermediate state}
%%%%%%%%%%%%%%%%%%%%%%%%%%%%%%%%%%%%%%%%%%%%%%%%%%%%%%%%%%%%%%%%%%%%%%%%%%%%%%%%%%%%%%%

\begin{figure}[th] 
   \centering
   \includegraphics[width=\columnwidth]{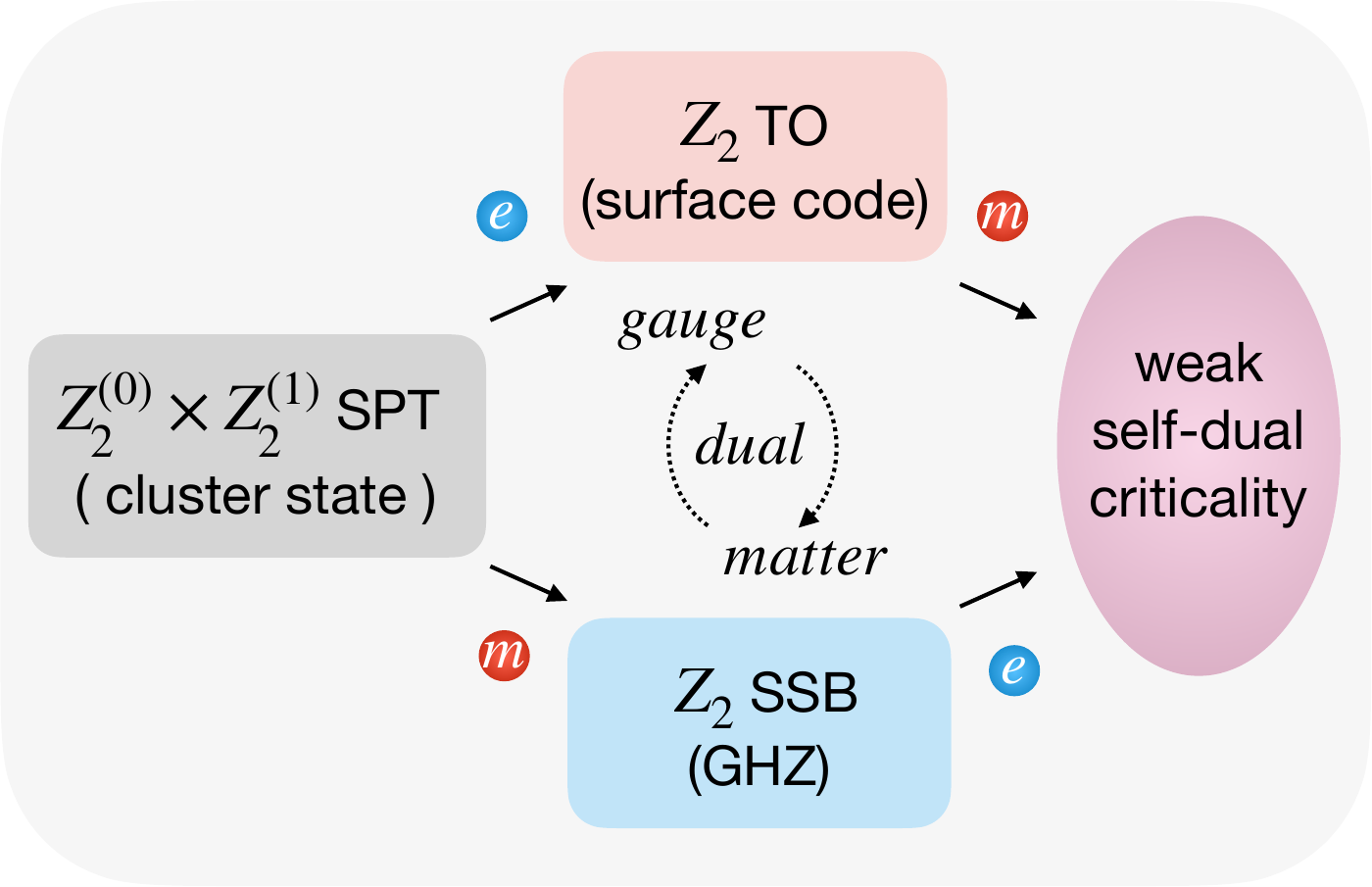} 
   \caption{{\bf Decoherence schematic}. 
   A $Z_2^{(0)}\times Z_2^{(1)}$ SPT can be realized as a cluster state on a Lieb lattice (with site and bond qubits), which can be viewed as a lattice gauge state with matter electric charge $e$- and gauge magnetic flux $m$-vortices, that are dual to each other. 
   When the site qubits are collapsed (decohered) / measured, 
   $e$ is uncondensed, resulting in average or correctable topological order as a deconfined gauge state that spontaneously breaks the 1-form $Z_2^{(1)}$ symmetry. 
   When the bond qubits are collapsed (decohered) / measured, 
   $m$ is uncondensed and leads to the GHZ-like state as an average $Z_2^{(0)}$ spontaneous symmetry breaking order. When both $e$ and $m$ particles are uncondensed by measurement,
   it can leave a boundary critical state with weak self-dual symmetry and weak $Z_2^{(0)}$ symmetry. 
  }
   \label{fig:clusterroute}
\end{figure}

For our later discussion it is useful to connect the states prepared by our protocol to the fundamental physics of a $Z_2$ gauge theory~\cite{Kogut79rmp}.
To do so, let us start by noting that the cluster state on the Lieb lattice satisfies the following two sets of stabilizers
\begin{equation*}
\includegraphics[width=.6\columnwidth]{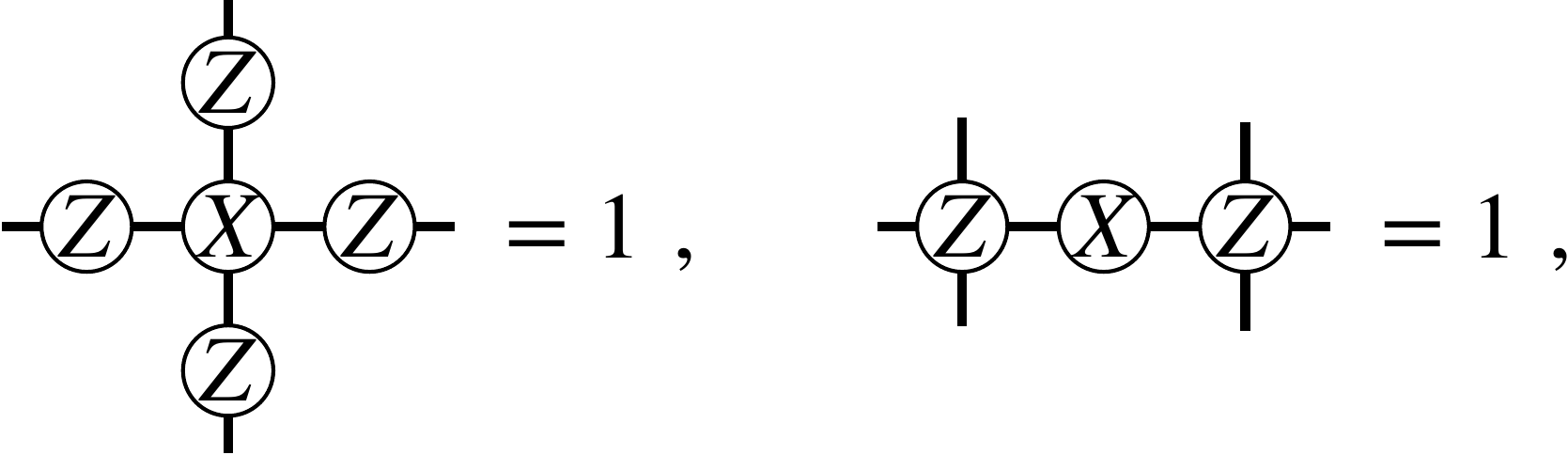} 
\end{equation*}
for the site and the bond qubits of the square lattice, respectively. The vertex stabilizers can be interpreted as a $Z_2$ Gauss law, where $\ket{X_j=-1}$ for the site qubit denotes a $Z_2$ {\it matter charge} usually labeled as $e$-vortex, and $Z_{ij}$ operator for the bond qubit denotes a corresponding electric field. Consequently, $\ket{X_{ij}=-1}$ labels a nonzero gauge field, and the bond stabilizer $Z_i X_{ij}Z_j=1$ describes the minimal coupling between matter and charge, where the matter hopping is coupled with the gauge connection $X_{ij}$ operator~\cite{Kogut79rmp}. When the matter charge hops around a plaquette, it picks up a negative sign factor if the Wilson loop $\prod_{\langle ij\rangle \in \square} X_{ij}=- 1$, which is usually dubbed an $m$-vortex, denoting a {\it magnetic vortex}. 

When one projectively measures the matter qubit in the $X$ basis, the matter charge is frozen into a classical configuration, leaving the bond qubits to form a pure gauge theory with random background charges but free of magnetic vortices. This is a random topological toric code state, see the top path in Fig.~\ref{fig:clusterroute}. 
When one instead projectively measures the gauge qubit in a tilted basis, it is the gauge field that is quenched, leaving the site qubits as matter degrees of freedom forming a Nishimori's cat state~\cite{NishimoriCat}.

%%%%%%%%%%%%%%%%%%%%%%%%%%%%%%%%%%%%%%%%%%%%%%%%%%%%%%%%%%%%%%%%%%%%%%%%%%%%%%%%%%%%%%%
\subsection{Tensor network state representation}
%%%%%%%%%%%%%%%%%%%%%%%%%%%%%%%%%%%%%%%%%%%%%%%%%%%%%%%%%%%%%%%%%%%%%%%%%%%%%%%%%%%%%%%
We can label the bulk classical bits after measurement in the following way: $e_j=0(1)$ if the site qubit at site $j$ is measured to be $X_j=\pm 1$ (after absorbing the $Z$ conditioned upon the adjacent bond qubit measurement outcomes); if the bond qubits at the center of the bond between site $i$ and $j$ are measured (after having been rotated by an angle $\theta$)  to be $Z_{ij}=- 1$, we associate that with an $m$ string crossing the bond, and the end point of a string with $Z_{ij}=-1$ all along the way is denoted by an $m$-vortex, $m_p=1$ (where $p$ denotes the plaquette). 

This post-measurement state exhibits, at its boundary, a 1D quantum state $\ket{\psi(em)}$, which can be represented, by assembling the building blocks in Eq.~\eqref{eq:tnderivation}, as an {\it exact} random tensor network state of the form
\begin{equation}
\includegraphics[width=\columnwidth]{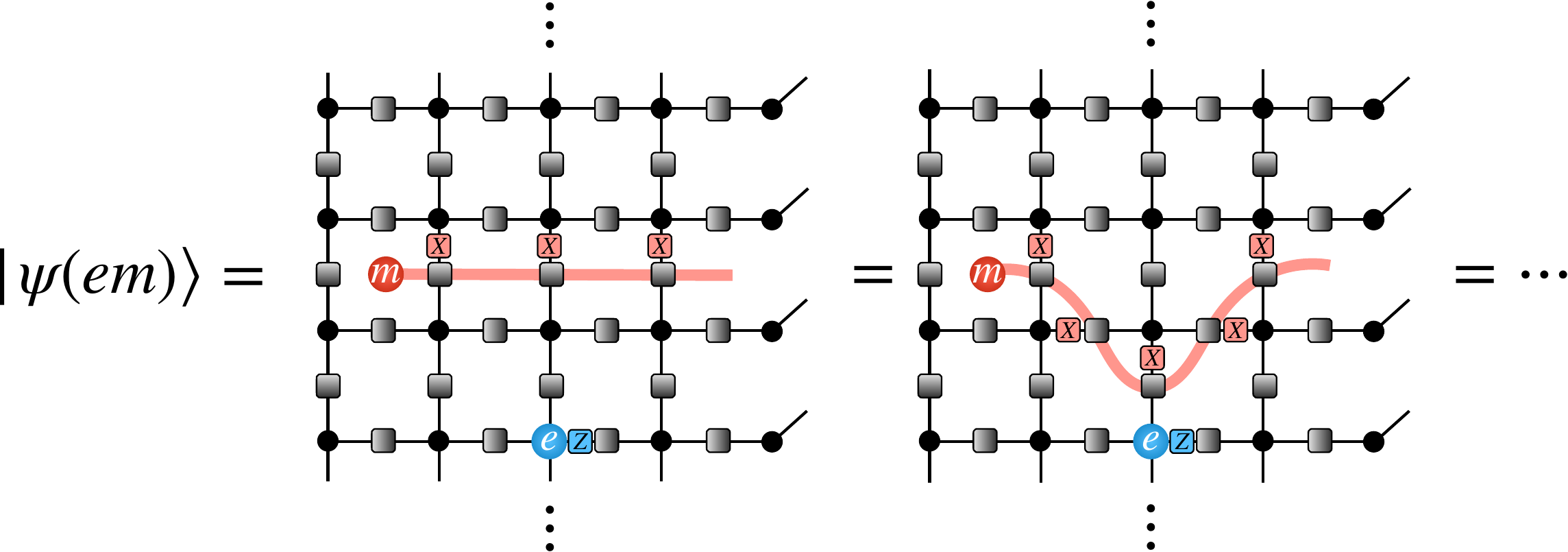} \ ,
\label{eq:gaugetransform}
\end{equation}
where each node at the vertex is a diagonal delta tensor~\footnote{A diagonal tensor of rank 4 can be expressed as $T_{i,j,k,l}=\delta_{i,j}\delta_{j,k}\delta_{k,l}$, where $\delta_{i,j}$ is the Kronecker delta function, and $i,j,k,l$ are the binary indices that take value $0$ or $1$. },
and each box on the bond is a 2-by-2 matrix $e^{\beta/2} + e^{-\beta/2} X$, with $\tanh\beta =\sin \theta$. 
A negative measurement outcome on the bond qubit injects an $X$ to the bond, while a negative measurement outcome on the site qubit injects a $Z$ to the vertex. 
 Note that this tensor network state is un-normalized, since its norm $P(em) = \langle \psi(em)|\psi(em)\rangle$ captures the measurement probability dictated by Born's rule. 

Because of the underlying $Z_2$ gauge symmetry, one can fluctuate the $m$-string across the lattice~\cite{NishimoriCat}, without altering the post-measurement boundary state. The latter is only determined by the {\it gauge invariant} $m$-vortex configuration~\cite{Preskill2002}. This can be explicitly verified by propagating the $X$-string through the tensor network~\cite{haegeman2015shadows}. 
Since a gauge transformation can be independently generated by creating a local $X$ loop around any vertex, one can count a total number of $2^{L_xL_y}$ gauge-equivalent configurations that share the same boundary quantum state $\ket{\psi(em)}$. We will therefore label the bulk bits only by their $em$-vortex configuration for brevity. 
Note that it might seem that if one starts from the clean toric code state, then $e$ and $m$ particles are constrained to form tightly bound pairs as shown in Fig.~\ref{fig:classicquantum}(b). However, we note that in an infinite large system, one can in general construct any configurations of $e$ and $m$ particles, and the probability is simply given by the tensor network. 

The final (unnormalized) state can then be compactly written as, as a special case of Eq.~\eqref{eq:PEPSmixedstate}
\begin{equation}
\rho = \sum_{em} \ketbra{em}_{C} \otimes \ketbra{\psi(em)}_{Q} \ ,
\label{eq:mixedstate}
\end{equation}
where $\ket{em}$ refers to the classical (orthonormal) states of the bulk bits, denoted by $C$, corresponding to the measurements record. The remaining quantum bits at the boundary are denoted by $Q$. 
If the 2D state is placed on a cylinder, i.e.\ one applies periodic boundary conditions in the vertical direction, then there is one additional (very big) plaquette at the left boundary with $L_x$ number of edges (see Fig.~\ref{fig:pureSvN}a inset for a schematic), where the absence (presence) of an $m$-vortex in this plaquette labels the even (odd) sector of the mixed state. 

\subsection{Symmetries}
The state~\eqref{eq:mixedstate} possesses a strong $Z_2$ Ising symmetry and a weak Kramers-Wannier self-duality symmetry. The former acts on the qubits on the boundary
\begin{equation}
\label{LabelEqIsing Symmetry}
\left(\prod_{j\in Q} X_j\right) \rho = \rho = \rho  \left(\prod_{j\in Q} X_j\right) \ ,
\end{equation}
which will be shown to exhibit a SW-SSB transition by tuning the measurement angle $\theta$. 
The duality symmetry is inherited from the pre-measurement state, which in the toric code or gauge theory is an electric-magnetic duality~\cite{Zhu19ToricCode,teleportcode}
\begin{equation}
{\rm KW}: \quad e \leftrightarrow m \ ,\quad \ \theta\leftrightarrow \pi/2-\theta \ ,
\label{eq:KW}
\end{equation}
where the operator ${\rm KW} $ will be explicitly written in the next section. 
The presence of the KW duality dictates a
KW symmetric
critical state at $\theta=\pi/4$, which is invariant under the duality transformation and can be interpreted as the gapless phase of matter protected by such a symmetry~\cite{Wen20categorysym, Ning24categorysym}. 
Under such a KW transformation, each post-measurement pure state at the boundary is mapped to its duality counterpart
\begin{equation}
{\rm KW} \ket{\psi(em; \theta)} = \ket{\psi(me; \pi/2-\theta)} \ ,
\end{equation}
which means the presence of $em$ disorder breaks the duality as a strong symmetry for the pure state. Nonetheless, the mixed state that encapsulates
all the weighted quantum trajectories
is invariant under a {\it weak} symmetry~\cite{Prosen12weaksym, Jiang14weaksym, Gorshkov20weaksym} version of the self-duality at $\theta=\pi/4$
\begin{equation}
{\rm KW} \rho \neq \rho \ ,\qquad {\rm KW} \rho {\rm KW} = \rho,
\ .
\end{equation}
as follows immediately from Eq.~\ref{eq:mixedstate}.
Note that the weak symmetry of the density matrix is also sometimes referred to as an ``average" symmetry common in open quantum systems described by a Lindbladian or density matrix~\cite{Prosen12weaksym, Jiang14weaksym, Gorshkov20weaksym, Wang23averspt, Lee23decoher, Wang23aversym, Xu24higherformweaksym, You24weaksym, Luo24weaksym}. 
As its strong symmetry counterpart, the weak self-duality here has the same predictive power: that it maps the SW-SSB phase at $\theta>\pi/4$ to the trivial phase at $\theta<\pi/4$, and thus the mixed quantum state at the self-dual $\theta=\pi/4$ point 
remain critical, assuming the two phases are separated by a continuous phase transition, in the presence of randomness. Namely, the otherwise {\it pure} Ising critical state with strong self-duality is turned into an intrinsically {\it mixed} quantum critical state with weak self-duality.

%%%%%%%%%%%%%%%%%%%%%%%%%%%%%%%%%%%%%%%%%%%%%%%%%%%%%%%%%%%%%%%%%%%%%%%%%%%%%%%%%%%%%%%
\subsection{Mixed state dynamics}
%%%%%%%%%%%%%%%%%%%%%%%%%%%%%%%%%%%%%%%%%%%%%%%%%%%%%%%%%%%%%%%%%%%%%%%%%%%%%%%%%%%%%%%
Finally, note that the object of interest in the following is the total mixed state of the bulk {\it and} the boundary, Eq.~\ref{eq:mixedstate},
-- we will show that this bulk-boundary mixed state can exhibit long-range entanglement. If, on the contrary, one considers only the boundary mixed state by simply tracing out the bulk (and thereby throwing out all bulk measurement outcomes), the
resulting boundary density matrix $\sum_{em} \ketbra{\psi(em)} = \mathbb{I} $ is a short-range entangled maximally-mixed state. Such a distinction between monitored dynamics (keeping the measurement outcomes) versus dissipative dynamics (averaging over measurement outcomes) 
is related to the fact that measurement-induced phase transitions~\cite{Li2018,Skinner2019} are only visible in the ensemble of quantum trajectories, but not in the associated quantum channel~\cite{Potter21review, Fisher2022reviewMIPT}. So it is crucial to keep a record of the measurement outcomes, which in our case are represented by the bulk qubits of the system.

\begin{figure*}[bt] 
   \centering
   \includegraphics[width=\textwidth]{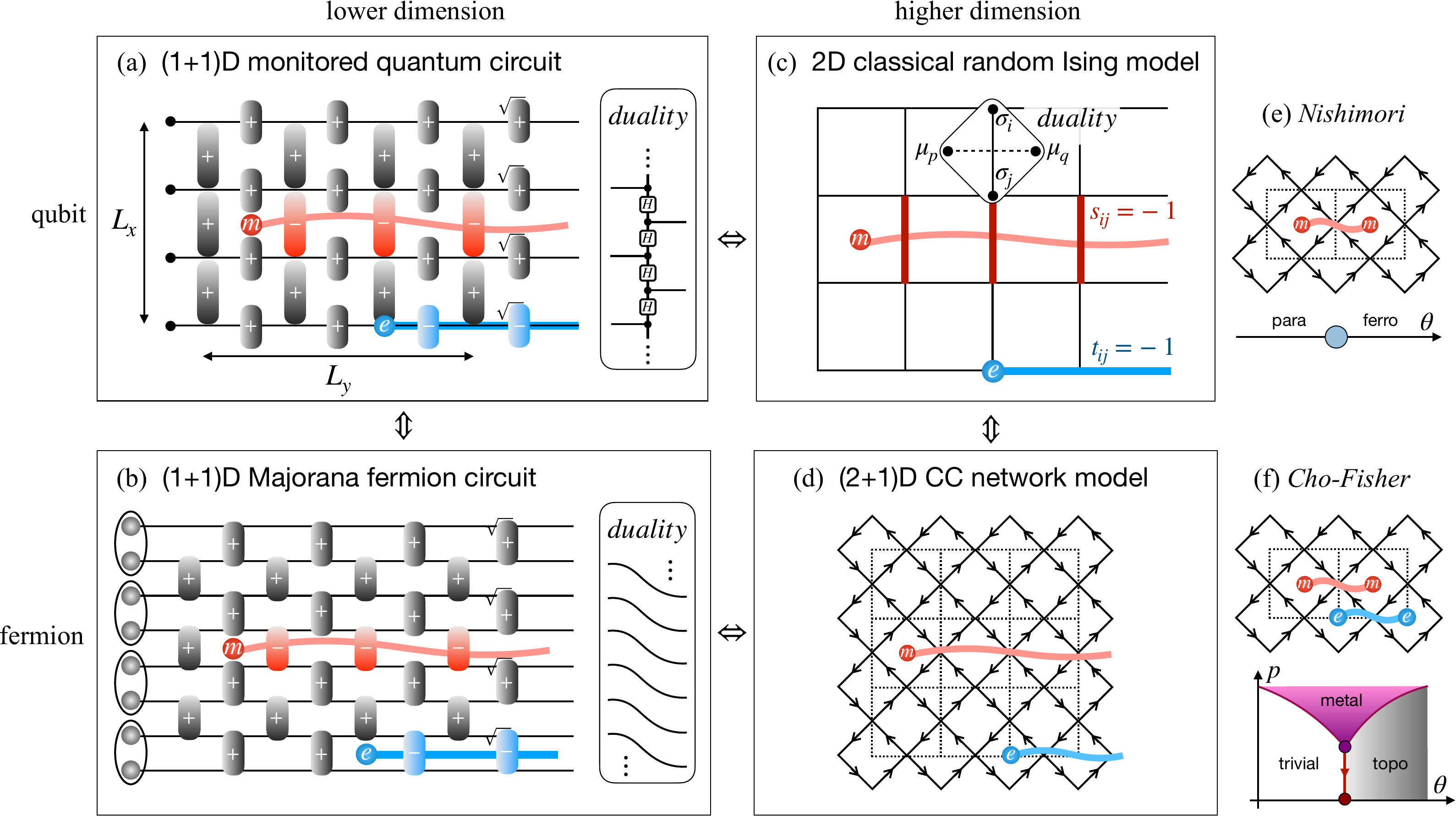} 
   \caption{
   {\bf Four equivalent model perspectives and their respective duality transformations.} 
   (a) {\bf (1+1)D monitored quantum circuit}, with $L_x$ qubits and circuit depth $L_y$, where each gate is an imaginary time evolution with random sign. 
   		The temporal kink of sign change of the $ZZ$ evolution defines an $m$-vortex, while that for the $X$ evolution defines an $e$-vortex. 
   		The duality transformation can be expressed in terms of a 
         matrix-product operator (MPO), with Hadamard gate matrix  $H$ sandwiched by the diagonal delta tensors~\cite{Verresen24measurespt, Hastings22honeycomb, Verstraete23dualitympo}. 
		(b) {\bf 2D classical statistical model} with defect strings pulled by the $e$ and $m$-vortices. 
   		The red defect string runs along the dual lattice and flips the sign of the Ising coupling from ferromagnetic to anti-ferromagnetic coupling, 
		while the blue defect string stretches along the original lattice and inserts an Ising spin operator at the end of the string (the location of the $e$-vortex) into the Boltzmann weight changing its {\it sign}. 
		Both strings can fluctuate across the lattice by gauge transformation, and do not change the gauge invariant partition function
		that is determined only by the $e$ and $m$ locations. 
   		The duality transformation replaces the vertex variable $\sigma$ by the plaquette variable $\mu$. 
   (c) {\bf (1+1)D Majorana quantum circuit}, by Jordan Wigner transforming (a). 
   		The duality transformation is a translation of the Majorana lattice.
(d) {\bf (2+1)D Chalker Coddington (CC) network model for symmetry class $D$}. 
The underlying dashed line illustrates the spin lattice as in (b). 
The solid lines with arrows denote the fermion propagator, where the vertex denotes the scatterer. 
There are two representative CC network models: (e) RBIM along Nishimori line that has only the $m$-vortices, which undergoes the Nishimori transition from a paramagnetic phase to the ferromagnetic phase, as shown schematically;
(f) The CF model that has $e$ and $m$ vortices, whose schematic phase diagram comprises a metal phase upon finite vortex density. Despite the equivalence under a fixed vortex configuration, the conventional CF model takes $N\to 0$ replica limit which is fundamentally distinct from our case where  the $N\to 1$ replica limit is taken. 
      }
   \label{fig:1dcircuit}
\end{figure*}

%%%%%%%%%%%%%%%%%%%%%%%%%%%%%%%%%%%%%%%%%%%%%%%%%%%%%%%%%%%%%%%%%%%%%%%%%%%%%%%%%%%%%%%
\section{Equivalent quantum circuits and statistical model} 
\label{sec:representations}
%%%%%%%%%%%%%%%%%%%%%%%%%%%%%%%%%%%%%%%%%%%%%%%%%%%%%%%%%%%%%%%%%%%%%%%%%%%%%%%%%%%%%%%

To understand the unusual phenomenon of our mixed-state criticality and particularly its manifestation in a critical boundary state, 
we adopt the strategy as follows:
(i) we first fix an arbitrary measurement outcome i.e. a classical snapshot of the 2D bulk bits, and investigate the conditional quantum state at the boundary. The conditional 1D quantum state will be shown to be generated by an effective (1+1)D quantum circuit akin to the MBQC. Then we show that such quantum circuit can be free fermionized into a monitored quantum circuit for a Majorana chain. 
(ii) we then discuss the distribution function, by mapping the random problem to a 2D classical statistical model. 
(iii) finally, via the fermionization, we can show that the statistical model can be understood as a 2D free fermion (Chalker Coddington) network model for symmetry class $D$, where the random measurement outcome acts as vortex disorder for the fermions. 
It allows us to apply our results to four physically different but mathematically equivalent systems:
a monitored $(1+1)$D quantum Ising chain and its mathematically equivalent Majorana counterpart, a 2D disordered classical statistical mechanics model, and the $(2+1)$D Chalker Coddington network model. 
We discuss the Kramers-Wannier duality in these representations of the problem.

%%%%%%%%%%%%%%%%%%%%%%%%%%%%%%%%%%%%%%%%%%%%%%%%%%%%%%%%%%%%%%%%%%%%%%%%%%%%%%%%%%%%%%%
\subsection{(1+1)D monitored quantum circuit}
%%%%%%%%%%%%%%%%%%%%%%%%%%%%%%%%%%%%%%%%%%%%%%%%%%%%%%%%%%%%%%%%%%%%%%%%%%%%%%%%%%%%%%%
Our first representation
follows the spirit of measurement-based quantum computation (MBQC)~\cite{Raussendorf2001oneway, briegel2009measurement}, 
where one views one spatial dimension of a resource quantum state (prior to any measurement) as a fictitious time dimension and a subsequent sequence 
of spatial measurements then allows to effectively induce a unitary evolution (which implements a quantum computation) along this fictitious time.
Here, in analogy, we can also view one spatial dimension of our bulk quantum state (prior to the measurements) as a fictitious time dimension, 
which then evolves under measurement and builds up the entanglement in the boundary quantum state (corresponding to the final time). 
Our circuit protocol, recast as tensor network diagram, can then be derived to be equivalent to a $(1+1)$D {\it imaginary-time} evolving transverse-field Ising model. 
The bulk classical bits serve as the ``register" that record the measurement outcomes, which determine the evolution of the quantum state, i.e.\ its 
quantum trajectory.

%%%%%%%%%%%%%%%%%%%%%%%%%%%%%%%%%%%%%%%%%%%%%%%%%%%%%%%%%%%%%%%%%%%%%%%%%%%%%%%%%%%%%%%
\subsubsection*{Clean Ising in a uniformly post-selected trajectory}
%%%%%%%%%%%%%%%%%%%%%%%%%%%%%%%%%%%%%%%%%%%%%%%%%%%%%%%%%%%%%%%%%%%%%%%%%%%%%%%%%%%%%%%
To discuss the physics of our protocol in this representation, let us first consider the post-selected case where all measurement outcomes are positive. By viewing each column slice of the network as a time step, 
the tensor network state~\eqref{eq:gaugetransform} can be viewed as a quantum chain evolved by a quantum circuit:
\begin{equation*}
\includegraphics[width=\columnwidth]{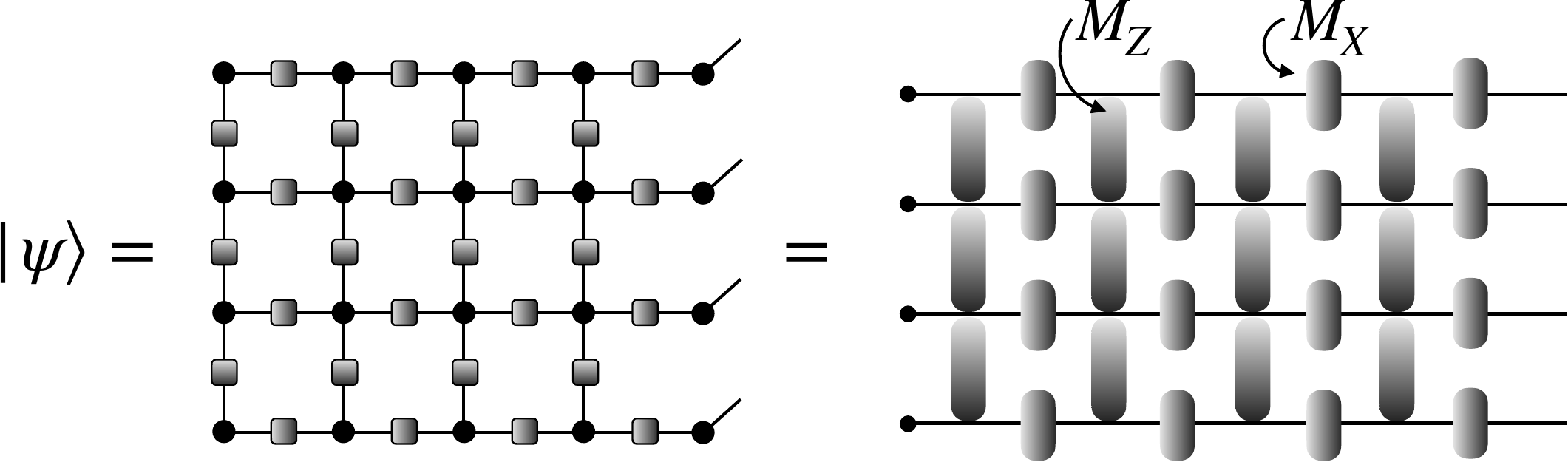} \ .
\end{equation*}
Here each matrix on the bond follows from Eq.~\eqref{eq:tnderivation}. 
The collection of gates for each time period is called transfer matrix in statistical mechanics, which can be derived as:
\begin{equation*}
M = \sqrt{M_X} M_Z \sqrt{M_X} \ , \
M_X = e^{\frac{\beta'}{2} \sum_j X_j}  \ ,\  M_Z = e^{\frac{\beta}{2} \sum_j Z_j Z_{j+1}} \ , 
\label{eq:M}
\end{equation*}
as an imaginary-time evolution of the transverse-field Ising model, with $\tanh\beta =\sin \theta$ and $\tanh\beta' =\cos \theta$. $\beta'$ is obtained by expressing the transfer as
%:
\begin{equation*}
\left(
\begin{matrix}
1 & e^{-\beta} \\
e^{-\beta} & 1 \\
\end{matrix}
\right) 
\propto \exp \left(\frac{\beta'}{2} X\right) \ ,
\end{equation*} 
where $\tanh\frac{\beta'}{2} = e^{-\beta}$ is the Kramers Wannier relation~\footnote{Note that our inverse temperature involves an extra factor of $1/2$ compared to the standard definition. }.
Note that the square root of $M_X$ is necessary to maintain the Hermitian~\footnote{Otherwise, the final state can be viewed as first preparing a critical Ising state described by unitary CFT and then subjecting that state to uniformly post-selected measurement,
which generally changes the entanglement structure~\cite{Garratt23measureising, Alicea23measureising, jian23measureising, Hsieh23channelcriticality, Ludwig24measurecritical, Puetz24}
} property of the transfer matrix $M$. This transfer matrix evolution is rigorously a discrete brickwall circuit, {\it without} any Trotterization approximation. 
In fact, such tensor network equivalence to a quantum circuit is essentially the key of the measurement based quantum computation, which usually designs the transfer matrix to be unitary, while in statistical mechanics the transfer matrix in general is non-unitary. 
In the tensor network language, the worldline of the quantum circuit lies on the {\it virtual} leg of the tensor network, rather than the physical leg that has been terminated by projective measurement. 
Physically, this means the effective circuit evolution evolves the state in the space direction, while the conventional circuit evolves the state in the time direction. 
One can view the final steady 1D state as a {\it boundary} state of the 2D state.

The KW duality transformation operator can be written as a cluster MPO~\cite{Verresen24measurespt, Hastings22honeycomb, Verstraete23dualitympo} in
%the 
Fig.~\ref{fig:1dcircuit}(a), which transforms
\begin{equation*}
{\rm KW}: \quad Z_j Z_{j+1} \to X_{j+1/2} \quad , \quad X_j \to Z_{j-1/2}Z_{j+1/2} \ .
\end{equation*}
It is equivalent to $\theta \leftrightarrow \pi/2-\theta$ or $\beta \leftrightarrow \beta'$, which are indeed related by the KW dual relation~\cite{KW1941}: 
$\tanh(\beta' /2) = \exp(-\beta)$ (here the Ising coupling strength is $\beta/2$).
At $\theta=\pi/4$, $\beta = \ln(1+\sqrt{2})=\beta'$, the model is (KW)
self-dual and the 1D quantum state evolves towards the ground state 
of the clean critical Ising model, with linear depth of the $(1+1)$D circuit i.e.\ $L_y=O(L_x)$. 
In the absence of $e$ and $m$ random vortices, such a pure state has the {\it strong} KW self-dual symmetry. That is, in the long wavelength limit, 
the resulting critical (boundary) state is effectively described by the 2D Ising CFT, protected by strong $Z_2$ symmetry and by strong KW self-dual symmetry. 
The non-invertibility~\cite{Rizi23selfdual, Seiberg24selfdual} of the transformation can be deduced when the MPO in Fig.~\ref{fig:1dcircuit}(a) is closed in periodic boundary condition, which inevitably maps $\prod_j X_{j+1/2} = 1$ and thus carries a global parity projector. 

Notably, the transfer operators $M_Z$ and $M_X$ above are {non-unitary} operators, 
which can be interpreted as the Kraus operators of effectively {\it  weak} measurements of the $Z_j Z_{j+1}$ and $X_j$ operators in the $(1+1)$D dynamics. 
Note that these tunable effective weak measurements should not be confused with the projective measurements of the 2D bulk. 

%

%%%%%%%%%%%%%%%%%%%%%%%%%%%%%%%%%%%%%%%%%%%%%%%%%%%%%%%%%%%%%%%%%%%%%%%%%%%%%%%%%%%%%%%
\subsubsection*{Measurement-induced random quantum trajectories}
%%%%%%%%%%%%%%%%%%%%%%%%%%%%%%%%%%%%%%%%%%%%%%%%%%%%%%%%%%%%%%%%%%%%%%%%%%%%%%%%%%%%%%%
Due to the uncertainty principle and the probabilistic nature of quantum measurement, our effective Ising circuit is subject to both bit-flip $X$ and phase-flip $Z$ errors,
manifesting themselves in the random measurement outcomes of the bond and the site qubits, respectively. 
Using a gauge transformation as illustrated in
Eq.~\eqref{eq:gaugetransform}, these random Paulis can be propagated to the final times, 
leaving an Ising circuit where each imaginary-time evolution step is characterized by a position-dependent
binary random number, resulting in a quantum chain evolved by a random quantum circuit:
\begin{equation*}
\includegraphics[width=\columnwidth]{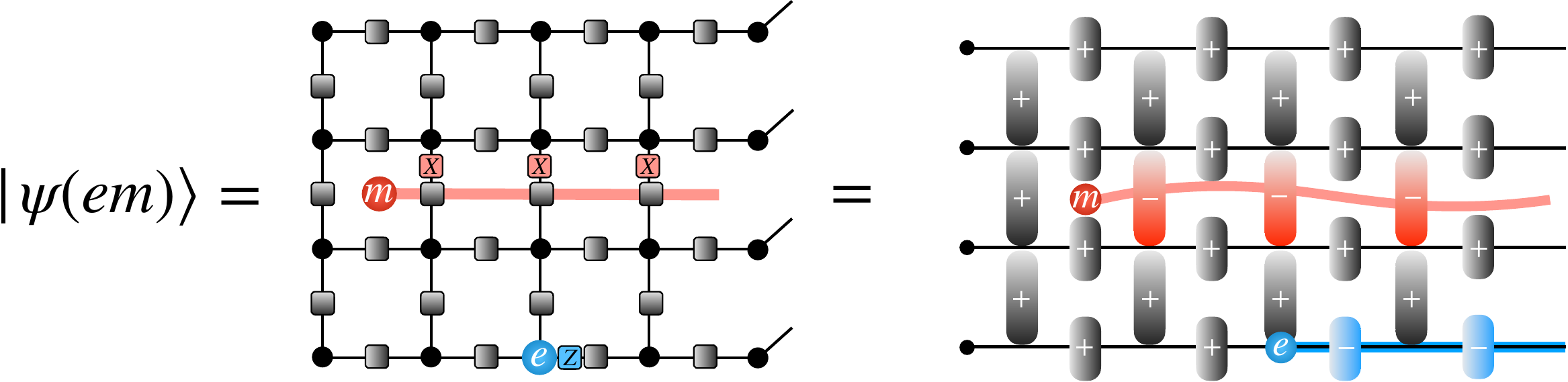} \ .
\end{equation*}
The locations of these gates with negative evolution time-steps form a string, whose end points correspond to fixed $e$ and $m$-vortices~\footnote{The Paulis at the final time slice can be annihilated
by acting with a single layer of exactly the same Pauli gates on 
the final quantum state, making use of the measurement outcome}.
In other words, an $e$-vortex necessarily pulls a string that changes the sign of $\beta'$, while $m$ pulls a string that changes the sign of $\beta$. 
The Ising evolution gates are then coupled to the random bond variables
\begin{equation}
\label{LabelEqDefsandtTransferMatrix}
M_Z = \exp\left( \frac{\beta}{2} \sum_j s_{j,y} Z_j Z_{j+1}\right) \ ,\ 
M_X = \exp\left( \frac{\beta'}{2} \sum_j t_{j,y} X_j \right) \ ,
\end{equation}
where the negative sign $s_{x,y}=-1$ occurs when an $m$ string goes across the bond, and $t_{x,y}=-1$ when an $e$ string goes along the bond, determined by the configuration $\ket{em}$, see Fig.~\ref{fig:1dcircuit}(a)~\footnote{Note that the final time $\beta'$ takes half of its value in order to implement $\sqrt{M_X}$}. 

\subsubsection*{Monitored Majorana chain}
\label{LabelSubsectionDeepMonitoredMajoranaCircuit}

The random Ising circuit introduced in the previous Section can be mapped to a Majorana quantum circuit
via a Jordan-Wigner transformation, see Fig.~\ref{fig:1dcircuit}(b). 
In detail, the $L_x$ qubits can be Jordan-Wigner transformed to $2L_x$ Majoranas $\gamma_j$, according to 
$X_j = i\gamma_{2j-1}\gamma_{2j}$ and $Z_j = \prod_{k=1}^{j-1}(i\gamma_{2k-1}\gamma_{2k}) \gamma_{2j-1}$. The non-unitary gates are transformed to the Gaussian evolution generated by the Majorana bilinears $Z_j Z_{j+1}=i\gamma_{2j}\gamma_{2j+1}$ and $X_j = i\gamma_{2j-1}\gamma_{2j}$. Then the GHZ state at $\theta=\pi/2$ in the spin representation is mapped to the topologically nontrivial Majorana state with Majorana zero modes $\gamma_1$ and $\gamma_{2L_x}$ on the edges~\cite{Kitaev2001}. 
The problem is then mapped to a 
measurement-only Majorana fermion chain weakly monitored by
local fermion parity measurements of the nearest-neighbor Majorana fermion bilinears above~\footnote{i.e., fermion parity measurements of the kind employed in \cite{Nahum23nlsmmajorana} and \cite{Jian23measurefreefermion}, where the second reference discussed only the case where unitary gates are also always present in addition to fermion parity measurements.}. 
The global Ising symmetry is mapped to the total fermion parity $\prod_j X_j \propto \prod_j i\gamma_{2j-1}\gamma_{2j} $, and the KW dual transformation is mapped to a single-site Majorana fermion translation
\begin{equation*}
{\rm KW}: \quad \gamma_j \to \gamma_{j+1} \ ,
\end{equation*}
see the box in Fig.~\ref{fig:1dcircuit}(b). 
The corresponding self-dual critical point is a 
monitored Majorana chain with statistical (``averaged'', or ``weak'') translational invariance
which separates trivial and topological phases upon dimerization~\cite{Kitaev2001} of the measurements,
and is described by the $N\to 1$ limit of the 2D  Non-linear Sigma Model (NLSM) in symmetry class $D$ with target space $SO(2N)/U(N)$ and topological angle-$\theta=\pi$~\cite{Nahum23nlsmmajorana, GRL2001}.
For a more detailed discussion of the description of this self-dual critical point in terms of this particular NLSM, see the third paragraph in part (b) of the following subsection.
This critical point  was also studied numerically in the context of a continuously weakly-monitored~\cite{kurt06continuous} 
Majorana chain in Ref.~\cite{Graham23majorana}~\footnote{As expected on general grounds, 
the continuously weakly-monitored chain falls into the same universality class as our spacetime isotropic model, see Appendix~\ref{sec:continuous}}.

The replica limit $N \to 1$ arises~\cite{Ludwig2020,Altman2020weak} 
because of the Born probability weighting: the physics of generic uncorrelated quenched disorder, corresponding to a replica limit $N \to 0$, is dramatically different and moreover exhibits a 
Majorana metallic phase~\cite{Read2000,ChalkerKagalovskyEtAlThermalMetalPRB2001}
which is absent in the $N\to 1$ limit -- see the following Section.

\subsubsection*{Comment on non-unitary measurement-based quantum computation}

We started our discussion of the $(1+1)$D monitored quantum circuit model with an analogy to the framework of measurement-based quantum computation (MBQC)~\cite{Raussendorf2001oneway, briegel2009measurement}. Let us close this discussion by pointing out some additional connections and distinctions in this context.
Similar to the MBQC approach, we initiate our protocol from a cluster state as the principal entanglement resource and perform projective measurements in the bulk to propagate a 1D quantum state on the edge by means of quantum teleportation, trading time with space~\cite{Browne2008, Altman2021measure, Chen2022measure, Harrow2022, google2023measurement}. 
In the standard MBQC approach, the transfer matrix that evolves the quantum state was shown to be {\it unitary} and can implement universal quantum computation when the measurement angles are carefully chosen to lie along the $Z$ direction or
are restricted to within the $XY$ plane~\cite{briegel2009measurement}. Then the measurement outcome is maximally random because each post-measurement unitary trajectory shares equal probability, as in the standard example of quantum teleportation with perfect entanglement resource~\cite{Bennett93teleport}. In the absence of noise one can always correct the measurement outcomes to obtain the same pure state. 
In contrast, in our protocol we tilt the measurement angle between $Z$ and the $XY$ plane~\cite{JYLee, Sela23nonunitary} (equivalent to an imperfectly prepared cluster state~\cite{NishimoriCat}), which turns the effective spatial propagation into a probabilistic {\it non-unitary}~\cite{Ueda05nonunitary} circuit. The measurement outcome is no longer maximally mixed but follows a highly correlated distribution. Not all post-measurement random pure states are guaranteed to be able to be corrected to the same pure state.

%%%%%%%%%%%%%%%%%%%%%%%%%%%%%%%%%%%%%%%%%%%%%%%%%%%%%%%%%%%%%%%%%%%%%%%%%%%%%%%%%%%%%%%
\subsection{2D Classical statistical model}
\label{sec:classical_statmech}
%%%%%%%%%%%%%%%%%%%%%%%%%%%%%%%%%%%%%%%%%%%%%%%%%%%%%%%%%%%%%%%%%%%%%%%%%%%%%%%%%%%%%%%

The probability distribution function of the boundary 1D states $\ket{\psi(em)}$, conditioned upon a given bulk classical configuration $\ket{em}$ in Eq.~\eqref{eq:mixedstate}, follows from Born's rule
\begin{equation}
\includegraphics[width=\columnwidth]{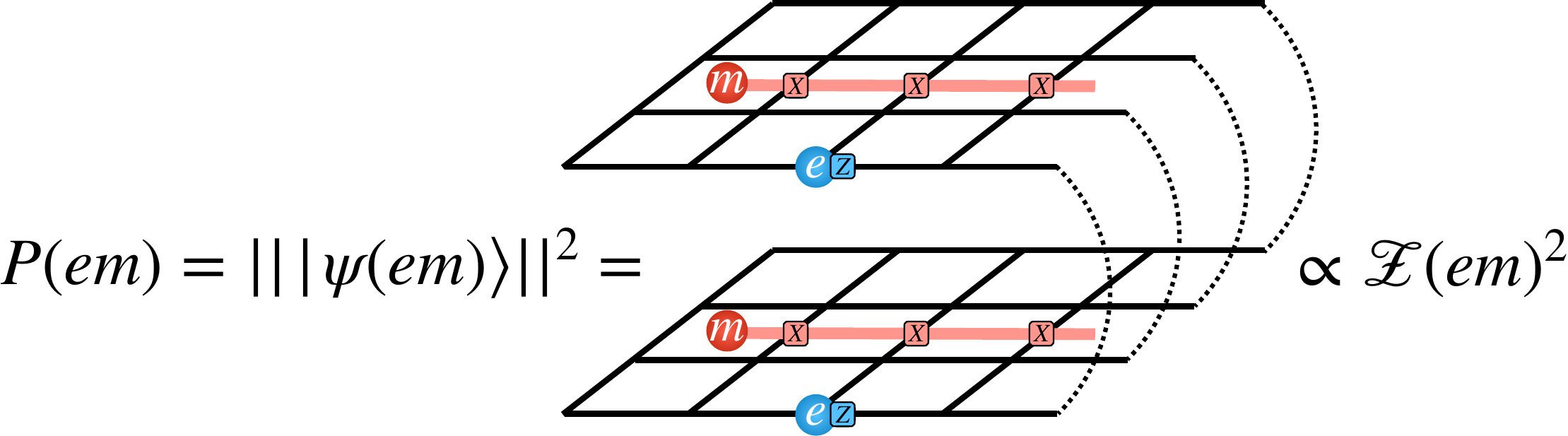} \ ,
\label{eq:Born}
\end{equation}
which is obtained by stacking two layers of the tensor network~\eqref{eq:gaugetransform} for the ket $\ket{\psi(em)}$ and the bra $\bra{\psi(em)}$ respectively, and gluing them at the boundary. By interpreting each virtual index (a dummy variable to be summed over) of the tensor network as a classical Ising spin~\cite{Zhu19ToricCode}, we can interpret this function as a partition function of a 2D classical statistical model. 
Since the two layers are decoupled in the bulk, we only need to write down the partition function of one layer:
\begin{equation}
\begin{split}
\mathcal{Z}(em) =\sum_{\mathbf{\sigma}} e^{\frac{\beta}{2} \sum_{ij} s_{ij} \sigma_i\sigma_j} \prod_{ij}\left(\sigma_i\sigma_j\right)^{\frac{1-t_{ij}}{2}}  \ , \\
\end{split}
\label{eq:statmech}
\end{equation}
where $\sigma_j$ is the classical Ising spin at the vertex $j$, also as the virtual index of the tensor network.
It expresses nearest neighbor
Ising model with random bond disorder on both the original lattice and its dual lattice (or imaginary bond coupling strength), indicated by the dashed line in Fig.~\ref{fig:1dcircuit}(b). Here $i,j$ denote the sites, and $\sigma=\pm 1$ corresponds to the 0 and 1 of the site qubit, which has been projected to the $\ket{\pm}=(\ket{0}\pm \ket{1})/\sqrt{2}$ state with a measurement-outcome-dependent sign factor. The sign factor is given by the $e$-vortex at the site, which is the end point of the string with $t_{ij}=- 1$. On the other hand, $s_{ij}=\pm 1$ is the random measurement outcome on the bond, and a string of $s_{ij}=-1$ on the dual lattice terminates at the $m$-vortex.  
The square $|\cdot|^2$ is because of the norm that contracts the ket and bra layers, or the forward and backward propagation of  spacetime, which also guarantees the positiveness of the probability. 
Upon a KW duality transformation for the 2D model that exchanges vertex and plaquette, one can verify that $s_{ij} \leftrightarrow t_{ij}$, and $\beta \leftrightarrow \beta'$~\cite{KadanoffCeva}. Upon a vertex plaquette duality~\cite{Kogut79rmp}, this random Ising model remains invariant. 
Note that if one starts from the clean toric code state, then $t_{ij}=s_{ij}$, which microscopic detail does not change the nature of the statistical model, that only depends on the gauge invariant $e$ and $m$ vortex configurations. 

If we now trace out one of the two types of
vortices, we are left with a random-bond Ising model (RBIM) in both cases
\begin{equation}
\begin{split}
P(m) &= \sum_e P(em) \propto \sum_\sigma e^{\beta \sum_{ij} s_{ij} \sigma_i\sigma_j}  \ ,\\
P(e) & = \sum_m P(em) \propto \sum_\mu e^{\beta' \sum_{pq} t_{pq} \mu_p\mu_q}  \ ,
\end{split}
\label{eq:Pm}
\end{equation}
each of which describes
the standard RBIM with gauge symmetrized random bond disorder~\cite{Nishimori1981, NishimoriCat}. 
Here $\mu_p$ is the dual Ising spin residing at the plaquette center, and with $t_{pq}=t_{ij}$, see Fig.~\ref{fig:1dcircuit}(c) for details
(where $ij$ and $pq$ share the same link). This turns into a random-bond Ising model on the dual lattice. 
One can derive the density of vortices by:
\begin{equation}
\begin{split}
1-2\langle m \rangle &= \langle \prod_{\langle ij\rangle\in \square} s_{ ij} \rangle = \tanh^4\beta= \sin^4\theta  \ ,\\
1-2\langle e \rangle &= \tanh^4\beta' = \cos^4\theta \ .
\end{split}
\label{eq:vortex_densities}
\end{equation}
Namely, at the self-dual point ($\theta=\pi/4$) we expect the two vortex densities to cross, i.e.\ $\langle e\rangle = \langle m \rangle$, which is precisely what we find also in our numerical sampling as shown in Fig.~\ref{fig:emflux}(a) below.

 \subsubsection*{Free energy and central charge}
 
All the classical statistical models that capture the transitions discussed in this paper exhibit transitions described by conformal field theories (CFTs) with central charge $c=0$~\cite{Ludwig05c0log, Pixley22MIPTCFT, Vasseur24boundarytrsf}. To explain this vanishing central charge, it is important to define precisely how a suitable notion of free energy is to be defined
for such models. For the Born probability measure $P(em)$, one natural way to do so is to define a family of replicated partition functions by summing over trajectories (measurement outcomes) and their corresponding free energy
\begin{equation}
\begin{split}
\mathcal{Z}_N =& \sum_{em} P(em)^{N} \, , \\
F_N =& -\ln \mathcal{Z}_N  \ .
\end{split}
\label{eq:ZNFN}
\end{equation}
Physical quantities are obtained in the replica limit $N \to 1$, corresponding to a Born weighting of the trajectories associated with the trivial partition function 
\begin{equation}
\mathcal{Z}_{N=1} = \sum_{em} P(em) =1 \ .
\end{equation}
Different replica numbers $N$ have transitions described (for replica number small enough) by CFTs with central charge $c(N)$, with $c(N\to1) = 0$ since the free energy in the replica limit $F_{N=1} = 0$ has trivial finite-size scaling. This central charge $c=0$ is a general feature of all mixed state (measurement-induced) phase transitions studied in this paper, even the so-called Ising+ transition we will encounter below where for some observables
$ O_{em}$, averages like $\sum_{em} P(em) O_{em}$ reduce to  Ising correlators (a CFT with central charge $c=1/2$). 

Since the CFTs of interest in this work have central charge $c=0$~\cite{GURARIE1993, Ludwig05c0log, GURARIE1999, Cardy2013log, Pixley22MIPTCFT, Vasseur24boundarytrsf} , it is useful to introduce instead the so-called {\it effective central charge}~\cite{LUDWIG1987687,Pixley22MIPTCFT}
\begin{equation}
	c_{\rm Casimir} = \lim_{N \to 1} \frac{d c(N)}{dN} \ , 
\end{equation}
which is a universal quantity that governs the finite-size scaling of the quenched average free energy
\begin{equation}
\begin{split}
F =& \lim_{N \to 1} \frac{d F_N }{dN} = - \sum_{em} P(em) \ln P(em) \\
=& \, {\rm const} \cdot  L_xL_y - c_{\rm Casimir} \cdot \frac{\pi}{6}\frac{L_y}{L_x}+ \ldots\ ,
\end{split}
\label{eq:Fscal}
\end{equation}
when being placed on a length-$L_y$ long cylinder of finite width $L_x\ll L_y$~\cite{Cardy86} (see the inset of Fig.~\ref{fig:pureSvN}(a) for a schematic). This formula follows from the general scaling form of the free energy $F_N$ at criticality expected from conformal invariance~\cite{Cardy86}. The universal number $c_{\rm Casimir}$ captures how the vacuum energy responds to the finite scale of the system, which is analogous to Casimir effect where the finite width of the system can reduce the ground state energy.
In this context, the free energy $F$ then acquires physical meaning as the Shannon entropy of the bulk ($S_C$ in Fig.~\ref{fig:classicquantum}b%Eq.~\eqref{eq:SAC}
) of the mixed 2D state, whose specific form considered in this paper is given in Eq.~\eqref{eq:mixedstate}, also called the entropy of the measurement record of the 1+1D monitored dynamics~\cite{Pixley22MIPTCFT}, or the frustration entropy of the random classical statistical model~\cite{Nishimori1981}.

\subsection{(2+1)D Chalker-Coddington network model}
\label{LabelSubsectionChalkerCoddington}

\begin{figure}[t!]
   \centering
   \includegraphics[width=\columnwidth]{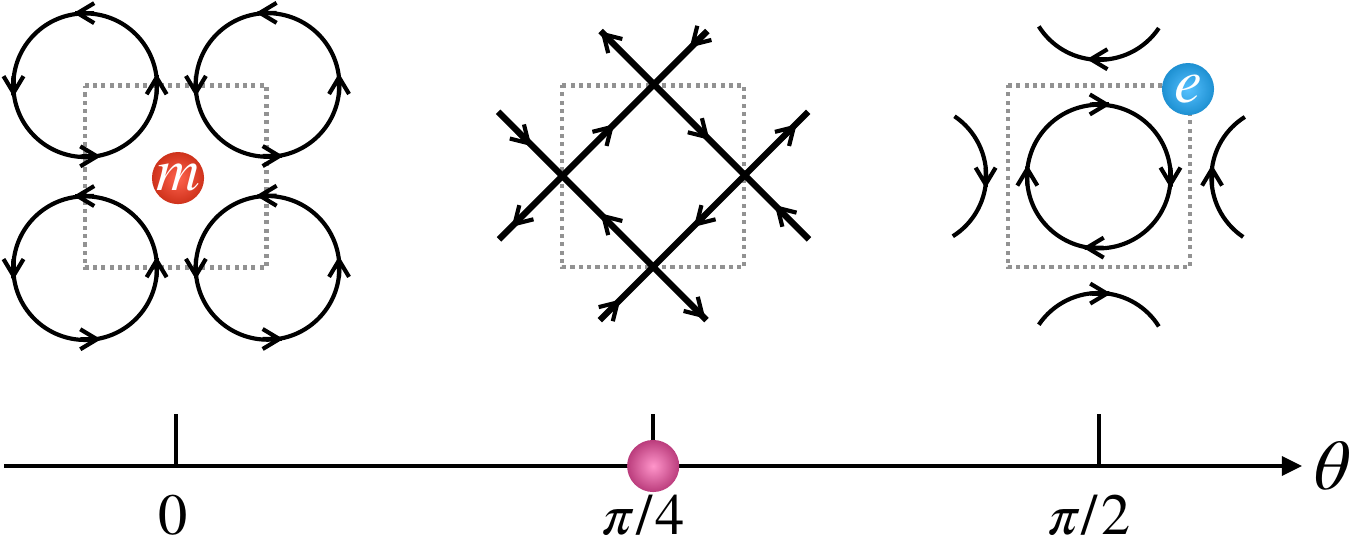}
   \caption{
      {\bf Measurement angle tunes the fermion network model}.
      Near $\theta=0$ or $\theta=\pi/2$, the fermion modes are localized
      to move in small circles, surrounding
      either (i) the vertex, or (ii) the plaquette center, insensitive to the presence of a $\pi$ vortex at the plaquette center in case (i), or 
       the vertex in case (ii).
      Consequently, at $\theta=0$, the $m$-particle as $\pi$-vortex for the Majorana fermions proliferates, while
      the $e$-particle is confined, and vice versa at $\theta=\pi/2$. 
      At the  self-dual angle $\pi/4$, the scattering node renders equal probability of tunneling or reflection of the Majorana fermions, which results in a critical state. 
   }
   \label{fig:networktransition}
\end{figure}

Making contact with the body of previous work on Anderson localization~\cite{Read2000,GRL2001,ChalkerKagalovskyEtAlThermalMetalPRB2001}, the $(1+1)$D Majorana circuit can be described~\cite{Jian22network},
for any  fixed configuration of $e$ and $m$ ``vortices'',  equally as the ground state of a system of non-interacting Majorana fermions in two spatial dimensions, i.e.\ in (2+1)D  spacetime dimensions at zero energy (chemical potential). 
For the system under consideration this is  a superconductor in two spatial dimensions in Altland-Zirnbauer \cite{AltlandZirnbauer} symmetry class D.

The whole network is shown in Fig.~\ref{fig:1dcircuit}(d), where the local unitary scattering matrix of the single-particle Majorana fermion modes is derived from the transfer matrix in the standard formulation~\cite{Chalker2002,Fisher1997},
to be the following:
\begin{equation}
\includegraphics[width=.8\columnwidth]{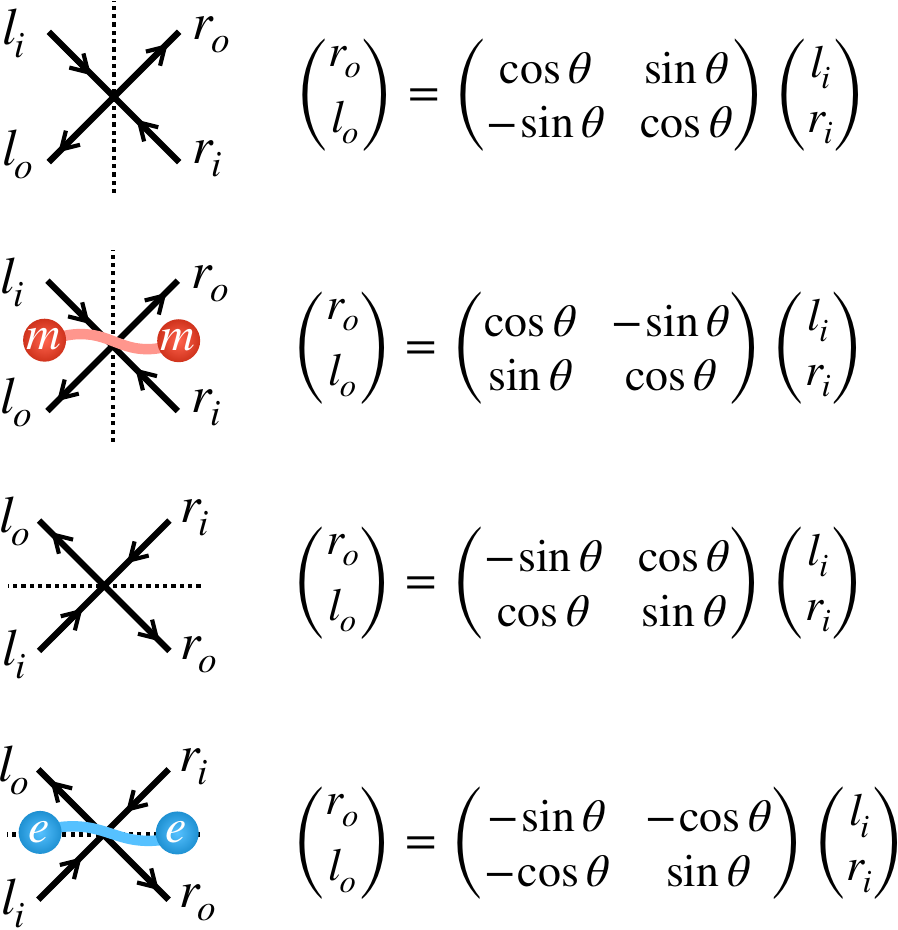} \ ,
\end{equation}
where the dashed line denotes the link of the original square lattice, such that the vertical and horizontal links are related by a 90 degree rotation. The $e$ or $m$ particle recide on the vertex or the plaquette center, acting as a $\pi$ vortex that pulls a branch cut, across which the Majorana fermion experiences a $\pi$ phase shift. The measurement angle $\theta$ determines the tunneling and reflection amplitudes, which is schematically showns in Fig.~\ref{fig:networktransition}.

As discussed in~\cite{Jian22network}, the two-dimensional space (Fig.~\ref{fig:1dcircuit}(d)) of the $(2+1)$D system represents the spacetime of the circuit depicted in Fig.~\ref{fig:1dcircuit}(a,c), and in a
formulation of the $(2+1)$D system on a 2D spatial lattice and discrete  time, this is referred to as a ``Chalker-Coddington'' model \cite{ChalkerCoddington1988}: In each  time-step of the discrete time the Majorana fermions propagate across a link of a square lattice [the so-called ``surrounding lattice''~\footnote{The ``surrounding lattice'', whose sites are the midpoints of the links in Fig.~\ref{fig:1dcircuit}(c), and on whose bonds
the Majorana fermions propagate, is not drawn explicitly drawn in Fig.~\ref{fig:1dcircuit}(d)}, 
whose sites are the midpoints of the links of the lattice in Fig.~\ref{fig:1dcircuit}(d)] whose plaquette-centers represent the locations of Ising spins ($\sigma_i$) on one sublattice, and dual Ising spins $(\mu_p)$ on the other sublattice of the bipartite lattice  of plaquettes, see Fig.~\ref{fig:1dcircuit}(c).
A spin ($\sigma_i$) and dual spin $(\mu_p)$ located at (the center of)  a plaquette corresponds to a 
$(2+1)$D vortex enforcing a minus sign for fermions encircling the plaquette in the time of the $(2+1)$D system.
In this formulation, 
a negative sign of the variable ``$s_{ij}$'' appearing in the transfer matrix, Eq.~\eqref{LabelEqDefsandtTransferMatrix},
 corresponds to a  pair $\sigma_i$, $\sigma_j$ of vortices on nearest-neighbor sites $i$ and $j$
of the sublattice of {\it spin} plaquettes, while 
a negative sign of the variable ``$t_{p,q}$'' in 
Eq.~\eqref{LabelEqDefsandtTransferMatrix}
corresponds to a pair
$\mu_p$, $\mu_q$ of vortices on nearest-neighbor sites $p$ and $q$
of the sublattice of {\it dual-spin} plaquettes. 
In this language, the square of the partition function of the Ising model in any fixed configuration of $e$ and $m$ vortices 
(in a particular gauge) is 
obtained (see~\cite{Read2000}) by
tracing over (two identical copies of Majorana, or equivalently a single copy of complex) fermions, which we denote by ${\cal Z}(e,m)$. (See  Eq.~(\ref{eq:statmech}) below, where the same trace is computed, equivalently, in the Ising spin formulation.) 

As mentioned, the so-described formulation using the Chalker-Coddington model is given in a {\it fixed} configuration of the vortices mentioned above, or equivalently (by a gauge choice) of $e$ and $m$ vortices.  
Historically, such a Chalker-Coddington model was (a) applied to various settings of  generic (``uncorrelated'') {\sl quenched} disorder averages over configurations of Majorana fermion zero 
modes ($e$ and $m$ vortices)~\cite{Read2000,GRL2001,ChalkerKagalovskyEtAlThermalMetalPRB2001}.
In a theoretical description using replicas, these situations correspond to the limit of the number $N$ of replicas going to zero, $N\to 0$.
In the present work we apply, (b),  the aforementioned Chalker-Coddington model to  settings where $e$ and $m$ vortices are {\sl measured}, and  we are thus interested in quenched {\sl measurement} disorder subjected specifically to the Born-rule probability distribution. In a theoretical description using replicas, 
these situations correspond~\cite{Ludwig2020,Altman2020weak} to 
the limit of the number $N$ of replicas going to unity, $N\to 1$.
In both applications, (a) and (b), Kramers-Wannier duality -- meaning the invariance of the
respective {\it probability distribution} under Kramers-Wannier duality, also called ``statistical'', or ``weak", or ``average'' Kramers-duality -- plays an essential role for the resulting physical behavior of the system.

\subsubsection*{Relationship with previous work}

The key conclusion of the prior work in Refs.~\cite{Read2000,GRL2001,ChalkerKagalovskyEtAlThermalMetalPRB2001},
referring to generic (``uncorrelated'') quenched disorder (corresponding to the replica limit $N \to 0$),
was that there are two fundamentally different situations, leading to very different physical properties: case (i), discussed below, which does not exhibit a metallic phase, 
as well as case (ii) discussed below which does exhibit a metallic phase. We discuss those now in turn, with support of the schematics in Figs.~\ref{fig:1dcircuit}(e,f):
\begin{itemize}
\item[(i)] In the case of the RBIM, in its entire phase diagram {\it at and away from the Nishimori line}, the vortices are allowed to appear only on {\it one} sublattice of plaquettes of the Chalker-Coddington model, namely on the sublattice of  (say) {\it spin} plaquettes. 
In this case, only two phases are possible~\footnote{besides an actual spin-glass phase appearing in the phase diagram of the RBIM at {\it zero temperature} (using the conventional classical statistical mechanics model coordinates of the phase diagram, which are temperature and negative exchange coupling disorder), which we do not address in the present work}, a ferromagnetic (topological) phase and a paramagnetic (topologically trivial) phase. In the RBIM these vortices occur in adjacent pairs. Each such pair corresponds to a negative sign of the random variable $s_{ij}$ described in the first paragraph of this section. This model, the RBIM case (i), {\it maximally} violates statistical (``average'')  Kramers-Wannier symmetry since vortices are only allowed  to occur on
{\it one} of the two sublattices of plaquettes~\footnote{related to
each other by Kramers-Wannier duality}, 
the (say) sublattice of {\it spin} plaquettes.\\
\item[(ii)] If, on the other hand, Majorana zero modes are also allowed to appear on the other sublattice of plaquettes of the Chalker-Coddington model, namely the sublattice of {\it dual-spin} plaquettes where they occur in adjacent pairs, then in addition to the ferromagnetic (topological) and paramagnetic (topologically trivial) phases, a third, {\it metallic} phase will occur
in the phase diagram. This is the so-called {\it Cho-Fisher} (CF) model, originally intended~\cite{Fisher1997} as a formulation of the Nishimori system (a), but later shown~\cite{Read2000,GRL2001} to actually describe a different system with the novel properties  
described in \cite{Read2000,ChalkerKagalovskyEtAlThermalMetalPRB2001}
and reviewed here. It should also be noted that it was proven in \cite{Read2000} that  a metallic phase is forbidden when vortices are only allowed to occur on one sublattice of plaquettes of the Chalker-Coddington model, as is the case in the Nishimori model, case (a).
A variant of the CF model was also discussed in \cite{ChalkerKagalovskyEtAlThermalMetalPRB2001} where vortices occur randomly with some probability on {\it any} site of the two sublattices of plaquettes of the Chalker-Coddington model, but with equal probability on {\it both} sublattices (dubbed the $O(1)$-model).
This system was shown to exhibit {\it only} the metallic phase throughout its entire phase diagram, and no ferromagnetic nor paramagnetic phases whatsoever survives infinitesimal disorder strength.
\end{itemize}

We now discuss the two cases of distinct replica limits:\\

{\it (a) $N\to 0$ replica limit, uncorrelated quenched disorder}.--
The phase diagram of the CF model, case (ii), exhibits a line
possessing average Kramers-Wannier symmetry along the phase boundary separating ferromagnetic (topological) and paramagnetic (non-topological) phases, and connecting the {\it non-random (pure)} Ising critical point with a multicritical point at which a transition to the metallic phase sets in, and continuing into the metallic phase. This self-dual line is depicted as the vertical line in Fig.~\ref{fig:1dcircuit}(f)~\cite{ChalkerKagalovskyEtAlThermalMetalPRB2001}. Recent numerical work~\cite{Gruzberg-Et-Al-2D-Class-D-2021} found that the  RG flow {\it emerges} from the multicritical point and flows {\it into}
the pure Ising critical point which (obviously) possesses ``strong'' as opposed to ``statistical'' (``weak'', or ``average" -- in recent jargon) KW symmetry.\\

{\it (b) $N\to 1$ replica limit, Born measurement disorder}.--
The previous work~\cite{Read2000,GRL2001,ChalkerKagalovskyEtAlThermalMetalPRB2001}, briefly summarized in item (a) above, 
raises the question about the nature of these various phases and transitions when the quenched randomness does not originate from generic, uncorrelated randomness, but rather from {\it measurements} of the $e$ and $m$ degrees of freedom (representing $(2+1)$D vortices) satisfying the Born-rule probability distribution. 
This question has, in fact, been the very motivation for the work that we report in the paper at hand and is what we will address in the following.

It is useful to begin with the KW self-dual system which, in the above-reviewed language of the Chalker-Coddington model, is a version of the KW-self-dual line of the CF model, where however now randomness arises from measurements of $e$ and $m$, and is thus subjected to the Born-rule probability distribution. 
In short, we will demonstrate that the KW self-dual model is
a ``measurement-version'' of the CF model discussed in (a) (ii) above, in which the generic quenched disorder is replaced by the intrinsic randomness of quantum mechanical measurement outcomes. In that sense, the KW self-dual system is a self-dual `cousin' of the Nishimori critical point (and it will be in a different universality class, as we also numerically confirm - see Table~\ref{tab:criticalpoints}).

As the self-dual circuit is a circuit of non-interacting fermions in Altland-Zirnbauer~\cite{AltlandZirnbauer} symmetry 
class D, it is generically described~\cite{Jian22network,LudwigNobelSymp2016,ZirnbauerSUSY1996,10FoldWayReviewNJPhys2010}
by a NLSM with target space $SO(2N)/U(N)$ where $N$ is the number of replicas. Moreover, following the same logic 
as that used in Ref.~\cite{Jian23measurefreefermion} for symmetry class DIII, the case of randomness arising from measurements satisfying the Born-rule probability distribution requires~\cite{Ludwig2020,Altman2020weak} taking the replica limit $N\to 1$. 
In addition, and importantly for the present context, Ref.~\cite{GRL2001} established averaged (i.e. statistical) KW duality of this NLSM.
As the topology of the target space $SO(2N)/U(N)$ allows~\cite{GRL2001} for a $\Theta$-term, 
criticality is known to occur when 
$\Theta=\pi$. [This is in complete analogy
with, e.g., the familiar  $O(3)$-NLSM, describing the 1D Heisenberg chain.] 
 In summary, this establishes the description of the KW self-dual point by the 
$SO(2N)/U(N)$ NLSM at  $\Theta$-angle $\Theta=\pi$ in the limit $N\to 1$. (We note that independent work in \cite{Nahum23nlsmmajorana} arrived at the same conclusion about this critical point, starting from the microscopic (`lattice')  formulation of Majorana fermion parity measurements.~\footnote{This NLSM was recently discussed again in the context of weakly monitored quadratic (``$q=2$'') Sachdev-Ye-Kitaev chains~\cite{Luca24monitoredSYK}.})
Moreover,  Ref.~\cite{GRL2001} also found that the  $SO(2N)/U(N)$
NLSM as $N\to 1$ is stable under RG to  a certain class of 
deformations (not necessarily related to the ones we study numerically in the present paper). 
It is worth noting that with Born-rule measurements there is, in contrast to generic, uncorrelated quenched randomness [discussed in part (a) above],  no stable metallic phase. This is easily understood as a consequence of the known RG beta function of the coupling constant of the NLSM on this symmetry class D target space (reproduced, e.g., in \cite{GRL2001}), which demonstrates stability of the metallic phase in the replica limit $N\to 0$, relevant for case (a),  while in the replica limit $N\to 1$, relevant for case (b), the metallic phase is unstable. The phase diagram that thus emerges for the measurement-version of the CF model with Born-rule measurements is that depicted in Fig.~\ref{fig:1dcircuit}  above (the same as Fig. 1 of Ref.~\cite{ChalkerKagalovskyEtAlThermalMetalPRB2001}),
except that the metallic phase is completely removed: There is then a vertical line describing the self-dual model connecting the non-random (pure) critical Ising point (bottom) with the KW self-dual fixed point (top). Moving to the right or to the left of this vertical line breaks statistical KW duality and leads to the ferromagnetic (topological) or paramagnetic (topologically trivial) phase. These are described by moving  the $\Theta$-angle of the NLSM away from $\pi$ in one direction or the other.

%%%%%%%%%%%%%%%%%%%%%%%%%%%%%%%%%%%%%%%%%%%%%%%%%%%%%%%%%%%%%%%%%%%%%%%%%%%%%%%%%%%%%%%
\section{The self-dual critical state} 
\label{sec:self-dual}
%%%%%%%%%%%%%%%%%%%%%%%%%%%%%%%%%%%%%%%%%%%%%%%%%%%%%%%%%%%%%%%%%%%%%%%%%%%%%%%%%%%%%%%

Let us start our discussion of mixed-state boundary criticality by characterizing the self-dual critical state first. 
Using numerical simulations, we first confirm its precise location at $\theta=\pi/4$ by calculating the coherent information for tuning the measurement angle $\theta$, see Fig.~\ref{fig:classicquantum}(d). 
Such scans for different system sizes also allow us to extract the correlation length exponent $\nu$ from a finite-size scaling analysis. 
We then sit on the critical mixed state with weak self-dual symmetry, and extract the comprehensive bulk and boundary universal conformal data from the entropy associated with the partitioning in Fig.~\ref{fig:classicquantum}b.

\begin{figure*}[tb!]
\centering
\includegraphics[width=\textwidth]{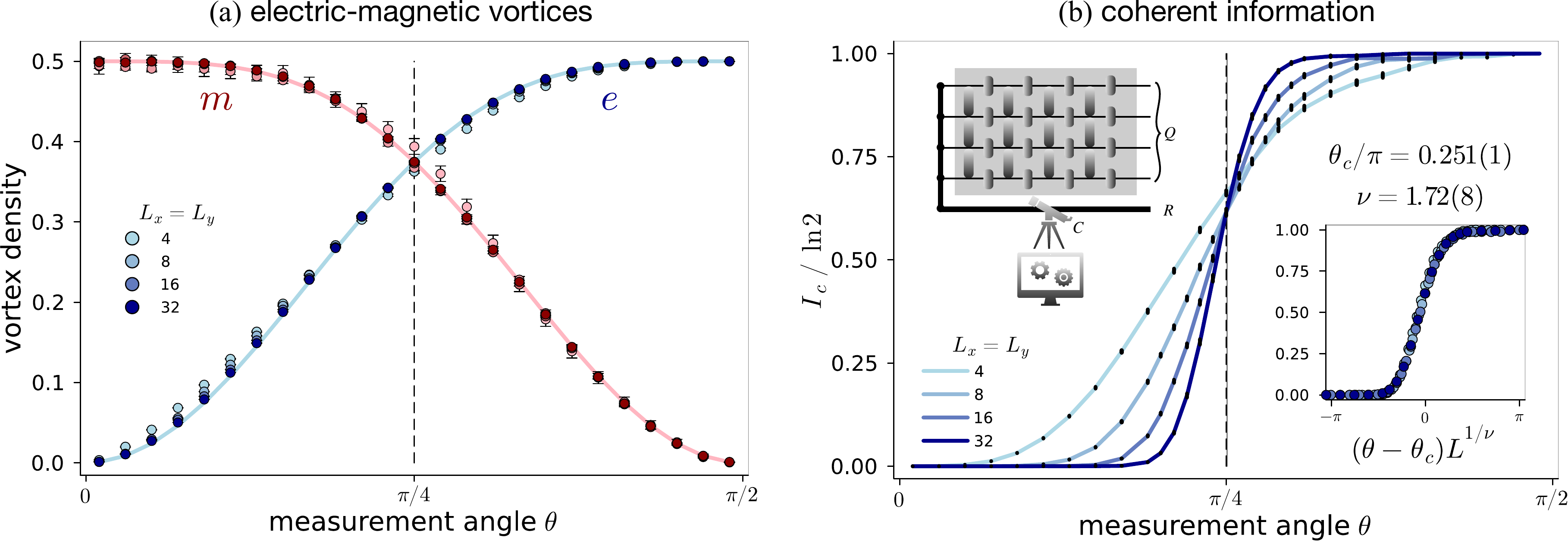}
\caption{{\bf Self-dual mixed state transition.}
(a)
{  Vortex densities.}
Shown are numerical (Monte Carlo) results for the $e$-flux density $\langle e \rangle$ and $m$-flux density $\langle m \rangle$ as a function of $\theta$ for different system sizes $L_x$. The solid lines are the analytic predictions of Eq.~\eqref{eq:vortex_densities}. It witnesses the electric-magnetic duality at $\theta=\pi/4$ where the density of the electric charges coincide with the density of the magnetic charges. The vertical line indicates the self-dual location, where the sampled $e$ and $m$-vortex density becomes identical.
(b)
{  Coherent information across the self-dual mixed state transition.}
   The coherent information can be computed as shown in the schematic inset on the upper left: 
   Logical information of $R$ is encoded into the quantum chain (Q), which then undergoes a monitored quantum evolution. 
   An external observer ``eavesdrops", attempting to recover the logical information -- with the coherent information reflecting the success of this recovery
   (with $I_c = \ln 2$ indicating perfect recovery).
   The solid vertical line indicates the critical point $\theta_c=0.251(1)\pi$ (fitted via the data collapse shown in the lower right inset), 
   which agrees with the exactly known self-dual location $\pi/4$ within its error bar. 
   In our numerical computation, $100\sim 1000$ independent $m$ configurations are sampled 
   and 1000 Monte Carlo sweeps for $e$ configurations are performed. 
   For $\theta< 0.2\pi$, we initiate from $e$-vortex-free for Monte Carlo sampling, while for $\theta>0.2\pi$ we initiate from random vortex configurations.
}
\label{fig:emflux}
\end{figure*}

%%%%%%%%%%%%%%%%%%%%%%%%%%%%%%%%%%%%%%%%%%%%%%%%%%%%%%%%%%%%%%%%%%%%%%%%%%%%%%%%%%%%%%%
\subsection{Coherent information}
\label{LabelSubsectionCoherentInformation}
%%%%%%%%%%%%%%%%%%%%%%%%%%%%%%%%%%%%%%%%%%%%%%%%%%%%%%%%%%%%%%%%%%%%%%%%%%%%%%%%%%%%%%%

The mixed-state phase transition of a long-range entangled phase upon decoherence can be diagnosed by the coherent information~\cite{Nielsen96coherentinfo, Gullans20scalabledecoder, Fan24coherinfo, teleportcode, Huang24coherinfofermion, Puetz24}, utilizing the fact that a long-range entangled phase can serve as a logical memory. 
Let $Q$ be a set of qubits of a general quantum system that is the focus of our interest. In order to determine the coherent information, which indicates whether this logical memory is still intact, one employs a  reference qubit $R$, whose state maximally entangles with the logical qubit state of the memory. 
For instance, when viewing the GHZ state as a repetition code, one would entangle the reference qubit together with the initial qubit chain  to form a joint GHZ state
$\ket{\psi_0}_{QR} =\ket{00\cdots 00}_Q\otimes\ket{0}_R + \ket{11\cdots 11}_Q\otimes\ket{1}_R$, as shown in Fig.~\ref{fig:emflux}(b).
When viewed from the Majorana fermion representation, this protocol can alternatively be interpreted as encoding and decoding the logical information in a Kitaev Majorana chain-based 
topological quantum computation setting~\cite{Nayak08tqc}. 
In the topologically nontrivial phase of the Majorana chain~\cite{Kitaev2001}, the unpaired Majoranas at the boundaries form, within their topologically degenerate 2-state manifold,
 a mixed state of even and odd parities. This mixed state corresponds to a 2-dimensional logical code space without any encoding yet. 
It can, however, be purified by maximally entangling it with a reference qubit $R$~\cite{Nielsen96coherentinfo}. 

The logical information being encoded into the system spreads through the bulk evolution, and meanwhile keeps being ``eavesdropped" by the local measurements of an observer~\cite{Vasseur2022qec, Ippoliti24learnability}, such that the spacetime evolution can be interpreted as a faulty channel. Whether the quantum logical information is stolen by the observer or whether it is still preserved in the system, is what is quantified by the quantum coherent information~\cite{Gullans20scalabledecoder, Fan24coherinfo, teleportcode, Huang24coherinfofermion}
\begin{equation}
I_c = S_{\rm vN}(\rho_{QC}) - S_{\rm vN}(\rho_{QCR}) \ ,
\label{eq:Ic1a}
\end{equation}
where we have included an additional index $C$ indicating the existence of classical information (obtained from measurements) accessible for a decoder. 
In our setup, $C$ refers to the bulk bits ``$(em)$'' that are {\it fully dephased} after the measurement(s), 
and $Q$ now refers to   a set of  ``quantum" qubits at the  boundary which are thus not dephased (see Fig.~\ref{fig:miptmixedstate}, where $Q=A$ in subpanel (b)), such that from Eq.~\ref{eq:mixedstate} [with $\rho_Q(em)$ normalized]
\begin{equation}
\rho_{QC} = \bigoplus_{em} P(em) 
\rho_Q(em)
\label{LabelEqDefRhoQC}
\end{equation}
is a block-diagonal matrix where each block corresponds to a different measurement outcome $(em)$, and a corresponding expression applies {\it in general} to any case of qubits $C$ and $Q$ where all the qubits contained in the {\it  set $C$}  of qubits are {\it fully dephased}. 
Thus the von Neumann entropies of the 2D mixed state are stemming from two parts -- the Shannon entropy of the bulk classical bits, and the trajectory-averaged von Neumann entropy of the boundary quantum bits~\footnote{The Rényi entropies of the mixed state in 
Eq.~(\ref{eq:mixedstate}), on the other hand, cannot be simply decomposed, but their description
rather involves an $n$-replica average of the ``$n$-th order purity' defined as
$tr(\rho^n)$, and reads
$S_{(n)}(\rho) =  \frac{1}{1-n} \ln {\rm tr} \rho^{n}=\frac{1}{1-n} \ln \sum_{em} P(em)^{n} e^{(1-n) S_{n}(em)}$
}
\begin{eqnarray}
\nonumber
&S_{\rm vN}(\rho_{QC}) & =-{\rm tr} \left(\rho_{QC} \ln \rho_{QC}\right) \\
\label{LabelEqSRhoQC}
&& = F + \sum_{em} P(em) 
S_{\rm vN}(\rho_Q(em)),
\end{eqnarray}
where $F$ was defined in Eq.~(\ref{eq:Fscal}) above.
Consequently, the coherent information~\eqref{eq:Ic1a} reduces to the quantum trajectory-averaged conditional entropy
\begin{equation}
I_c = \sum_{em} P(em) \ 
\Bigl
 (S_{\rm vN}(\rho_Q(em)) - S_{\rm vN}(\rho_{QR}(em))
\Bigr )
\, .
\label{eq:Ic1}
\end{equation}
Under the measurement channel, each trajectory $em$ yields a pure post-measurement state. Then $S_{\rm vN}(\rho_{QR}(em)) = 0$ and $S_{\rm vN}(\rho_Q(em)) = S_{\rm vN}(\rho_R(em))$. As a result, Eq.~\eqref{eq:Ic1} reduces to a measurement average of the entanglement entropy between the system and the reference qubit~\cite{Gullans20scalabledecoder} 
\[
	I_c =  \sum_{em} P(em) S_{\rm vN}(\rho_R(em)) \,.
\] 
Cast in the language of the random Ising statistical model, 
the reference qubit $R$ determines the boundary condition of the Ising layers. The density matrix element of $R$ is determined by the partition function under the corresponding boundary conditions, which can be represented by the tensor network as follows:
\begin{equation*}
\includegraphics[width=.66\columnwidth]{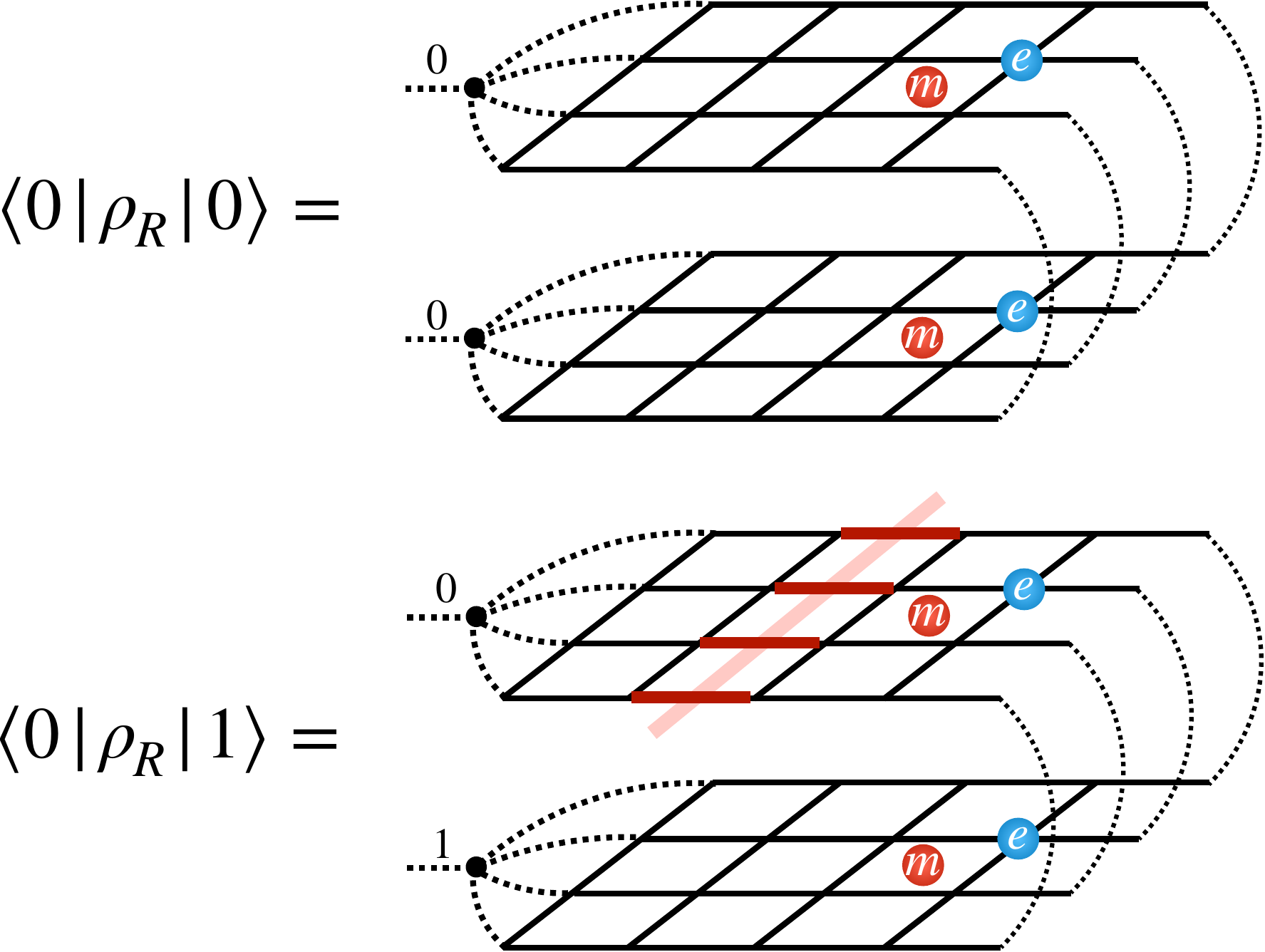} \ ,
\label{eq:refrho}
\end{equation*}
for a given typical snapshot of $em$. 
Here for $\bra{0}\rho_R\ket{0}$ the boundaries of both layers are pinned to product state $0$; in contrast, for $\bra{0}\rho_R\ket{1}$ the boundaries of the two layers are pinned to the opposite directions, which means a non-contractible domain wall or twist-defect line is inserted to the bulk, indicated by the shaded red string in the diagrammatic equation. The global Ising symmetry of the network guarantees that $\bra{0} \rho_R \ket{0} = \bra{1}\rho_R \ket{1}$ and $\bra{0} \rho_R \ket{1} = \bra{1}\rho_R \ket{0}$. 
The resultant coherent information, equal to the entropy of $R$, is determined by the {\it absolute} ratio $|\bra{0}\rho_R\ket{1}/\bra{0}\rho_R\ket{0}|$, which is simply the expectation value of inserting a domain wall defect. In the ordered phase $\theta\to\pi/2$, the domain wall decays exponentially with system size, $\bra{0}\rho_R\ket{1}/\bra{0}\rho_R\ket{0}\propto \mathcal{O}(e^{-L_x})$, which means that $R$ is almost diagonal and maximally mixed, thereby giving rise to a maximal coherent information $I_c\to \ln 2$. In the disordered phase $\theta\to 0$, $\bra{0}\rho_R\ket{1}/\bra{0}\rho_R\ket{0}\propto \mathcal{O}(1)$, and $\rho_R$ converges to a purified state with $I_c\to 0$. 
For the non-perturbative regime between these two limits, we perform a hybrid Monte Carlo and Gaussian fermion numerical computation of the coherent information $I_c$ while sweeping $\theta$, which is found to exhibit a clear level crossing near the self-dual point $\theta_c=\pi/4$ as shown in Fig.~\ref{fig:emflux}(b). 
The clean finite-size collapse indicates the critical length exponent $\nu=1.72(8)$. 

\begin{figure*}[t!] 
   \centering
   \includegraphics[width=\textwidth]{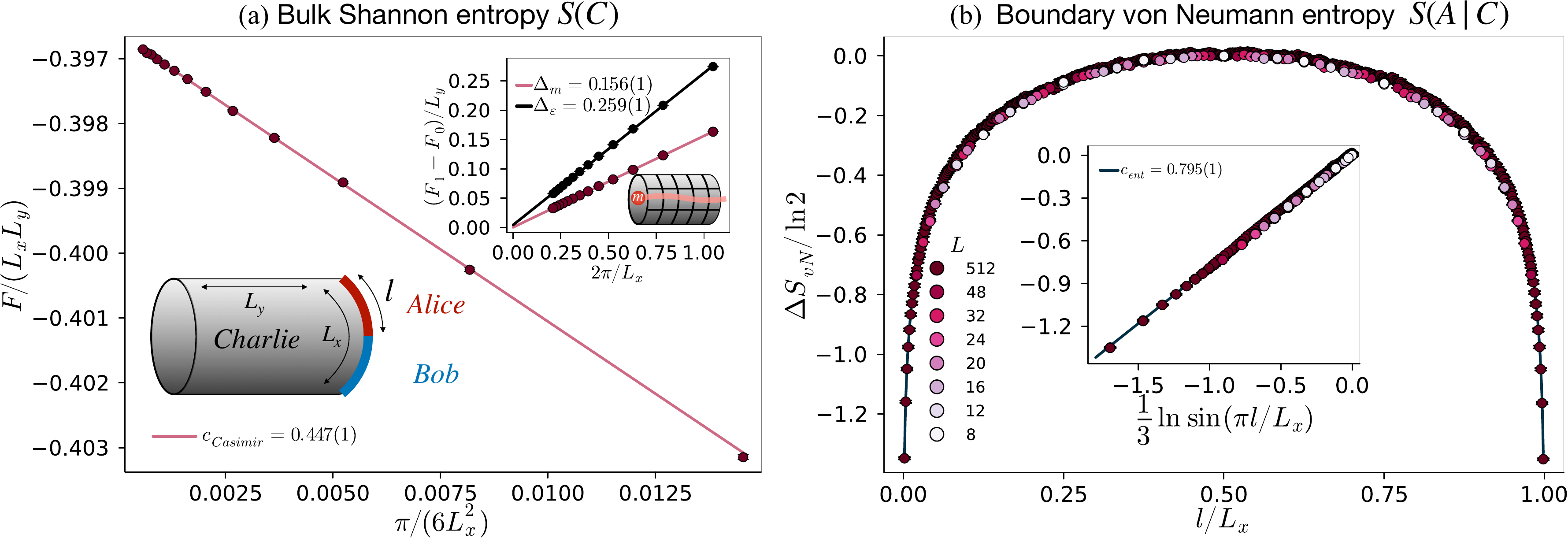} 
   \caption{
   {\bf Characterization of the self-dual critical theory} via the effective central charge and entanglement scaling at the point $\theta_c=\pi/4$. 
   For the system placed on a cylinder, we label the decohered bulk as ``Charlie" (C)
   and the two segments of the bipartition of the boundary quantum chain as ``Alice" (A) and ``Bob" (B). 
   (a) The Shannon entropy density of Charlie is equal to the free energy density of the classical statistical model~\eqref{eq:statmech}, 
   	which is governed by the {\it effective} central charge $c_{\rm Casimir}= 0.447(1)$. 
	Inset: twisting the boundary condition yields the scaling dimension of the vortex operator being $\Delta_m = 0.156(1)$. For comparison, the black line in the inset shows the scaling dimension $\Delta_\epsilon=0.259(1)$ of the first excitation without the twist i.e. in the even sector. The numerical sampling is performed for long cylinder of sizes $L_x=6-30$ and $L_y=100L_x$ with $30,000$ Monte Carlo samples, where the energy density is extracted deep in the bulk. 
  (b) The von Neumann entropy of Alice conditioned upon Charlie
  scales logarithmically with the chord length, 
  from which we extract the prefactor $c_{\rm ent}^{\rm vN}=0.795(1)$ as the scaling dimensions of the boundary twist operator. To show the data collapse, we further subtract the half-cut entropy: $\Delta S_{vN} \equiv S_{vN}(l)- S_{vN}(l=L_x/2)$ to get rid of the $L_x$ dependence. 
  The numerical calculation was performed for cylinder of sizes up to $L_x=512$ and $L_y\geq 2L_x$ in periodic spatial boundary condition, 
  with $2,000$ Monte Carlo sweeps for $L_x=512$ and $500, 000$ sweeps for $L_x=8-32$.  
  }
   \label{fig:pureSvN}
\end{figure*}

At $\theta>\pi/4$ the mixed 1D quantum state is a $Z_2$ SW-SSB phase. The strong Ising symmetry $(\prod_{j\in Q} X_j) \rho = \rho$ is spontaneously broken as evidenced by the exponential cost of the domain wall, or the long-range fidelity correlator~\cite{Wang24strtowksym}:
\begin{equation*}
\text{tr} \sqrt{
\sqrt{\rho} Z_i Z_j \rho Z_i Z_j \sqrt{\rho}}
=
\sum_{em} P(em) |\langle Z_i Z_j\rangle_{em}| \neq 0\ ,
\end{equation*}
where the overlap between the charge-neutral mixed state and its charged counterpart is reduced to the disorder average of the {\it absolute} value of the spin-spin correlation function, akin to an Edwards-Anderson correlation studied for the same model in Ref.~\cite{NishimoriCat}. The resultant state still preserves the {\it weak} $Z_2$ symmetry because of vanishing long-range correlation ${\rm tr} (\rho Z_i Z_j) = \sum_{em} P(em) \langle Z_i Z_j\rangle_{em} = 0$. 
Therefore the weak self-dual critical state ($\theta=\pi/4$) separates the SW-SSB ordered phase ($\theta>\pi/4$) from the symmetric trivial phase ($\theta<\pi/4$). 

%%%%%%%%%%%%%%%%%%%%%%%%%%%%%%%%%%%%%%%%%%%%%%%%%%%%%%%%%%%%%%%%%%%%%%%%%%%%%%%%%%%%%%%
\subsection{Bulk Shannon entropy}
%%%%%%%%%%%%%%%%%%%%%%%%%%%%%%%%%%%%%%%%%%%%%%%%%%%%%%%%%%%%%%%%%%%%%%%%%%%%%%%%%%%%%%%

Having precisely located the self-dual critical point at $\theta=\pi/4$ also in our numerical simulation, we compute the bulk Shannon entropy at this point
and find that it indeed follows the predicted~\cite{Pixley22MIPTCFT}
scaling law~\eqref{eq:Fscal} of a CFT, as shown in Fig.~\ref{fig:pureSvN}(a). 
Fitting our numerical data we extract an estimate of the effective central charge 
\[
	c_{\rm Casimir} = 0.447(1),
\] 
for the self-dual point. 

On a more technical note, we calculate this effective central charge estimate for the ``vacuum state", i.e.\ a state in the even sector for which 
we enforce the absence of an $m$-vortex through the hole of the cylinder by restricting $\prod_x s_{x,y=1}=+1$ 
as a Wilson loop surrounding the cylinder along the first column, which is akin to the periodic boundary condition of the clean Ising model without disorder. 
For this vacuum state, the typical scaling dimension of the $m$-vortex determines the typical correlation between two far separated $m$-vortices~\cite{Chalker02negative}. It can be deduced from the energy cost of threading an $m$-vortex through the hole of the long cylinder, which pulls a semi-infinite long line defect through the cylinder, see the inset of Fig.~\ref{fig:pureSvN}(a) for a schematic, equivalent to changing the boundary condition from periodic to antiperiodic. 
From the perspective of the 2D mixed state, this energy cost is equivalent to the relative entropy between the even and odd sector of the mixed state, when the density matrix is divided into two blocks according to the absence or presence of an $m$-vortex through the hole of the cylinder:
\begin{equation}
\frac{F_1 - F_0}{L_y}= -\sum_{em} P(em) \ln \frac{P_1(em)}{P(em)} \propto \frac{2\pi\Delta_m}{L_x}  \, ,
\end{equation}
where $P_{1}(em)$ is the partition function with an extra $m$-vortex through the cylinder hole, given the same $em$ configuration in the bulk. 
The additional entropy scales with a universal scaling dimension of the vortex, which from our numerical simulations summarized in the inset of Fig.~\ref{fig:pureSvN}(a),
can be extracted to be 
\[
	\Delta_m\approx 0.156(1) \,.
\] 
Due to the self-duality, the scaling dimension of the $e$-vortex in the bulk shall be identical to $\Delta_e = \Delta_m$. Note that the $e$-vortex corresponds to the spin operator $\sigma$ in Eq.~\eqref{eq:statmech}, while the $m$-vortex corresponds to the dual spin operator in the statistical model, which is often labeled by $\mu$~\cite{Chalker02negative}. Then the typical spin-spin correlation function in the bulk is governed by the same scaling dimension $\Delta_e$. We also compute the scaling dimension of the first excitation in the even sector, which is found to be 
\[\Delta_\epsilon\approx 0.259(1) \ .\]

\subsubsection*{Lyapunov spectrum}

Besides looking solely at the finite-size scaling of the vacuum energy, Eq.~(\ref{eq:Fscal}),
one can also look, more generally, at the full (single-particle) Lyapunov 
spectrum~\cite{Pixley22MIPTCFT,Vasseur24boundarytrsf, Fuji24monitoredmajoranalyapunov},
which reveals more information about the $(1+1)$D dynamics.
In particular, we compare clean Ising, Nishimori, and the weak self-dual criticality, see Fig.~\ref{fig:spec} below. 
As a consequence of the underlying 10-fold way~\cite{AltlandZirnbauer,10FoldWayReviewNJPhys2010} 
(Altland-Zirnbauer) symmetry class D, the
single-particle spectrum has particle-hole symmetry, being symmetric with respect to the zero energy level. The even and odd sectors are labeled by the black and red colors. The spectrum becomes denser upon increasing the system size $L_x$, which can be viewed as forming two ``bands" in the thermodynamic limit where the gap closes. 
The vacuum of the CFT is obtained by fixing the even sector and filling the lower band. Pumping an $m$-vortex through the cylinder hole traps a Majorana mode, drawing two levels from the band into the middle of the ``gap". They lie at exactly zero energy for the clean Ising critical point, resulting in  a two-fold degeneracy of the many-body spectrum due to the Majorana zero mode~\cite{Kitaev2001}. However, they split in the Nishimori critical state, which we attribute to the hybridization with the background $m$-vortices in the bulk. 
For Nishimori we numerically verify that $\Delta_m \approx 0.341(1)$, which roughly agrees with the scaling dimension $2\pi\Delta_m/L_x \approx 0.691(2)\pi/L_x$
reported in Ref.~\cite{Chalker02negative, Pujol2001, Picco2006}, see Appendix~\ref{sec:spec} for more details.
For the weak self-dual critical state with not only proliferated $m$-vortices but also proliferated  $e$-vortices, the midgap modes appear to converge to the zero levels, restoring the Majorana zero modes. Consequently, the many-body spectrum exhibits a double degeneracy in the odd sector akin to the clean Ising critical state, see Appendix~\ref{sec:spec}. 
\\

\begin{figure}[htbp]    
	\centering
   \includegraphics[width=\columnwidth]{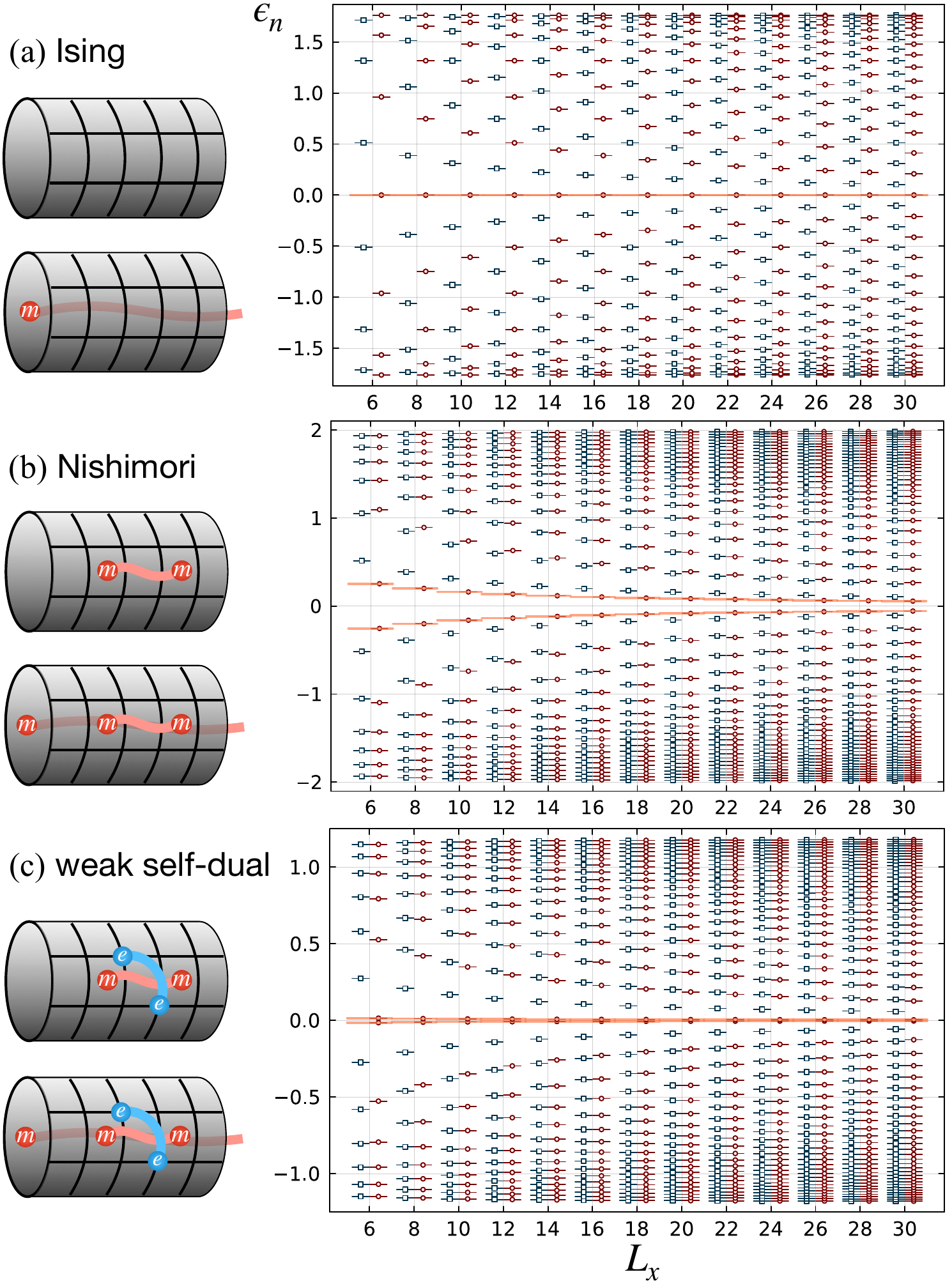} 
   \caption{{\bf Energy or Lyapunov spectra.}
   Shown are the single-particle free-fermion energy 
   %or 
   Lyapunov spectra
   for a cylinder of finite width for the Ising, Nishimori and weak self-dual critical states, respectively. 
   For each case, we show schematically a typical vortex configuration in a cylinder in the even sector, compared with its counterpart in the odd sector with an $m$-vortex threaded into the big hole of the cylinder.  The spectrum for the even sector is denoted by the black levels with squared marker, while that for the odd sector is denoted by the red levels with circle markers. The orange lines in the middle of the spectrum highlight the ``midgap" levels as created by the $m$-vortex threaded to the cylinder hole. 
   (a) For the clean Ising critical state placed on a cylinder of finite width (a quasi-1D system), 
   	an $m$-vortex pulls two fermion states from the band onto the mid-gap zero-energy level, 
	forming a pair of exact Majorana zero modes, akin to the topological Majorana chain~\cite{Kitaev2001}. 
   (b) A hybridization with the background $m$-vortices results in the energy splitting of the zero modes. 
   (c) In the presence of $e$-vortices, the Majorana zero modes are restored. 
   }
   \label{fig:spec}
\end{figure}

%%%%%%%%%%%%%%%%%%%%%%%%%%%%%%%%%%%%%%%%%%%%%%%%%%%%%%%%%%%%%%%%%%%%%%%%%%%%%%%%%%%%%%%
\subsection{Boundary entanglement entropy}
\label{LabelSubsectionBoundaryEntanglementEntropy}
%%%%%%%%%%%%%%%%%%%%%%%%%%%%%%%%%%%%%%%%%%%%%%%%%%%%%%%%%%%%%%%%%%%%%%%%%%%%%%%%%%%%%%%

For the  2D mixed state at hand we have identified in our discussion above of the coherent information the fact that its von Neumann entropy stems from two parts -- the Shannon entropy of the bulk classical bits, which we have discussed in the previous Section, and a trajectory-averaged von Neumann entropy of the boundary quantum bits. The latter is what we will turn to now. The key notions  for this discussion were developed in 
Eqs.~(\ref{LabelEqDefRhoQC}, \ref{LabelEqSRhoQC}) of Sect.~\ref{LabelSubsectionCoherentInformation} above.

The von Neumann entropy of the boundary quantum state exhibits CFT scaling 
(``entanglement arcs'') of the form
\begin{equation}
\begin{split}
	S_{AC}-S_{C} &= \sum_{em} P(em) S_{\rm vN}(\rho_A(em)) \\
	&= 
	\frac{1}{3} c_{\rm ent}^{\rm vN} \cdot \ln\left(\frac{L_x}{\pi} \sin\frac{\pi l}{L_x}\right) + \ldots \,,
	\label{eq:CalabreseCardy}
	\end{split}
\end{equation}
when considering the {\it conditional entanglement entropy}
$S_{AC}-S_C$, with $A$ (standing for ``Alice'') denoting a segment (of length $l$) of the quantum state at the boundary (of length $L_x$ - considering here periodic boundary conditions), where we abbreviate
$S_{\rm vN}(\rho_{AC})$ by $S_{AC}$, using the notation introduced in
Eq.~(\ref{LabelEqDefRhoQC}), and  the statement in Eq.~(\ref{LabelEqSRhoQC}) -- compare
%see
the caption of Fig.~\ref{fig:pureSvN}
and the drawing
in panel (a) of this Figure.
Here $S_C$ equals the bulk quantity  denoted by $F$ in
Eq.~(\ref{eq:Fscal}), that is
\begin{equation}
\label{LabelEqSCFExplicit}
S_C=F=
- \sum_{em} P(em) \ln P(em).
\end{equation}
The so-defined variant $S_{AC}$ of the entanglement entropy is the quantity mentioned 
in Fig.~\ref{fig:classicquantum}b
of the  introductory section of this paper, and is equivalent to the measurement trajectory-averaged von Neumann entropy between Alice and Bob, as depicted in the drawing in the inset of Fig.~\ref{fig:pureSvN}(a).
Fitting the numerically obtained
``entanglement arcs'' of Fig.~\ref{fig:pureSvN}(b) we can determine the 
universal number $c_{\rm ent}^{\rm vN}$
(which characterizes the typical
critical exponent of the boundary condition changing (bcc) twist operator - see below)
\[
	c_{\rm ent}^{\rm vN}\approx 0.795(1) \,.
\]%
Technically, we perform a free-fermion evolution for a fixed random trajectory $em$, and use  Born's rule to sample the ensemble of trajectories. 
In this numerical context, we can easily generalize from the von Neumann entropy to the family of higher-order Rényi entropies and calculate their respective 
universal numbers $c_{\rm ent}^{(n)}$ for the $n^{\rm th}$ R\'enyi entropies for the Born-averaged ensemble of trajectories (as above). We find that these universal numbers monotonically decrease with the 
%order 
R\'enyi order $n$ towards the limiting value
\[	
	c_{\rm ent}^{(\infty)} = 0.484(1),
\]
 in the $\infty$-Rényi-order limit, see Appendix~\ref{sec:renyi} for supporting numerical data.
 
While, as already discussed in the introductory section of the paper,
$c_{\rm ent}$ is identical to the 
usual
central charge $c$
in the case of translationally invariant, not-monitored (``non-random'')
%for 
unitary CFTs~\cite{Cardy86, PhysRevLett.56.746},
for the monitored (or: in general random) non-unitary CFTs at hand it captures the
(typical) scaling dimension of 
so-called boundary condition changing (bcc) operators. 
The latter describe the scaling of bipartite entanglement of the non-unitary boundary CFT~\cite{Ludwig2020, Pixley22MIPTCFT, Vasseur24boundarytrsf, Ludwig24clifford}, 
and should not be confused with $c_{\rm Casimir}$, a universal finite-size scaling amplitude at criticality, which can be extracted, as discussed in the previous Section, from the Shannon entropy of the measurement record~\cite{Pixley22MIPTCFT, Ludwig24clifford, Vasseur24boundarytrsf}. 

The universal number $c_{\rm ent}$  also determines the conditional mutual information (CMI), a non-linear correlation related to the reversibility of the quantum channel, which can signal the mixed state phase transition point~\cite{Hsieh24markovlength}. For this quantity we can partition the 1D chain into 4 segments, denoted by $A, B, D, E$, respectively, with the length being $|A|=|B|=a$, $|D|=l-a$ such that the distance between the centers of $A$ and $B$ is $l$:
\begin{equation*}
\includegraphics[width=.8\columnwidth]{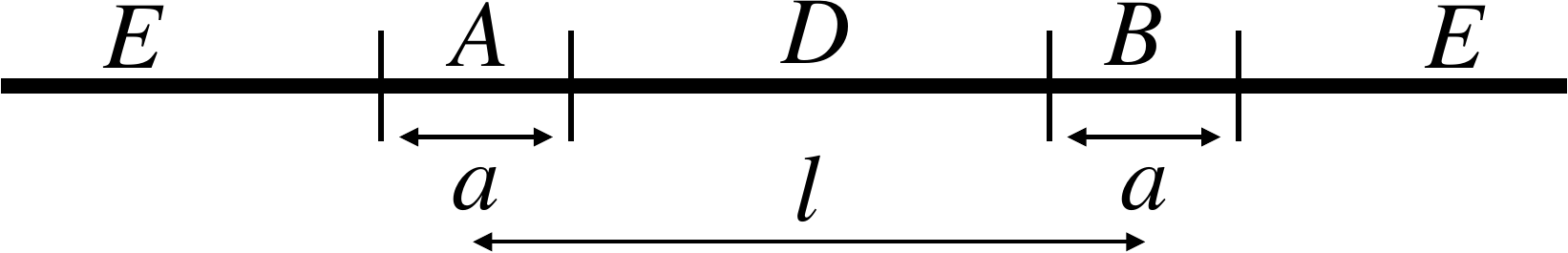} \ .
\end{equation*}
Then we investigate the CMI scaling with distance as a function of $L_x$. 
By making use of the block diagonal property of the density matrix, the Shannon entropy is subtracted from the CMI of the total system, which is reduced into the Born average of the CMI of each post-measurement pure state
\begin{equation}
I(A:B|D) = \sum_{em}P(em) \left(S_{AD}+S_{BD} - S_D - S_{ABD}\right)_{em} \ .
\end{equation}
Each term is the Born-averaged entanglement entropy of a contiguous block, which is a function of its length only, due to its average translation symmetry. 
We consider the regime $a\ll l$. In the area law phase, $S$ saturates to a constant $\sim \ln \xi$ that depends only on the characteristic length scale but not the subsystem size, and thus $I(A:B|D) \sim 0$, signaling a non-zero Markov gap and finite Markov length. 
At the critical point, using the numerically confirmed scaling law Eq.~\eqref{eq:CalabreseCardy}, we find
\begin{eqnarray}
I(A:B|D)\rvert_{\theta=\pi/4} & = & \frac{1}{3}c_{\rm ent}^{\rm vN} \cdot \ln\frac{\sin^2(\frac{\pi l}{L_x})}{\sin\left(\frac{\pi(l-a)}{L_x}\right)\sin\left(\frac{\pi(l+a)}{L_x}\right)} \nonumber \\[1mm]
&\sim &
\begin{cases}
\frac{1}{3} c_{\rm ent}^{\rm vN} \cdot \frac{a^2}{l^2} & \ ,\ l\ll L_x\\[2mm]
\frac{1}{3} c_{\rm ent}^{\rm vN} \cdot \frac{\pi^2a^2}{L_x^2}  & \ ,\ l=L_x/2\\
\end{cases}
\end{eqnarray}
which decays in an inverse square law with a universal prefactor $c_{\rm ent}^{\rm vN}/3$. Physically, the divergence
of the Markov length implies  the {\it irreversibility} of the channel which occurs at the critical point that separates two distinct mixed state phases~\cite{Hsieh24markovlength,Hsieh24spacetimemarkov}.

%%%%%%%%%%%%%%%%%%%%%%%%%%%%%%%%%%%%%%%%%%%%%%%%%%%%%%%%%%%%%%%%%%%%%%%%%%%%%%%%%%%%%%%
\section{Mixed-state phase diagram and RG flows}
\label{sec:phase_diagram}
%%%%%%%%%%%%%%%%%%%%%%%%%%%%%%%%%%%%%%%%%%%%%%%%%%%%%%%%%%%%%%%%%%%%%%%%%%%%%%%%%%%%%%%

\begin{figure}[t!] 
   \centering
   \includegraphics[width=\columnwidth]{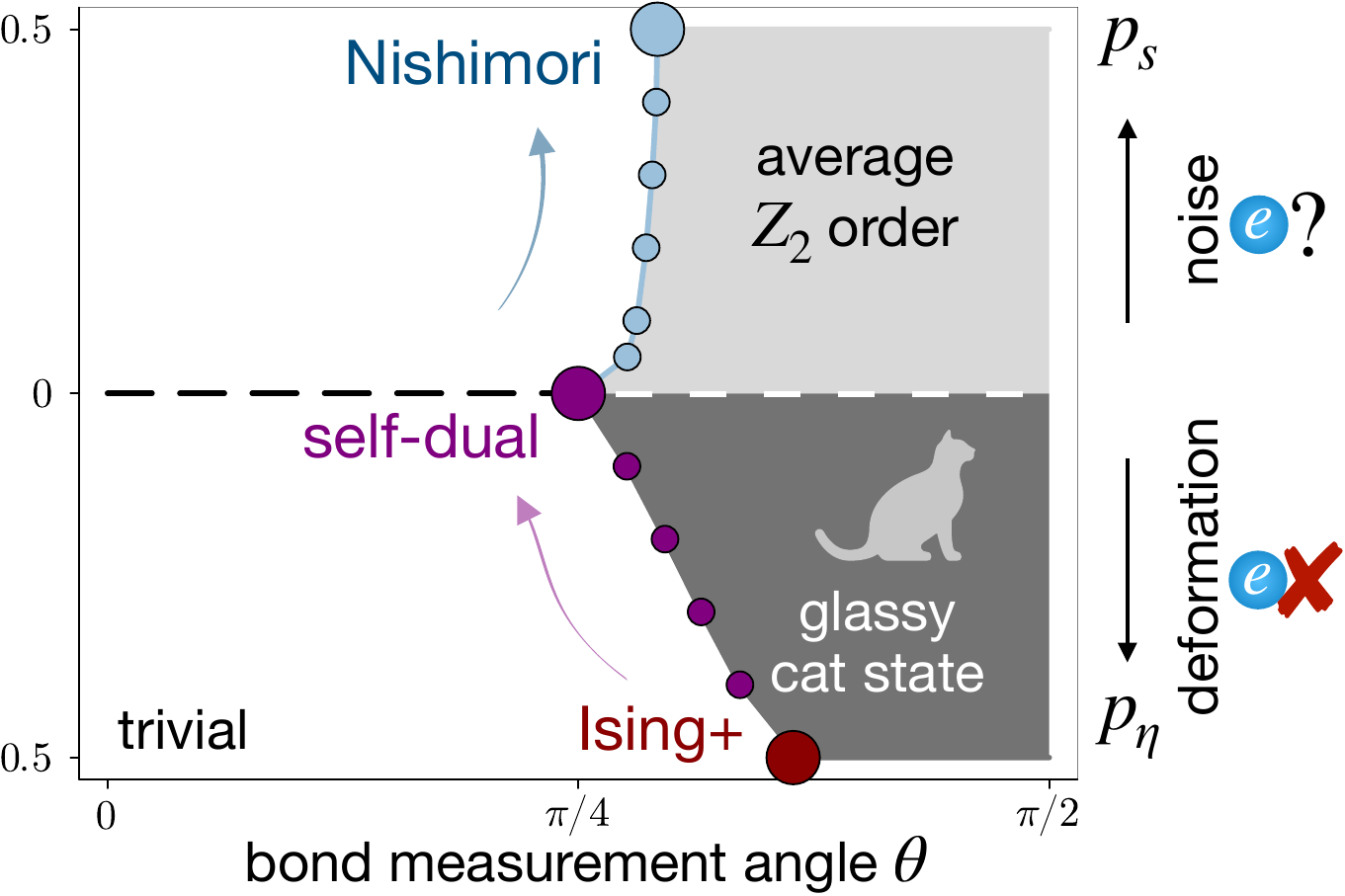} 
   \caption{{\bf Noisy phase diagrams with measurement noise and deformation}. 
   Different perturbations can be used to
   tune the $e$-vortices away from duality: the upper panel uses the (measurement) noise to mask and blur
   the existence of an $e$-vortex, while the lower panel uses the deformation to damp it. 
   Within the phase diagram, the white region is the trivial state; the dark gray region indicates the strong-to-weak spontaneous-symmetry-breaking (SW-SSB) of the
   $Z_2$ ordered phase with long-range glassy correlation $[|\langle Z_i Z_j\rangle|]\neq0$ and $[\langle Z_i Z_j\rangle]=0$ (easily verfied at $\theta=\pi/2$); the light gray region has only weak $Z_2$ symmetry due to the noise, but exhibits the same long-range correlation. There are 3 paradigmatic critical points in this phase diagram: the Nishimori critical point, the weak self-dual critical point, and the Ising+ critical point.
    The arrows in the phase diagram outline the RG flow indicated by our numerical results. 
  Here we unfold the two perpendicular phase diagrams (by tuning $p_s$ and $p_\eta$, respectively) into the plane, and their crossing line $p_s=0=p_\eta$ is not a phase transition generically. 
  }
   \label{fig:noisyphasediagram}
\end{figure}

Let us now move away from the self-dual line ($p_s=p_\eta=0$)  in our phase diagram of Fig.~\ref{fig:noisyphasediagram} and consider  two perturbations that both explicitly break self-duality.
First in the form of incoherent noise (upper half in our phase diagram), then in form of a coherent deformation (lower half in our phase diagram) 
and, last but not least, we restore 
%the
strong self-duality by a duality-preserving coherent deformation. 

%%%%%%%%%%%%%%%%%%%%%%%%%%%%%%%%%%%%%%%%%%%%%%%%%%%%%%%%%%%%%%%%%%%%%%%%%%%%%%%%%%%%%%%
\subsection{Breaking self-duality by measurement noise}
\label{sec:incoherent_noise}
%%%%%%%%%%%%%%%%%%%%%%%%%%%%%%%%%%%%%%%%%%%%%%%%%%%%%%%%%%%%%%%%%%%%%%%%%%%%%%%%%%%%%%%

\begin{figure*}[t!] 
   \centering
   \includegraphics[width=\textwidth]{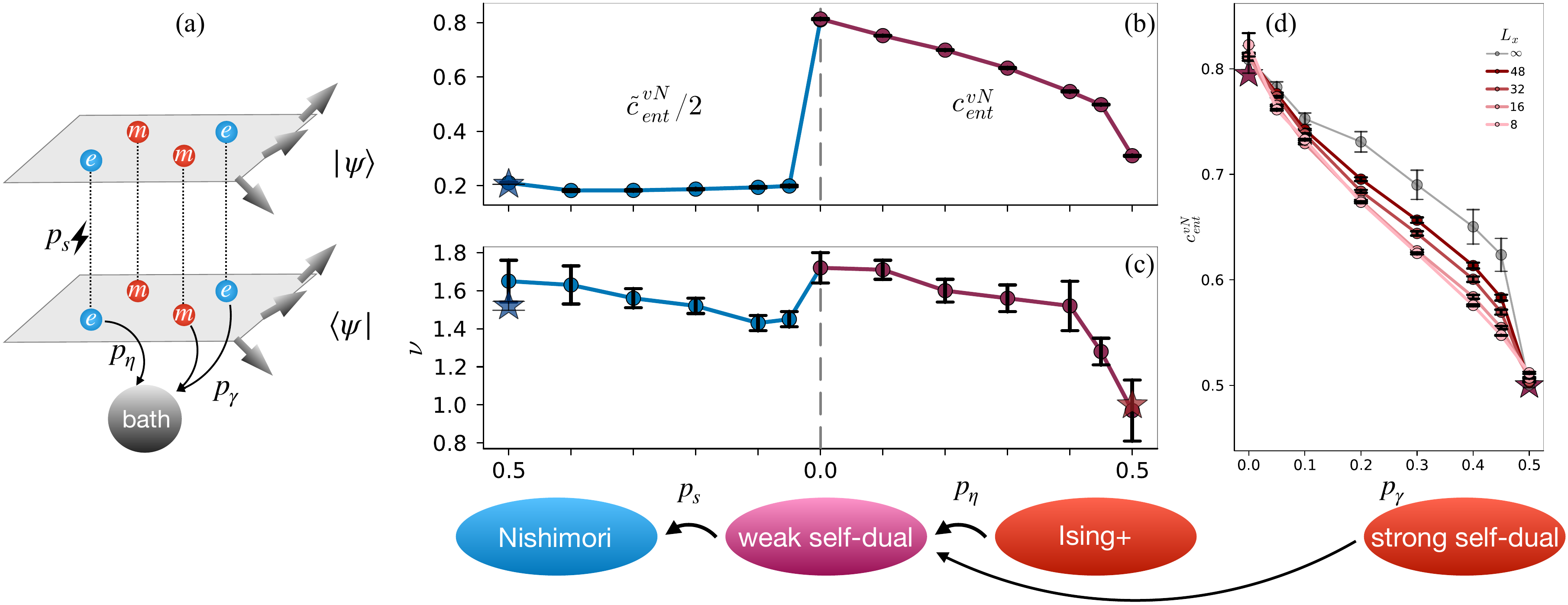} 
   \caption{
   {\bf RG flow between the critical states}, tracking the critical line in the phase diagram in Fig.~\ref{fig:noisyphasediagram}.
   (a) Schematic of the three types of disturbances.
   Each layer denotes one tensor network as the spacetime evolution of the $(1+1)$D quantum state, 
   subjected to random $e$- and $m$-vortices. 
   The site measurement noise $p_s$ blurs the $e$-vortex configuration, 
   the deformation $p_\eta$ absorbs the $e$-vortices to the bath, 
   while the coherent deformation $p_\gamma$ absorbs the $e$- and $m$-vortices together, maintaining the self-duality. 
   (b) $\tilde{c}_{\rm ent}^{\rm vN}$ is the scaling exponents for the von Neumann entanglement entropy of the $(1+1)$D noisy quantum state, viewed as a Choi pure state, where the disorder is averaged according to the Born's rule. $c_{\rm ent}^{\rm vN}$ governs the trajectory averaged pure state entanglement entropy under the coherent deformation. 
   (c) $\nu$ is the critical length exponent extracted from FSS for coherent information.
   At $p_s=50\%$ the Nishimori critical point, $\nu=1.65(11)$, $\tilde{c}_{\rm ent}^{\rm vN}=0.421(5)$ based on our calculation of sizes up to $L_x=32$. For comparison, the star markers denote the data using larger system size $O(10^3)$ calculation, adapted from Puetz {\it et al.}~\cite{Puetz24}: $\nu=1.52(2)$, $\tilde{c}_{\rm ent}^{\rm vN}= 0.410(1)$. At $p_\eta=0.5$ the Ising+ critical point, $\nu=1.00(5)$, $c_{\rm ent}^{\rm vN}=0.310(2)$. The data is computed by finite size scaling of size $L_x=4,8,16,32$, whose data collapse is shown in Appendix Sec.~\ref{sec:fss}. 
   Here we plot $\tilde{c}$ by half because it expresses the entropy of the Choi state in the double Hilbert space. 
   (d) Flow from strong to weak self-dual symmetry, witnessed by the entanglement entropy scaling exponent $c_{\rm ent}^{\rm vN}$. 
   Referring to all subpanels (a)-(d), the computation is performed 
   %in 
   with open boundary conditions 
   for $p_s$ noisy regime, and periodic boundary conditions
   for $p_\eta$ and $p_\gamma$ coherent deformation regime.  
}
   \label{fig:criticalexp}
\end{figure*}

The presence of $em$ asymmetric
measurement
``readout'' noise explicitly breaks the self-duality. Here we consider measurement noise for the site qubit, controlled by a parameter $p_s$. For the intermediate topological toric code, the measurement noise of the $e$-vortices introduces an effectively finite temperature which, in two spatial dimensions, immediately destroys its topological order~\cite{Castelnovo2007,Hastings11finiteT}.
Nonetheless, the $m$ vortex keeps being subjected to
quenched (Born-rule measurement) disorder, such that the phase transition persists even in the presence of this noise and turns into
the Nishimori transition
for $p_s = 50\%$. In this limit,
the noise is equivalent, as we show, to
 tracing out the site qubits and can be mapped to the RBIM in the replica limit $N \to 1$, thus rigorously corresponding to the Nishimori transition~\cite{NishimoriCat, JYLee, Chen24nishimori}. 
Upon close inspection of the
finite-size scaling behavior
at the phase boundary appearing
upon introducing the noise (see Fig.~\ref{fig:noisyphasediagram}), we find that its universality qualitatively changes
immediately (Fig.~\ref{fig:criticalexp}), indicating an RG flow from self-dual criticality to Nishimori as we will discuss in the following.
Similar considerations can, by duality, be applied to $m$-vortex noise. 

In the presence of a measurement error the state corresponding to each of the bulk classical states $\ket{em}$ becomes a noisy mixed state
\begin{equation}
\ket{\rho(em)}\rangle = \sum_{e'} P(e'|e) \ket{\psi(e'm)}\rangle \ ,
\label{eq:noisystate}
\end{equation}
where $e$ is the measurement record, while $e'$ is the vortex that the Majorana fermion truely experiences, whose correlation is determined by the noise probability: $P(e'|e)=(1-p_s)\delta_{e',e} + p_s (1-\delta_{e',e}) = P(e|e')$ for each individual site independently. 
Here we use the double-ket notation to denote the density matrix as a purified state in the doubled Hilbert space~\cite{Lee23decoher, Altman23errorfielddouble, Wang23aversym, Xu24higherformweaksym, You24weaksym, Luo24weaksym, Fan24coherinfo, Yang24purifiedmpo}, which is sometimes referred to as ``Choi state", due to the Choi-Jamiolkowski isomorphism~\cite{choi1975completely, jamiolkowski1972linear} that maps a channel to a state. For example, for a pure state $\ket{\psi(em)}\rangle = \ket{\psi(em)}\bra{\psi(em)}$ its Choi state is simply a tensor product of the two states, which can be further glued by the noise Kraus operators, see Fig.~\ref{fig:criticalexp}(a) for schematic. 
In detail, it can be represented as a bilayer tensor network state, where the two layers correspond to the {\it double} Hilbert space:
\begin{equation}
\includegraphics[width=.8\columnwidth]{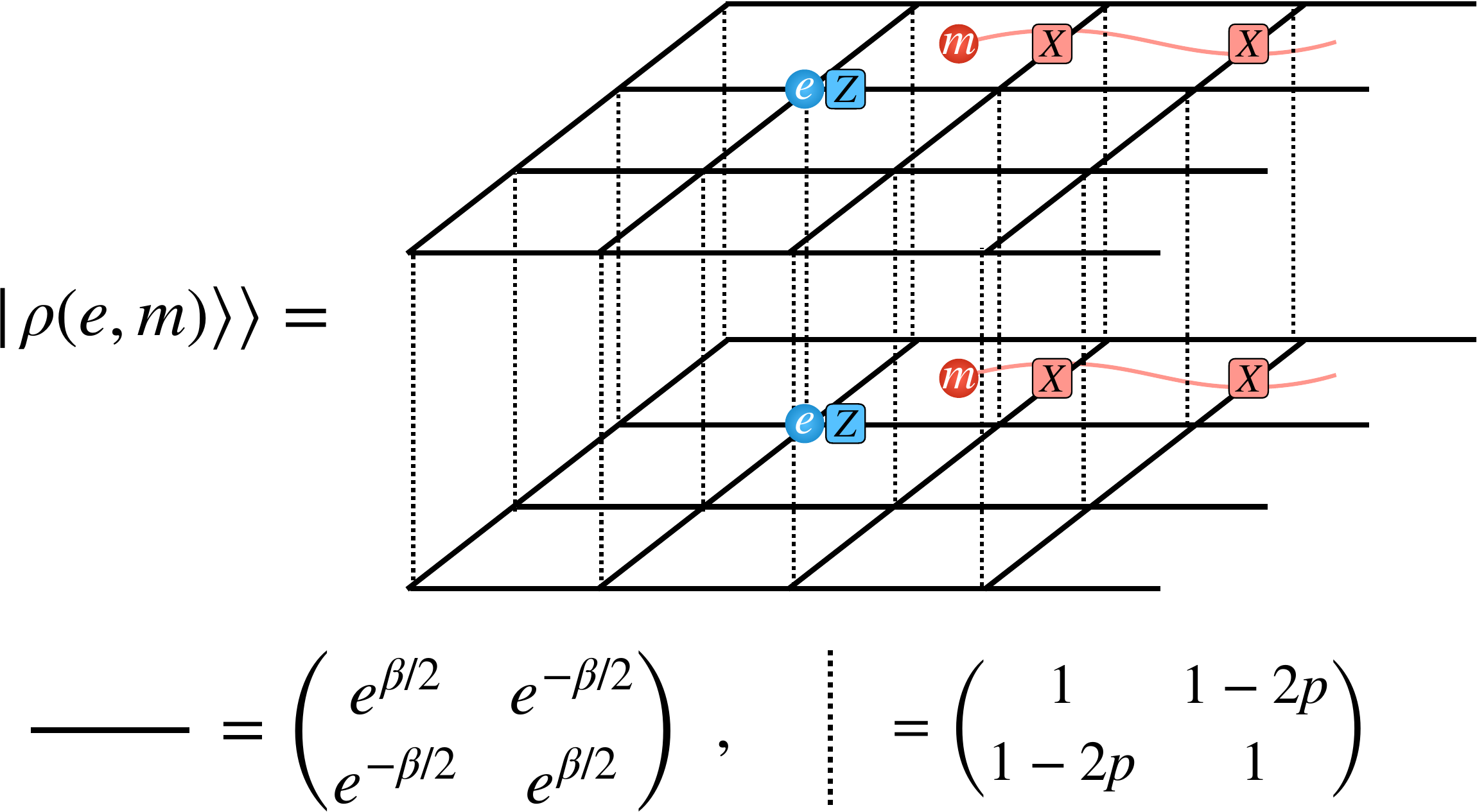}  \ ,
\label{eq:choistate}
\end{equation}
where each bond is a 2-by-2 matrix as written. The intra-layer solid bond originates from the pure state Eq.~\ref{eq:gaugetransform}, and the inter-layer dashed bond is glued by the noisy Kraus operators, see Appendix Sec.~\ref{sec:ChoiTN} for derivation. 
The trace of the original physical qubits is equivalent to an overlap with a Bell state between the ket and bra: $\text{tr}\rho = \langle \bra{\text{Bell}}\ket{\rho}\rangle$, while the purity becomes the norm of the Choi state: $\text{tr}\rho^2 = \langle \braket{\rho}\rangle$. 
Thus the corresponding Born's probability is also accordingly distorted by the noise: $\tilde{P}(em)  = \langle \bra{\text{Bell}}\ket{\rho(em)} \rangle = \sum_{e'} P(e'|e) P(e'm)$. 

It is important to note that the effective 2D statistical model thus becomes a bilayer Ising model, where the noise probability tunes the interlayer Ising coupling, and the random measurement outcome injects vortices to both layers. The (1+1)D dynamics is thus a quantum {\it ladder} of Ising spins or Majorana fermions, with interchain interactions beyond the non-interacting fermion regime. 
Explicitly, in the language of  the statistical mechanics model described in 
Eq.~(\ref{LabelEqDefsandtTransferMatrix}), (\ref{eq:statmech}), the partition function for the circuit  reads in the presence of measurement noise, in a fixed quantum trajectory (fixed configuration of $e$ and $m$ vortices, or equivalently  of the signs $s_{ij}$ and $t_{ij}$)
\begin{eqnarray}
\label{LabelEqStatMechWeightMeasurementNoiseParitionFct}
\left [
{{\cal Z}(em)}^2
\right]_{\beta_p}
:=
\sum_{\{\sigma_j\}}
\sum_{\{\tau_j\}} \exp
\left (
- {\cal H}[\{\sigma_j\}, \{\tau_j \}
]
\right ) \,,
\end{eqnarray}
where
\begin{eqnarray}
\nonumber
&&
- {\cal H}[\{\sigma_j\}, \{\tau_j\}
]
=
{\beta\over 2}
\sum_{\langle i,j \rangle}
s_{i,j} (
\sigma_i \sigma_j + \tau_i \tau_j)+
\\ 
\label{LabelEqStatMechWeightMeasurementNoise}
&&
+i \pi \sum_{\langle i j \rangle}
{1 - t_{ij}\over 2} \ 
{1 - \sigma_i\sigma_j \tau_i \tau_j \over 2}
+ \beta_p \sum_j \sigma_j \tau_j \,,
\end{eqnarray}
where $\beta_p$ denotes the strength of the measurement noise related to $p_s$ by $e^{-2\beta_p}= 1-2p_s $, and $\sigma_j$ and $\tau_j$ denote the Ising spins in the ket and bra parts of the circuit, respectively. The `vortices' (signs) $s_{ij}$ and $t_{ij}$ act  both simultaneously on ket and bra parts. Note that in the absence of measurement noise, $\beta_p=0$, the expression  in 
Eq.~\eqref{LabelEqStatMechWeightMeasurementNoiseParitionFct}
 above is the square of the partition function ${\cal Z}(em)$ from Eq.~(\ref{eq:statmech}), expressing the fact that $s_{ij}$ and $t_{ij}$ act simultaneously on the ket and the bra part. The term multiplying $\beta_p$ represents a coupling between ket and bra.
In the maximally noisy limit $p_s=1/2$, and $\beta_p=+\infty$, the ket and bra are locked to have identical configurations. The original bilayer Ising model then reduces to a single layer of RBIM with the same $m$ vortices, where the Ising interaction strength is doubled from $\beta/2$ to $\beta$. This can be derived using the tensor network representation~\eqref{eq:choistate}, or by using Eq.~\eqref{eq:statmech} to see that
\begin{equation*}
\sum_e \mathcal{Z}(em)^2 
\propto \sum_{\mathbf{\sigma}} e^{\beta \sum_{ij} s_{ij} \sigma_i\sigma_j} \ .
\end{equation*}

It is important to stress that the noisy mixed state for $0<p_s<1/2$ is no longer Gaussian in the Majorana representation, and the noise amounts to interactions between the Majorana fermions, which are non-interacting in the limits $p_s=0, 1/2$~\cite{GRL2001}.
A small value of $p_s$ amounts to RG-relevant fermion interactions which break statistical (`weak'') KW symmetry, while these fermion interactions are numerically found RG-irrelevant in the vicinity of the infrared Nishimori point at $p_s=1/2$.
The lack a Gaussian Majorana representation for $0 < p_s < 1/2$ also implies that numerical calculations (e.g. of the Neumann coherent information, etc.) are generically hard to perform for those values of $p_s$.
However, we can turn to a Born measurement average of the second Rényi coherent information by adapting Eq.~\eqref{eq:Ic1}, as it is a {\it linear} physical observables w.r.t. the Choi state
\begin{equation}
\begin{split}
I_c^{(2)} =& \sum_{e,m} \tilde{P}(em) [S_Q^{(2)}(em) - S_{QR}^{(2)}(em)] \\
=& \sum_{e,m} \tilde{P}(em)  [-\ln \langle\langle \hat{B}_R \rangle\rangle] \ ,
\end{split}
\end{equation} 
where $\hat{B}_R = 2\ket{\text{Bell}}\rangle_R\langle\bra{\text{Bell}}$ is two times of the Bell projector of the reference qubit. Its expectation value can be efficiently calculated by our tensor network representation~\footnote{Note that we supplement the finite-size random tensor network computation by a Markov chain Monte Carlo sampling as in Ref.~\cite{NishimoriCat}}, see Appendix Sec.~\ref{sec:ChoiTN} for details. 

For a {\it generic mixed} quantum state, the location of the
phase transition in parameter space and its universality class could depend on the order of the Rényi quantity, as different copies of the density matrices are involved and are mapped to statistical models with different layers~\cite{Fan24coherinfo} which often become singular at different parameter values for different R\'enyi quantities. 
However, in our case here, the limit $p_s=0$ is an
{\it $em$-trajectory-resolved pure state} whose different Rényi entropies are related to different scaling operators within the {\it same} critical 
theory~\cite{Ludwig2020}.
Analogously, in the maximally noisy limit $p_s=50\%$,
it becomes an {\it $m$-trajectory-resolved, fully dephased, diagonal density matrix}, whose  Rényi entropies all diverge, as in the $p_s=0$ case,
at the {\it same} transition point~\cite{Ludwig2020,GRL2001}. 
Specifically, our $p_s=50\%$ point can be mapped to the 2D RBIM along its Nishimori line~\cite{NishimoriCat, JYLee, Chen24nishimori}. Indeed, the numerical exponents we compute in Fig.~\ref{fig:criticalexp} are consistent with those obtained for the Nishimori criticality in Refs.~\cite{Puetz24, Pujol2001,Chalker2002, Picco2006, Queiroz2006squarelatt, sasagawa2020entanglement}. 
As shown in Fig.~\ref{fig:noisyphasediagram}, for $p_s>0$ and $\theta>\theta_c$, we expect the boundary state to be dephased in the thermodynamic limit into a classical glassy Ising-ordered mixed state. This is evidenced by the numerically computed noisy coherent information $I_c\to 0$ under finite-size scaling (see Fig.~\ref{fig:enoise} in Appendix~\ref{sec:fss}). The onset of measurement noise leads to a dephasing effect that immediately collapses the logical qubit into a logical classical bit in thermodynamic limit. In the Majorana representation, this means the topological protected logical qubit composed of the two edge Majoranas is unstable against the (noise-induced) proliferation of $e$-vortices in spacetime. 
This does not contradict the topological protection because the $e$-vortices are highly {\it non-local} topological defects. 
If a bond measurement error is turned on in addition, the ordered phase and the Nishimori criticality survive until a finite threshold, as realized in a noisy digital quantum processor~\cite{Chen24nishimori}. 

\subsubsection*{RG flow}
To understand the renormalization group (RG) flow for the manifold of critical points for $0 \leq p_s \leq 50\%$, we have calculated the entanglement entropy for the Choi state $\ket{\rho(em)}\rangle$. The entropy is averaged over the Born ensemble given by $\tilde{P}(em)$, and agrees with the logarithmic scaling as in Eq.~\eqref{eq:CalabreseCardy} with a scaling prefactor denoted by $\tilde{c}_{\rm ent}^{(n)}$, which should be distinguished from the pure state ones, $c_{\rm ent}^{(n)}$. 
Only in the noiseless limit, where the Choi state is simply a tensor product of the ket and bra wave functions, are the two trivially related by $\tilde{c}_{\rm ent}^{(n)} = 2c_{\rm ent}^{(n)}$ due to the additivity property of the entanglement entropies. 
Estimates in the presence of noise $0 \leq p_s \leq 50\%$ are shown in Fig.~\ref{fig:criticalexp}(a) for the von Neumann entropy prefactor  $\tilde{c}_{\rm ent}^{\rm vN}$.
In addition, we plot the scaling exponents for the Rényi entanglement entropy of the Choi state in Fig.~\ref{fig:criticalexp}(b). 

Looking at the data in Fig.~\ref{fig:criticalexp}(b,c), we see a sudden drop of the von Neumann entropy prefactor $\tilde{c}_{\rm ent}^{\rm vN}$ upon introducing a finite noise
$p_s > 0$, as well as an immediate (though less dramatic) change of the critical exponent $\nu$. What these changes signal, is that the universality of
the critical behavior instantaneously changes in the presence of noise -- indicating an RG flow towards a different universality class. 
Specifically, our results imply a decoherence-driven renormalization group flow from the self-dual criticality to  Nishimori criticality. 

The instability of the self-dual critical point in the presence of noise can be rationalized due to the effect of noise-induced Kramers-Wannier duality breaking decoherence, which induces 
as already mentioned above, unavoidable interactions between the two layers of non-interacting Majorana fermion theories, numerically established to induce renormalization group relevant perturbations.
This raises the question of whether there is a way to break the self-duality while preserving the {\it purity} of the trajectory-resolved Majoranas, which we will now address by considering a {\it coherent} deformation instead of incoherent noise.

%%%%%%%%%%%%%%%%%%%%%%%%%%%%%%%%%%%%%%%%%%%%%%%%%%%%%%%%%%%%%%%%%%%%%%%%%%%%%%%%%%%%%%%
\subsection{Breaking self-duality by deformation}
\label{sec:coherent_deformation}
%%%%%%%%%%%%%%%%%%%%%%%%%%%%%%%%%%%%%%%%%%%%%%%%%%%%%%%%%%%%%%%%%%%%%%%%%%%%%%%%%%%%%%%

To break self-duality without breaking purity of each trajectory, one can introduce a bath to absorb the $e$-vortices, or to damp the Pauli $X$ operator of the site qubits. To be concrete, we deform the quantum state by applying a finite-depth local {\it non-unitary} circuit $\exp(-\eta\sum_j (1-X_j)/2)$ onto the site qubits, before their measurement. 
This non-unitary gate has previously been denoted as a ``wave function deformation"~\cite{Henley04RK, Fradkin04RK, Troyer10topocrit, Zhu19ToricCode}.
If not directly implemented in a circuit this deformation can, in principle, also be realized as the ground state of a 2D Rokhsar-Kivelson type Hamiltonian \cite{DynamicsConformalQCP}. 
Consequently, the statistical model~\eqref{eq:Born} picks up an additional weight factor
\begin{equation}
\begin{split}
P(em)' \propto  \mathcal{Z}(em)^2 
\times 
\exp(\frac{\eta}{2} \sum_{+} \prod_{\langle ij\rangle \in +}t_{ij} )\ , \\
\end{split}
\label{eq:statmechEmass}
\end{equation}
up to an $em$ independent normalization constant such that $\sum_{em}P(em)'=1$. 
Note that this extra term is reminiscent of a gauge coupling, with $\eta$ endowing a mass to the $e$-vortex. By tuning $\eta$ from $0$ to $\infty$, we can thereby control the density of the $e$-vortices, which completely 
vanish~\footnote{The product of $t_{ij}$  in the exponential is negative when a vortex is present.} 
at $\eta\to\infty$. This expression can alternatively be viewed as a ``reweighting" of the post-measurement Born ensemble. 
The corresponding $(1+1)$D circuit is composed of a deterministic $\exp(\beta' X/2)$ and a monitored (non-deterministic) $\exp(\pm \beta Z_jZ_{j+1}/2)$ imaginary-time evolution. However, we caution that the probability of the random gate $\exp(\pm \beta Z_jZ_{j+1}/2)$ at a given time is not only conditioned upon the ``past" but also the ``future" of the $(1+1)$D evolution, which is computed by tracing out the whole $(1+1)$D spacetime sheet. 

For convenience we compactify the phase diagram within $p_\eta\in[0,0.5]$ where $\eta = \ln\frac{1+2p_\eta}{1-2p_\eta} $. 
In the limit $\eta\to\infty$, the $e$-vortices vanish, $P(em)'\propto P(m)'\delta_{e,0}$. In the absence of $e$, the $m$-dependent partition function $P(m)'$ reduces to two random bond Ising layers that share the same $m$-vortex disorder:
\begin{equation}
\begin{split}
&P(m)' \propto  \mathcal{Z}(m)^2  = \sum_{\sigma,\tau} \exp\left(\frac{\beta}{2} \sum_{\langle ij\rangle}s_{ij}(\sigma_i\sigma_j + \tau_i\tau_j)\right)  \\
&=\sum_{\sigma,\tau} \prod_{\langle ij\rangle}
[
1+\tan^2\left(\frac{\theta}{2}\right) \sigma_i\sigma_j\tau_i\tau_j + s_{ij} \tan(
\frac{\theta}{2}) (\sigma_i\sigma_j+\tau_i\tau_j) 
] \ ,
\end{split}
\end{equation}
where we use the fact that $\tanh(\beta/2) = \tan(\theta/2)$, and in the second line we perform a high-temperature expansion. 
When the disorder is traced out, the $s$ terms drop out, so we find the normalization constant of the Born probability to be $\sum_m P(m)' 
\propto \sum_{\sigma\tau}\exp[\tanh^{-1}\tan^2(\theta) \sum_{\langle ij\rangle}\sigma_i\sigma_j\tau_i\tau_j]$, which is a non-random Ising model describing the disorder average correlations of the joint Ising spin $\sigma_j'\equiv\sigma_j\tau_j$. A transition point can then be readily determined via Kramers-Wannier duality 
\[
	\tan^2\left(\frac{\theta_c}{2}\right)= \frac{1}{1+\sqrt{2}} \,,
\] 
which yields $\theta_c/\pi=0.364 \ldots$. 

Notably, at this critical point all  {\it disorder-independent} operators are described by the {\it unitary} Ising CFT, which obeys this self-duality. For instance, the two-point correlation $\langle \sigma_i' \sigma_j'\rangle \sim |i-j|^{-1/4}$ because $\sigma'$ picks up the scaling dimension $1/8$ from the Ising CFT. 
Nevertheless, measurement-induced quenched randomness (which also breaks self-duality) causes (random) quantum trajectory resolved quantities to be no longer described by simple unitary Ising CFT, which is the reason why we have dubbed this critical theory ``Ising+'' theory.
As shown in Fig.~\ref{fig:criticalexp}(b), the critical length exponent $\nu=1.00(5)$ extracted from coherent information agrees with 
that of the simple Ising CFT. 
If one looks into the entanglement entropy scaling of the (1+1)D quantum state formed by the joint spins, $\sigma_j'$, or its ground state energy scaling, one would recover the central charge $1/2$. 
In contrast, the 
measurement-averaged
entanglement entropy for our quantum state is expressed in terms of
$\tau$ spins alone, which yields 
\[
	c_{\rm ent}^{\rm vN}\rvert_{\eta=\infty} = 0.310(2) \,,
\]
which is a universal number that is not captured by the unitary Ising CFT. Besides, the normalization condition $\sum_{m}P(m)'=1$ ensures that this critical point has vanishing central charge following Eq.~\eqref{eq:ZNFN}.

Properties characteristic of quenched disorder (including the intrinsic randomness of quantum mechanical measurements, governed by the Born-rule distribution)
are described by derivatives with respect to replicas, analogous to the effective central charge 
-- just for different observables. (See, e.g., Ref.~\onlinecite{Ludwig2020}.)
For instance, when formulated within the replica formalism, $c_{\rm ent}^{\rm vN}\rvert_{\eta=\infty} = 0.310(2)$ above should be a derivative with respect to $N$ as $N\to 1$ of the corresponding replicated boundary condition changing (``twist'')  operator correlation function computed with $N$ replicas;
this  is needed to generate an average of a {\sl logarithm} in the  entropy. The result $c_{\rm ent}^{\rm vN}= 0.310(2)$ comes from the extra $(N-1)$ replicas and the derivative.

As shown in Fig.~\ref{fig:criticalexp}(b)(c), when deviating from the $\eta=\infty$ limit, $\nu$ and $c_{\rm ent}$ quickly cross over
to the exponents for the self-dual criticality. This again indicates an RG flow, now from Ising+ criticality to self-dual criticality. (Numerical data for the finite-size collapse is shown in Fig.~\ref{fig:emass} in Appendix~\ref{sec:fss}.)

\subsection{Strong to weak (explicit) self-dual symmetry breaking}
\label{sec:flow}

%%%%%%%%%%%%%%%%%%%%%%%%%%%%%%%%%%%%%%%%%%%%%%%%%%%%%%%%%%%%%%%%%%%%%%%%%%%%%%%%%%%%%%%

If we further add a 
term to the probability weight that not only 
penalizes~\footnote{The product of $t_{ij}$ and of $s_{ij}$ is negative when a respective vortex is present.}
configurations with $e$ but also with $m$-vortices,
we can maintain the self-dual symmetry while suppressing both $e$ and $m$ on equal footing. Such a scenario is captured by a statistical model given by
\begin{equation}
\begin{split}
P(em)'' \propto  \mathcal{Z}(em)^2 
\times 
\exp[\frac{\gamma}{2} \left(\sum_{+} \prod_{\langle ij\rangle \in +}t_{ij} +\sum_\square \prod_{\langle ij\rangle \in \square}s_{ij} \right)]\ , \\
\end{split}
\label{eq:statmechEMmass}
\end{equation}
where $\gamma = \ln\frac{1+2p_\gamma}{1-2p_\gamma} \geq 0 $. 
In the limit $p_\gamma=1/2$, $\gamma=\infty$, all $e$ and $m$ vortices are suppressed and our mixed-state self-dual criticality reduces to the non-random Ising criticality, whose $(1+1)$D quantum state exhibits self-duality as a strong symmetry  
\begin{equation*}
{\rm KW} \ket{\psi}\rvert_{\gamma=\infty} = \ket{\psi}\rvert_{\gamma=\infty} \, ,
\end{equation*}
up to  boundary condition terms~\footnote{For periodic boundary conditions,
${\rm KW}\ket{\psi}\rvert_{\gamma=\infty} = \frac{1+\prod_j X_j}{2}\ket{\psi}\rvert_{\gamma=\infty}$, the self-dual state is invariant locally but experiences a global parity projector, which can be verified in the MPO form of the symmetry in Fig.~\ref{fig:1dcircuit}a. In a twisted boundary condition it becomes the odd parity projector.}. This can be visualized in the Majorana representation where the state inherits the Majorana translation symmetry from the Majorana quantum circuit. 
For any finite $p_\gamma$, the self-duality is respected and the point at $\theta=\pi/4$ remains always critical, distinct from the deformation critical line in Fig.~\ref{fig:noisyphasediagram}. Therefore we sit at $\theta=\pi/4$ and perform our Born sampling computation for the Born-averaged entanglement entropy to obtain the scaling exponents $c_{\rm ent}^{\rm vN}$ as a function of $p_\gamma$. As shown in Fig.~\ref{fig:criticalexp}(d), the finite-size fit $c_{\rm ent}^{\rm vN}$ for the critical point closer to the clean Ising significantly drift upwards, while that closer to the mixed critical point converges faster. This indicates a crossover from the unitary Ising CFT, with strong self-dual symmetry, to the mixed state self-dual criticality, with only weak self-dual symmetry. 
For the measurement-version of the Cho-Fisher model at hand  the RG flow is thus towards {\it decreasing} mass, i.e. flowing from pure Ising to the KW-self dual point. Note that this is the {\it opposite} direction of the RG flow within the conventional Cho-Fisher model in the $N\to 0$ replica limit~\cite{Gruzberg-Et-Al-2D-Class-D-2021},
which
is schematically shown in Fig.~\ref{fig:1dcircuit}(f). 

%%%%%%%%%%%%%%%%%%%%%%%%%%%%%%%%%%%%%%%%%%%%%%%%%%%%%%%%%%%%%%%%%%%%%%%%%%%%%%%%%%%%%%%
\section{Discussion and outlook}
\label{sec:discussion}
%%%%%%%%%%%%%%%%%%%%%%%%%%%%%%%%%%%%%%%%%%%%%%%%%%%%%%%%%%%%%%%%%%%%%%%%%%%%%%%%%%%%%%%

\subsubsection*{Summary of RG flows}

Stepping back, one result of our study is a cascade of RG flows between different critical theories, see Fig.~\ref{fig:criticalexp}.
Putting together all the numerical evidence, one arrives at a flow within the mixed-state critical line from Ising+, to self-dual criticality, and finally towards Nishimori criticality.
In other words, when noise sets in, every fixed point could flow to Nishimori criticality -- elevating it to a remarkably stable universality class in this context. 
This adds to the mounting evidence that Nishimori criticality might be an ubiquitous phenomenon in the  quantum dynamics of monitored cluster states -- it naturally arises from Born's rule \cite{NishimoriCat}, it emerges both in the presence of incoherent noise \cite{Preskill2002} and coherent deformations \cite{NishimoriCat} as well as when both are present simultaneously \cite{Chen24nishimori}. 
The self-dual criticality can be viewed, in its own right, as a quantum parent state of the Nishimori criticality, which maintains the self-duality. 
It is interesting that in the language of the non-interacting fermion description of the KW self-dual point this noise amounts to an interaction amongst the Majorana fermions during the crossover, while  the two endpoints of the RG flow, the ultraviolet KW self-dual as well as the infrared Nishimori critical point, are both described in terms of {\sl non-interacting} fermion systems.

\subsubsection*{Relationship with measurements performed on quantum critical ground states}

The type of (1+1)D systems we discuss in the present paper belong to the setups appearing in the context of measurement induced phase transitions (MIPT) in deep 1D quantum circuits 
\cite{Li2018,Skinner2019,Huse2020qec,Pixley20mipt,Ludwig2020,Altman2020weak,Potter21review,Fisher2022reviewMIPT}.
The systems we discuss are a particular ($e$- and $m$-) measurement-only~\cite{Vedika2021measure} version 
of those, that surround the key question of weak self-duality symmetry, 
and which we write in the formulation of the mixed state of
Eq.~\ref{eq:mixedstate} in 2D space. In these setups measurements are performed in
the 2D bulk spacetime of the quantum circuit. In such a circuit, including at the MIPT, the 2D bulk  lacks  space and time translational invariance in a fixed quantum trajectory. Only observables averaged over quantum trajectories (measurement outcomes) are described by (rather rich) non-unitary
``random'' CFTs. Examples are the self-dual and the Nishimori critical points discussed in this paper. The quantum state at the final time (the boundary) of the 
circuit is described by boundary critical properties of this 
``random'' non-unitary bulk CFT [dictated by a particular scale-(and conformally)-invariant boundary condition].

This setup should be contrasted with another
(technically more tractable) type of measurement setups that has been discussed in the more recent literature, where measurements are performed on a quantum critical ground state of a 1D translationally invariant critical Hamiltonian of  a unitary (1+1)D CFT including, e.g., a gapless Luttinger liquid~\cite{Garratt22} or a critical transverse field Ising model in the examples first discussed in the literature,
with natural generalizations to higher dimensional analogs (see, e.g.,~\cite{Lee23decoher, LuitzGarratt2024MeasStatesHigherDim}). 
When formulated in path integral language in (1+1)D spacetime~\cite{Garratt22}, the quantum critical ground state is generated
at the imaginary time $\tau=0$ time-slice by  an infinite imaginary time evolution from infinity with the unitary CFT Hamiltonian. Measurements are performed only at this $\tau=0$  time-slice, representing a line-defect in (1+1)D space time. Away from the defect line, the 2D bulk of spacetime is completely translationally invariant.
There are a number of variations
of this setup, including e.g. those discussed in~\cite{Garratt22, Garratt23measureising, Alicea23measureising, jian23measureising, Hsieh23channelcriticality, 
Alicea24teleportation,liu2024boundarytransitions, Ludwig24measurecritical} 
to  briefly mention only a small subset. In many cases, including those where {\it uniform measurement outcomes} are
{\it postselected}, the
defect is described by a perfectly unitary defect/boundary CFT (thus, in those cases, bulk and defect/boundary are {\it both} unitary). In cases where {\it no postselection} (or bias) of measurement outcomes on the quantum critical ground state is performed, a {\it defect} can appear that itself needs to be described by 
{\it non-unitary (`random')
``measurement-dominated'' defect/boundary fixed point} CFT (exhibiting  scaling behavior whose richness is qualitatively similar to that exhibited at general MITPs and by the systems discussed in the paper at hand). Examples of this are provided, e.g.,  in \cite{Ludwig24measurecritical}.

\subsubsection*{Connection to decoherence transitions of the toric code}

The self-dual decoherence of the toric code here is due to the non-Clifford Kraus operator $\rho \to \rho + \frac{Z+X}{\sqrt{2}}\rho \frac{Z+X}{\sqrt{2}}$, which involves some terms like $X\rho Z$ that create $m$ vortices in the ket space but $e$ vortices in the bra space, contributing to off-diagonal elements of the density matrix in the anyon basis. 
This should be distinguished from the self-dual Clifford noise where
both bit-flip and phase-flip appear with the same probabilities: $\rho \to (1-p)^2 \rho + p(1-p)X\rho X + p(1-p)Z\rho Z + p^2 Y\rho Y)$, which is purely block diagonal in the anyon basis. The latter case results in a trivial state at $p=50\%$, separated from the toric code by a decoupled Nishimori$\times$Nishimori transition at $p_c\approx10.9\%$, as the statistical model can be {\it factorized} into two independent RBIM along the Nishimori line. The physical reason for the factorization is that the $e$ and $m$ vortices always appear in pairs in both the ket and the bra space at the same location, with trivial braiding statistics~\cite{Fan24coherinfo}. Similarly, the Clifford depolarizing noise is also mapped to the Nishimori transition of a random bond Ashkin Teller model~\cite{Bombin12depolarize}. 
It is interesting to note that another way of Clifford self-dual decoherence by proliferating the fermions with Kraus operator $Z_r X_{r+(1/2,1/2)}$ can lead to a non-critical intrinsic mixed state topological order~\cite{Prem25mixedstateto, Wang25decoherfermion, Ichinose25decoherfermion} at the maximal decoherence limit.

Previously, an $(N\to 0)$-replica limit was discussed in randomly deforming the toric code~\cite{Tsomokos2011}, while the $(N=2)$-replica case was discussed more recently in Refs.~\cite{Grover24selfdual, Moon24surfacecodecohererr, Fan24coherinfo, Lee23decoher, Luo24weaksym}. 
These different replica limits all result in fundamentally distinct universality classes, pointing to the necessity to study the device-relevant $(N\to 1)$-replica limit 
of Born-measurement induced randomness in its own right (and despite its significant numerical effort) for any decoherence channel of interest. 

\subsubsection*{Non-unitary measurement-based quantum computation}

On a conceptual level, one way to think about the relation between bulk and boundary states is within the framework of measurement-based quantum computation (MBQC)~\cite{Raussendorf2001oneway, briegel2009measurement}. There, a 2D bulk cluster state serves as a (short-range) entanglement resource for
a deep $(1+1)$D quantum circuit -- propagating along one the two space dimensions instead of a real time dimension~\cite{Browne2008, Altman2021measure, Chen2022measure, Harrow2022, google2023measurement} -- to prepare (compute) a nontrivial $(1+1)$D quantum state.
The universality of MBQC guarantees that any {\it pure} state can be prepared, in this way, at the boundary of a cluster state 
by an effective {\it deterministic} (1+1)D unitary circuit with sufficient space resource and classical communication.
What we discuss here is a seemingly very different setting with an effective circuit that is {\it probabilistic} non-unitary~\cite{Sela23nonunitary, Ueda05nonunitary} 
and leads to a {\it mixed} state~\cite{Hsieh23mixedstatecriticality, NishimoriCat, JYLee, Hsieh24markovlength}.
But if one thinks of MBQC as a measurement-based approach to implement conventional unitary state preparation (quantum annealing~\cite{Nishimori98anneal}),
then one can cast our setting as a generalization of MBQC to non-unitary state preparation. This in turn points to a much broader application landscape of MBQC,
as one can now imagine to prepare entangled mixed-states of matter that have no counterpart in the pure-state framework. 

\subsection*{Outlook}

\subsubsection*{Decoding}

When decoding is employed and a conditional unitary based on the bulk classical bits (measurement records) is applied to the boundary quantum state, one can trace out the bulk leaving a stand-alone boundary quantum state. An {\it optimal} decoder is one that can correct the whole SW-SSB phase at $t>\pi/8$ with glassy long-range order into an SSB phase with long-range order. Such a decoder would leave the KW self-dual critical location invariant, and could possibly preserve the self-duality at the transition point, turning our weak self-dual criticality into a standalone $(1+1)$D mixed state: $\rho_Q = \sum_{em} U(em)\ketbra{\psi(em)}_Q U(em)^\dag$, which should be amenable to near-term experimental probe~\cite{Chen24nishimori}. 

\subsubsection*{Higher dimensions \& Kitaev spin liquids}

Moving onto higher dimensions, it is well-known that the 3D Ising criticality is {\it not} self-dual, but can rather be dualized to the 3D classical gauge theory~\cite{Wegner71duality}. 
However, the fermion perspective points to another potential generalization in three dimensions: a Majorana translation symmetric model with random vortex disorder. This could potentially be realized in cleverly designed 3D Kitaev spin liquids~\cite{Kitaev2006, OBrien2016, Eschmann2020}. It is worth noting that a randomized version of monitored Kitaev spin liquids~\cite{Zhu23structuredVolumeLaw, Zhu23qubit, klocke2024entanglement} is also closely related to the Floquet code of Hastings and Haah \cite{Haah21honeycomb} -- a new family of dynamical quantum error correcting codes~\cite{Hastings22honeycomb, Nat24floquetcode3dkitaev}. This points to another possible application of our duality-enriched mixed-state quantum criticality, to engineer the thresholds of  dynamical quantum error correcting codes. 
\\

\subsubsection*{Generalized symmetries}
Another versatile direction is to explore systematically a weak-symmetry generalization~\cite{Wang23averspt, Lee23decoher, Xu24higherformweaksym, You24weaksym, Luo24weaksym} of the more generic non-invertible or categorical symmetries~\cite{Wen20categorysym, Verstraete21mposym, Rizi23selfdual, Tachiwawa23cftnoninvertible, Seiberg24selfdual, Tachikawa24noninvertible, Ning24categorysym, You24noninvert}, when a quantum theory is subjected to decoherence. On the other hand, our self-dual decoherence could be generalized to more generic quantum error correction codes enriched with duality~\cite{Jian24dualities}, or gauge theory with matter as for the (good) qLDPC codes~\cite{Khemani23ldpc}. 

\section*{Note added}
After our work was posted on the arXiv, a related new preprint~\cite{Grover25entropy} also discusses the bulk Shannon entropy density of certain noisy mixed states. 
%%%%%%%%%%%%%%%%%%%%%%%%%%%%%%%%%%%%%%%%%%%%%%%%%%%%%%%%%%%%%%%%%%%%%%%%%%%%%%%%%%%%%%%
% Data
%%%%%%%%%%%%%%%%%%%%%%%%%%%%%%%%%%%%%%%%%%%%%%%%%%%%%%%%%%%%%%%%%%%%%%%%%%%%%%%%%%%%%%%

\section*{Data availability}

The numerical data shown in the figures and the data for
sweeping the phase diagram are available on Zenodo \cite{zenodo_duality}.

%%%%%%%%%%%%%%%%%%%%%%%%%%%%%%%%%%%%%%%%%%%%%%%%%%%%%%%%%%%%%%%%%%%%%%%%%%%%%%%%%%%%%%%
% Acknowledgments
%%%%%%%%%%%%%%%%%%%%%%%%%%%%%%%%%%%%%%%%%%%%%%%%%%%%%%%%%%%%%%%%%%%%%%%%%%%%%%%%%%%%%%%

\begin{acknowledgments}

GYZ would like to especially thank Samuel Garratt for inspiring discussions on the preparation of $(1+1)$D critical states, as well as Shang-Qiang Ning and Chenjie Wang for discussions of the categorical symmetry and quantum anomaly. The authors would like to thank Malte Pütz for helpful discussions on the numerics of Nishimori criticality, to be published in a joint work~\cite{Puetz24}. AWWL thanks Chao-Ming Jian for  collaboration on several previous works in the area of quantum circuits of non-interacting fermions.
GYZ acknowledge the support of %NSFC-Young Scientists Fund (No. 12504181) and 
Start-up Fund of HKUST(GZ) (No. G0101000221).
This research was supported in part by grant no. NSF PHY-2309135 to the Kavli Institute for Theoretical Physics (KITP). % learning fine structure program
ST acknowledges partial funding from the Deutsche Forschungsgemeinschaft (DFG, German Research Foundation)
under Germany's Excellence Strategy -- Cluster of Excellence Matter and Light for Quantum Computing (ML4Q) EXC 2004/1 -- 390534769 
and within the CRC network TR 183 (Project Grant No.~277101999) as part of subproject B01. R.V. acknowledges partial support from the US Department of Energy, Office of Science, Basic Energy Sciences, under award No. DE-SC0023999.
 \\
\end{acknowledgments}

%%%%%%%%%%%%%%%%%%%%%%%%%%%%%%%%%%%%%%%%%%%%%%%%%%%%%%%%%%%%%%%%%%%%%%%%%%%%%%%%%%%%%%%
\bibliography{measurements}
%%%%%%%%%%%%%%%%%%%%%%%%%%%%%%%%%%%%%%%%%%%%%%%%%%%%%%%%%%%%%%%%%%%%%%%%%%%%%%%%%%%%%%%

\clearpage 

%%%%%%%%%%%%%%%%%%%%%%%%%%%%%%%%%%%%%%%%%%%%%%%%%%%%%%%%%%%%%%%%%%%%%%%%%%%%%%%%%%%%%%%
\appendix
%%%%%%%%%%%%%%%%%%%%%%%%%%%%%%%%%%%%%%%%%%%%%%%%%%%%%%%%%%%%%%%%%%%%%%%%%%%%%%%%%%%%%%%

\tableofcontents
\clearpage

\section{Technical details}

%%%%%%%%%%%%%%%%%%%%%%%%%%%%%%%%%%%%%%%%%%%%%%%%%%%%%%%%%%%%%%%%%%%%%%%%%%%%%%%%%%%%%%%
\subsection{Derivation for the random tensor network state}
\label{sec:derivetns}
%%%%%%%%%%%%%%%%%%%%%%%%%%%%%%%%%%%%%%%%%%%%%%%%%%%%%%%%%%%%%%%%%%%%%%%%%%%%%%%%%%%%%%%
Prior to the measurements, the cluster state can be written as a tensor network:
\begin{equation*}
\includegraphics[width=.36\columnwidth]{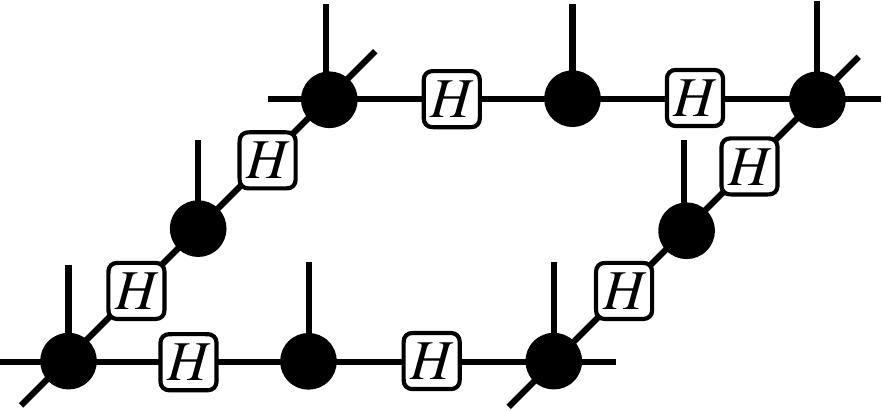} \ ,
\end{equation*}
which shows one plaquette as a unit-cell of the tensor network state. Each vertex is a diagonal delta tensor, originating from the initial $\ket{+}$ state of each qubit, and each bond is attached a 2-by-2 Hadamard matrix, which originates from the 4-by-4 CZ gate matrix: $\text{diag}(CZ)=[1,1,1,-1] \Leftrightarrow H=[1,1;1,-1]$. 
When a site qubit is measured in the $X$ basis, the physical leg is terminated, and a Pauli $Z$ is injected to the vertex if the measurement outcome is negative: $X=-1$. 
When a bond qubit is measured in the rotated basis, the physical leg is terminated, leaving a Boltzmann weight in the form of a 2-by-2 matrix associated with the bond
\begin{equation}
\includegraphics[width=\columnwidth]{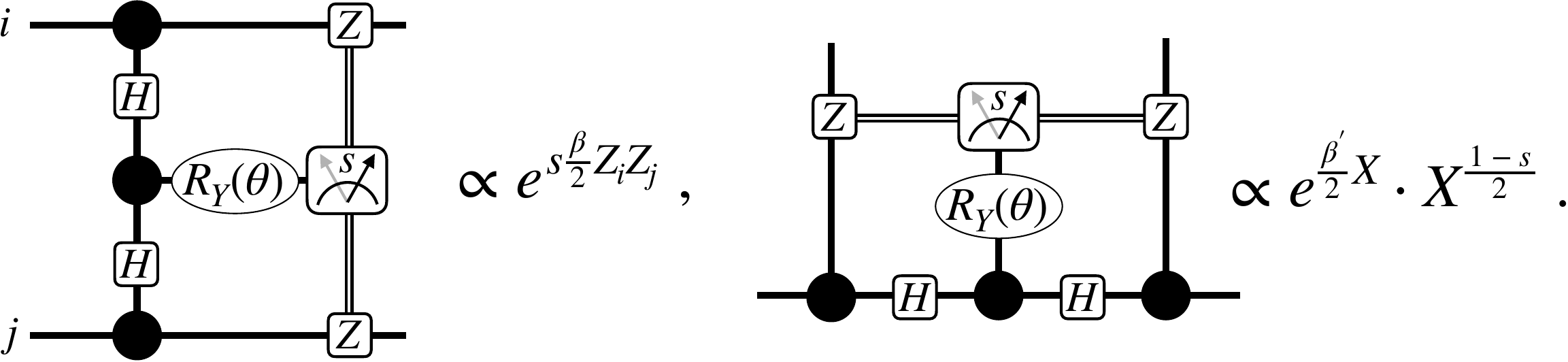} \ .
\label{eq:gate}
\end{equation}

\subsection{Cylindrical boundary conditions}
Here we discuss the boundary condition of the statistical model in the fermion representation when we place the 2D quantum state on a cylinder, i.e.\ implementing a periodic boundary condition along the X direction. If the measurement outcomes are post-selected to be positive, the statistical model is a clean 1+1D Ising model in the periodic boundary condition. And when the boundary term is twisted to be antiperiodic boundary condition, it excites an Ising primary field that costs energy. Nevertheless, in the presence of random bond disorder which breaks the translation symmetry, it is less clear to determine the periodic or anti-periodic boundary condition by directly looking at a given bond. Essentially, the boundary condition is related to the global flux pumped to the cylinder, which is still well defined in the disordered system. Consider the generators of the dynamics by taking into account a random bond, which under Jordan-Wigner transformation becomes
\begin{equation}
\begin{split}
s_{x,y} Z_x Z_{x+1} =& s_{x,y} i\gamma_{2x} \gamma_{2x+1}  \, ,\\
s_{x,y+\frac{1}{2}} X_x =& s_{x,y+\frac{1}{2}} i\gamma_{2x-1} \gamma_{2x}  \, ,
\end{split}
\end{equation}
where at the boundary $\gamma_{2L_x+1}=-P \gamma_1$. Here there are two independent global quantum numbers 
\begin{equation}
\begin{split}
W=& \prod_{x,y=1} s_{x,y} =\pm 1 \,, \\
P =& \prod_{j=1}^{L_x} X_j= \prod_{j=1}^{L_x} i\gamma_{2j-1}\gamma_{2j}=\pm 1 \ ,
\end{split}
\end{equation}
where $W$ determines the global flux pumped through the cylinder being 0 or $\pi$, and $P$ specifies the parity of the spin or the fermion. A schematic of the two sectors distinguished by the global flux is illustared below
\begin{equation*}
\includegraphics[width=.6\columnwidth]{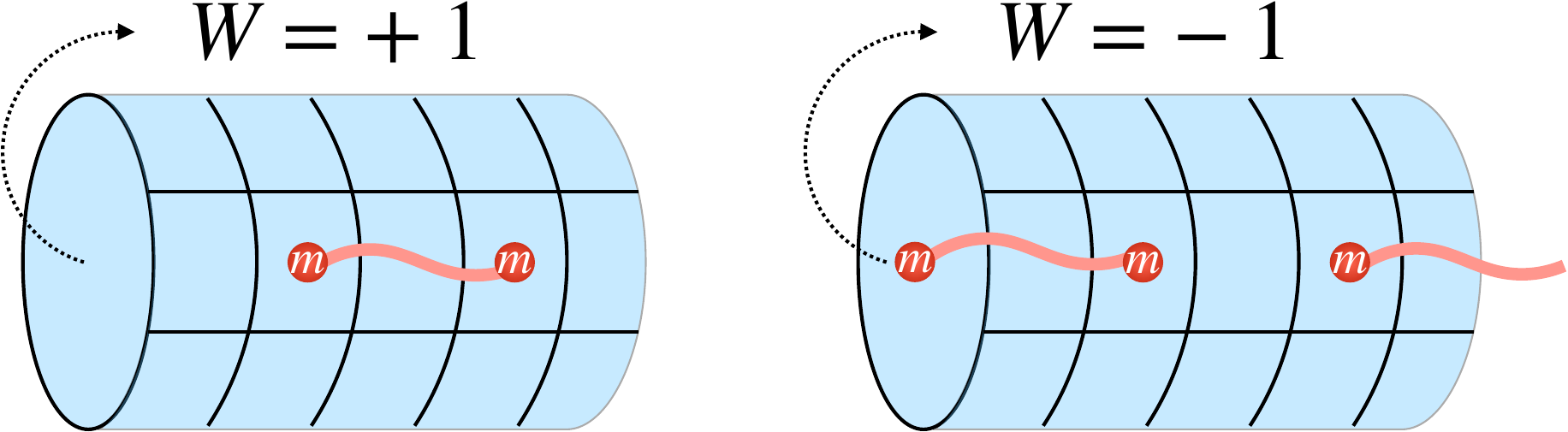} \ ,
\end{equation*}
where the bulk configurations are kept the same. 
In the clean 1+1D Ising, the vacuum state lies in $W=+1 \ , P=+1$ and the lowest excited Ising primary field lies at $W=+1 \ , P=-1$, which has a dual counterpart, i.e.\ a domain wall excitation that twists the boundary condition $W=-1 \ , P=+1$ with the same scaling dimension $1/8$. This degeneracy is related to the exact zero modes when the fermion is placed at periodic boundary condition such that its momentum can take 0 value. 
In the disordered system, the global flux that the fermion experiences is determined by $-WP$ taking into account the flux induced by the random bond disorder and the opposite of the fermion parity. Once the boundary condition is fixed, the state is further divided into the two sectors determined by $P$. Technically, we will perform the Born measurements to get the Born ensemble as a grand canonical ensemble of the global flux, and then we divide the ensemble according to $W=\pm 1$, and fill the ``Fermi sea" of the single-particle fermion eigenenergy levels to get the many-body state energy of the even fermion parity $\tilde{P}=+1$ in the eigen-fermion basis (caution that the fermion parity of the eigenmodes could differ from the original fermion modes~\cite{Chalker2002} by a fixed sign). Then we can generate the many-body Lyapunov spectrum of even parity by increasing the double number of eigen-fermion modes, and that of the odd parity $\tilde{P}=-1$ by increasing an odd number of eigen-fermion modes. \\[2cm]

\subsection{Lyapunov spectrum}
%%%%%%%%%%%%%%%%%%%%%%%%%%%%%%%%%%%%%%%%%%%%%%%%%%%%%%%%%%%%%%%%%%%%%%%%%%%%%%%%%%%%%%%
\label{sec:spec}
%%%%%%%%%%%%%%%%%%%%%%%%%%%%%%%%%%%%%%%%%%%%%%%%%%%%%%%%%%%%%%%%%%%%%%%%%%%%%%%%%%%%%%%

\begin{figure*}[htbp]    
	\centering
   \includegraphics[width=\textwidth]{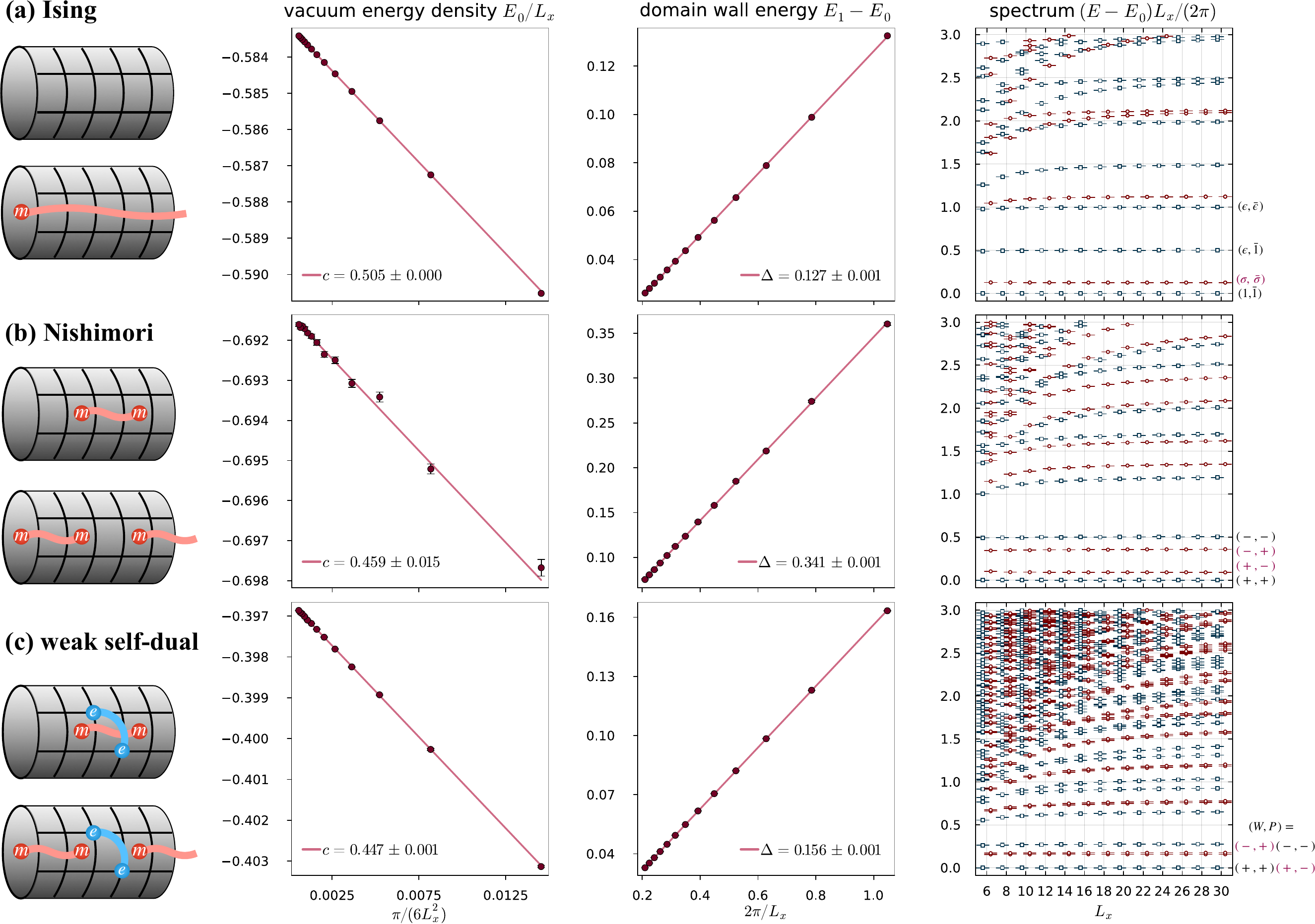} 
   \caption{{\bf Numerical computation for the domain wall energy, ground state energy and the Lyapunov spectrum for the many-body states in the second quantization picture, comparing Ising, Nishimori, and weak self-dual critical states}. The system size ranges $L_x=6\sim 30$. 
   The first column of the data plots lists the vacuum energy density in the even sector. 
   The second column is the domain wall energy, obtained by the energy difference between the ground states of the even sector and the odd sector. 
   The third column is the dimensionless many-body energy 
   Lyapunov spectrum normalized by the width of the cylinder $L\equiv L_x$. These levels are obtained by filling the energy levels of the single-particle fermion modes in Fig.~\ref{fig:spec}, and they are shifted by a constant such that the ground state level lies at 0. 
   For the spectrum, the black levels with squared markers are for antiperiodic $-WP=-1$ boundary conditions of the fermions, which are composed by the periodic spin chain in the even parity sector $(W,P)=(+,+)$ or the antiperiodic spin chain in the odd parity sector $(W,P)=(-,-)$; the red levels with circle markers include $(W,P)=(+,-)$ and $(W,P)=(-,+)$. 
   The three rows are Ising, Nishimori and weak self-dual, respectively. 
   Here clean Ising is for benchmarking, which exhibits the correct $c=1/2$ for Casimir energy and $\Delta_{(\sigma, \bar{\sigma})} = 1/8$ for the Ising primary field, and $\Delta_{(\epsilon, \bar{1})}=1/2$ for the fermion primary field.  
   The sound velocity or the spacetime anisotropy factor is found to be $1$ from the clean Ising benchmark, which is consistent with the fact that we work in spacetime isotropic lattice model instead of the continuously weak monitoring limit. 
   For Nishimori we find $E_1 - E_0 \propto 2\pi/L*0.341(1)$ with sizes up to $L=30$, which roughly agrees with Merz and Chalker's $\propto 0.691(2)\pi/L$ from the scaling dimension of the vortex-vortex correlation with fermion calculation of sizes up to $L=22$~\cite{Chalker02negative}, and the estimated window in the figures in Ref.~\cite{Pujol2001, Picco2006}. 
   }
   \label{fig:ceff}
\end{figure*}

Akin to the numerical approach to Chalker-Coddington network models in Ref.~\cite{Chalker2002}, we evolve the (1+1)D quantum chain with $2L_x$ Majorana fermions up to long times $L_y \gg L_x$, which relaxes to a ``steady" state, getting rid of the temporal boundary effect. In order to extract the universal {\it bulk} information, we discard the initial dynamics at early times, and perform statistics for the norm change of each fermion mode from now on. After a layer of gates, $L_x$ independent, normalized fermion modes linearly evolve to a set of un-normalized fermion modes, 
whose norms change by a factor $r_n$ for $n=1,\ldots,L_x$ in descending order. 
Numerically, $r_n$ is the diagonal element of the R matrix after the QR decomposition for the transfer matrix composed by a layer of gates. Here $Q$ is the orthogonal matrix and $R$ is the upper triangular matrix according to the standard linear algebra convention. 
In numerical practice, the transfer matrix of a layer can be further decomposed to smaller chunk of gates in order to suppress the rounding error when $L_x$ is large. $r_n$ can be treated as a random variable that fluctuates as time evolves. 
The typical average $\epsilon_n =-\overline{\ln r_n}$ for $n=1, ... , L_x$ can be interpreted as the single-particle fermion energies of the (1+1)D quantum chain. 
Back to the second quantization picture, the many-body energies, or the Lyapunov exponents are
\begin{equation}
E_m = \frac{1}{2}\sum_{n=1}^{L_x} \nu_n \epsilon_n  \propto \frac{2\pi}{L_x} \left(\Delta_m -\frac{c_{\rm Casimir}}{12}\right) + \ldots \ ,
\end{equation}
where $\nu_n=\pm1$ specifies the parity of the $n$-th eigenfermion mode, and by enumerating all the possibilities we have $m=1,\cdots,2^{L_x}$ many-body eigen-energy levels. Here the many-body levels can be further divided according to the global fermion parity. 

The spacetime bulk of the evolution as a transfer matrix converges to $\prod_y M(y) = \sum_m e^{-E_m L_y}\ketbra{E_m}$~\cite{Chalker2002}. Consequently, the 1D quantum ground state energy $E_0$ dominates in the limit $L_y\gg L_x$ in contributing to the 2D free energy (Shannon entropy of the measurement record) according to $F = E_0 L_y +\ldots$. The excitations $E_m - E_0 \propto 2\pi \Delta_m /L_x$ are expected to obey a scaling according to the underlying CFT, and due to the state-operator correspondence, their energy gaps capture certain operator correlation functions in spacetime. 
We have performed such calculations for the self-dual critical theory for different system sizes, resulting in the ``spectra"
of Fig.~\ref{fig:ceff}, where both Ising and Nishimori criticality are also included for comparison. 

\clearpage
%%%%%%%%%%%%%%%%%%%%%%%%%%%%%%%%%%%%%%%%%%%%%%%%%%%%%%%%%%%%%%%%%%%%%%%%%%%%%%%%%%%%%%%
\subsection{Derivation for the noisy Choi state}
\label{sec:ChoiTN}
%%%%%%%%%%%%%%%%%%%%%%%%%%%%%%%%%%%%%%%%%%%%%%%%%%%%%%%%%%%%%%%%%%%%%%%%%%%%%%%%%%%%%%%

%%%%%%%%%%%%%%%%%%%%%%%%%%%%%%%%%%%%%%%%%%%%%%%%%%%%%%%%%%%%%%%%%%%%%%%
\subsubsection*{ Choi state as a tensor network state}
%%%%%%%%%%%%%%%%%%%%%%%%%%%%%%%%%%%%%%%%%%%%%%%%%%%%%%%%%%%%%%%%%%%%%%%
The noisy state for each trajectory in Eq.~\eqref{eq:noisystate} is represented by a {\it bilayer} tensor network state Eq.~\eqref{eq:choistate} where the noise Kraus operator glues the ket layer and the bra layer, by partially tracing out the $e$ particles. 
The intra-layer solid bond matrix is the same as above: $[1, e^{-\beta}; e^{-\beta}, 1]$ (up to a negligible constant prefactor $e^{\beta/2}$).
The dashed bond matrix connecting each site of the ket layer to its counterpart in the bra layer is obtained by tracing out the noise Kraus operations:
\begin{equation}
\includegraphics[width=.9\columnwidth]{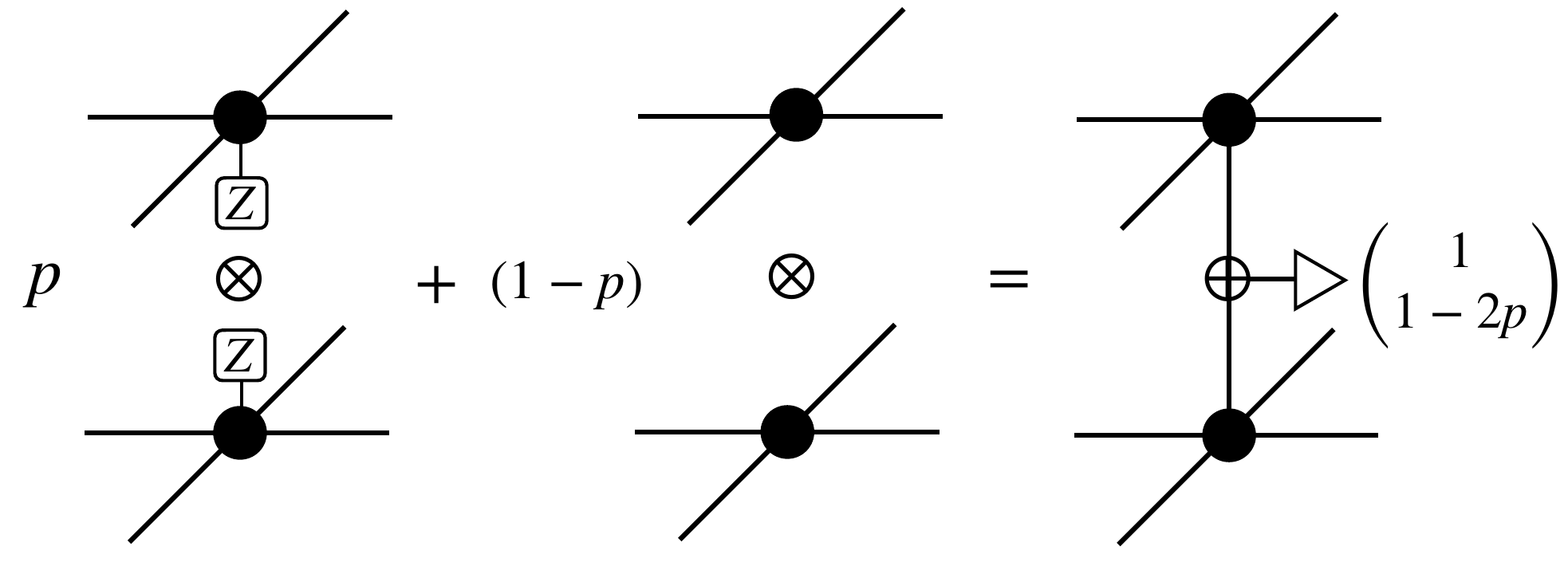} \ ,
\end{equation}
which yields a weight matrix $[1, 1-2p; 1-2p, 1]$ attached to the interlayer dashed bond. 
With this we can use MPS evolution by MPO to obtain the purity as the norm of the Choi state:
\begin{equation}
\includegraphics[width=\columnwidth]{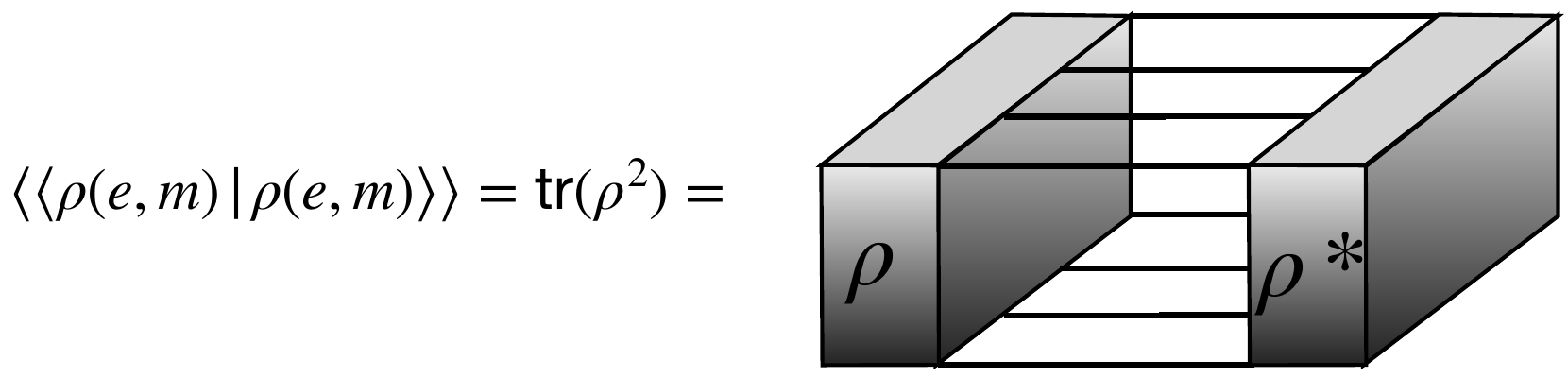} \ .
\end{equation}
Note that this bilayer tensor network also expresses the corresponding classical statistical model - a bilayer interacting Ising model with random $e$ and $m$ disorder. 
See Eqs.~(\ref{LabelEqStatMechWeightMeasurementNoiseParitionFct}), 
\ref{LabelEqStatMechWeightMeasurementNoise}) in the main text.
The noise probability tunes the interlayer Ising coupling strength. 

%%%%%%%%%%%%%%%%%%%%%%%%%%%%%%%%%%%%%%%%%%%%%%%%%%%%%%%%%%%%%%%%%%%%%%%
\subsubsection*{Second Rényi coherent information}
%%%%%%%%%%%%%%%%%%%%%%%%%%%%%%%%%%%%%%%%%%%%%%%%%%%%%%%%%%%%%%%%%%%%%%%
In the setting for coherent information, one of the physical legs is the reference qubit $R$, and the second Rényi coherent information (for each trajectory $em$) is reduced to the logarithm of the expectation of 2 times of the {\it Bell projector} that glues the $R$ qubit in the ket and the bra layer
\begin{equation}
\includegraphics[width=.7\columnwidth]{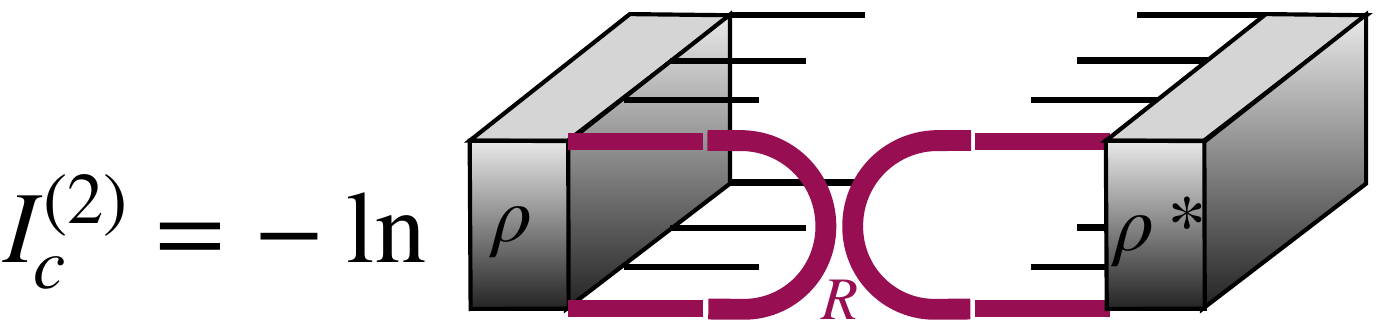} \ ,
\end{equation}
evaluated over the {\it normalized} Choi state (by noting that $e^{-S_{BR}^{(2)}}$ is the purity and thus the norm of the Choi state). 

For a more detailed analysis of the contribution of the coherent information, note that the Choi state can be decomposed into 4 states labeled by 2 logical quantum numbers (accounting for the ket and bra space, respectively)
\begin{equation}
\ket{\rho}\rangle_{BR} =\frac{1}{2} \sum_{\mu\nu=\uparrow,\downarrow}\ket{\rho_{\mu\nu}}\rangle_B \otimes \ket{\mu\nu}\rangle_R \ .
\end{equation}
The purity is reduced to 
\begin{equation}
\begin{split}
\text{tr}\rho_{BR}^2 = \frac{
\langle\braket{\rho_{00}}\rangle + \langle\braket{\rho_{01}}\rangle + \langle\braket{\rho_{10}}\rangle + \langle\braket{\rho_{11}}\rangle
}{
\left(
\text{tr}\rho_{00} + \text{tr}\rho_{11} 
\right)^2
} \\
\text{tr}\rho_{B}^2 = \frac{
\langle\braket{\rho_{00}}\rangle + \langle\bra{\rho_{00}}\ket{\rho_{11}}\rangle + \langle\bra{\rho_{11}}\ket{\rho_{00}}\rangle + \langle\braket{\rho_{11}}\rangle
}{
\left(
\text{tr}\rho_{00} + \text{tr}\rho_{11} 
\right)^2
} \ .
\end{split}
\end{equation}
Therefore, we arrive at the second Rényi coherent information of the noisy state for a fixed $em$ configuration:
\begin{equation}
\begin{split}
&I_c^{(2)} = S_B^{(2)} - S_{BR}^{(2)} = \ln \frac{\text{tr}\rho_{BR}^2}{\text{tr}\rho_B^2} 
= -\ln \langle\langle \Pi_R\rangle\rangle\\
&= \ln \frac{
\langle\braket{\rho_{00}}\rangle + \langle\braket{\rho_{01}}\rangle + \langle\braket{\rho_{10}}\rangle + \langle\braket{\rho_{11}}\rangle
}{
\langle\braket{\rho_{00}}\rangle + \langle\bra{\rho_{00}}\ket{\rho_{11}}\rangle + \langle\bra{\rho_{11}}\ket{\rho_{00}}\rangle + \langle\braket{\rho_{11}}\rangle
} \\
&= \ln \frac{
\langle\braket{\rho_{00}}\rangle + \langle\braket{\rho_{01}}\rangle 
}{
\langle\braket{\rho_{00}}\rangle + \text{Re}\langle\bra{\rho_{00}}\ket{\rho_{11}}\rangle 
} \ ,
\end{split}
\end{equation}
where in the third line we use the global Ising symmetry and the replica symmetry to reduce half of the terms: $\langle\braket{\rho_{00}}\rangle = \langle\braket{\rho_{11}}\rangle$, and $\langle\bra{\rho_{\mu\nu}}\ket{\rho_{\kappa\eta}}\rangle = \langle\bra{\rho_{\nu\mu}}\ket{\rho_{\eta\kappa}}\rangle$. The final result depends on only three overlaps of the Choi state in the logical space:
\begin{itemize}
\item $\langle\braket{\rho_{00}}\rangle$ is the vacuum amplitude; 
\item $\langle\braket{\rho_{01}}\rangle$ captures the off-diagonal elements of the density matrix of the logical qubit, which suffers from the dephasing error; 
\item $\langle \bra{\rho_{00}}\ket{\rho_{11}}\rangle $ describes the bit-flip error for the logical qubit that tunnels the diagonal $0$ state to the $1$ state. 
\end{itemize}
The four noise scenarios are summarized in Table~\ref{tab:coherinfo} below.
\begin{table}[th]
\centering
\caption{{\bf Coherent information determined by the logical state density matrix elements}. }
\begin{tabular}[t]{c | c  c | c}
\toprule
logical noise & $\langle\braket{\rho_{01}}\rangle$ & $\langle\bra{\rho_{00}}\ket{\rho_{11}}\rangle$ & $ I_c^{(2)}$\\
\midrule
none & 1 & 0 & $\ln 2$ \\
dephase & 0 & 0 & 0 \\
bit-flip & 1 & 1 & 0 \\
bit-flip and dephase & 0 & 1 & $-\ln 2$ \\
\\
\bottomrule
\end{tabular}
\label{tab:coherinfo}
\end{table}

%%%%%%%%%%%%%%%%%%%%%%%%%%%%%%%%%%%%%%%%%%%%%%%%%%%%%%%%%%%%%%%%%%%%%%%
\subsubsection*{Nishimori criticality in the maximally noisy limit} % 
%%%%%%%%%%%%%%%%%%%%%%%%%%%%%%%%%%%%%%%%%%%%%%%%%%%%%%%%%%%%%%%%%%%%%%%
In the maximally noisy limit ($p_s = 50\%$), the records of the $e$-vortices are destroyed, the decoherence erases the off-diagonal elements from the density matrix leaving only diagonal elements. Concretely, the site indices of the ket and the bra layers are locked together, reducing the bulk of the bilayer tensor network into a single layer, except the final time slice. However, the Choi state differs from the $(1+1)$D boundary MPS of the 2D classical RBIM with Nishimori disorder by only an additional {\it isometry} tensor that doubles the Hilbert space:
\begin{widetext}
\begin{equation}
   \includegraphics[width=\textwidth]{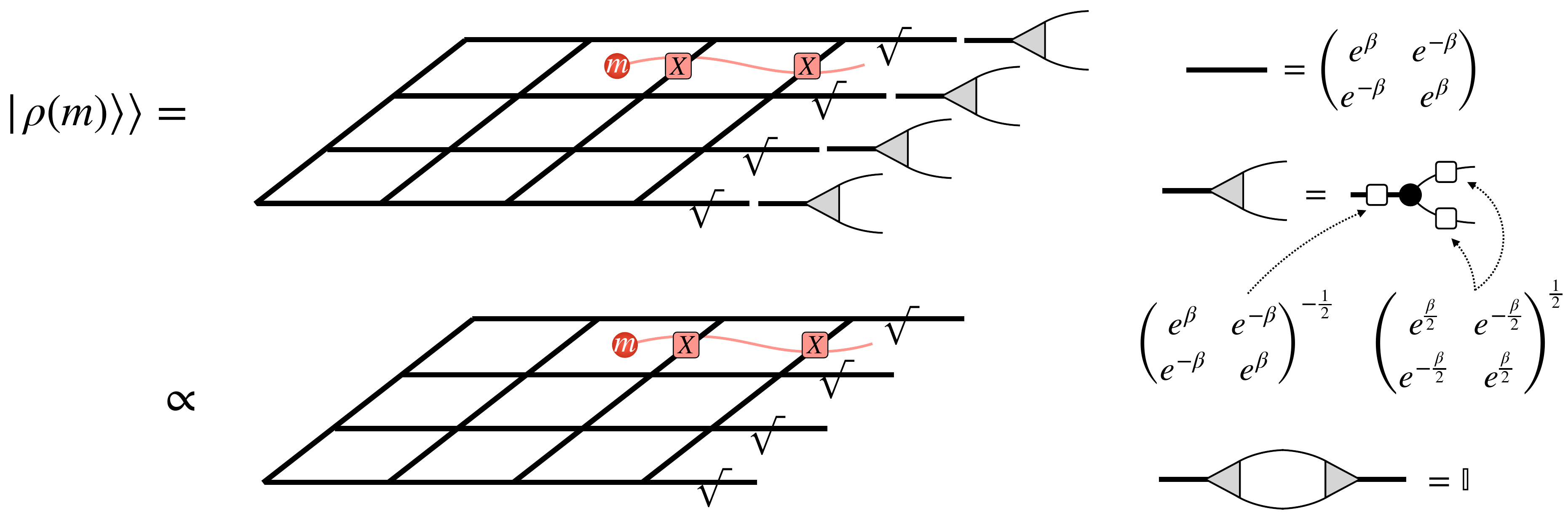} 
   \label{eq:maxnoise}
\end{equation}
\end{widetext}
where the rank-3 triangle tensor at the boundary is the isometry operator, which does not change the entanglement entropy. Therefore we can conclude that our maximally dephased boundary (1+1)D mixed state $\ket{\rho(m)}\rangle$ shares the same entanglement entropy as the (1+1)D Nishimori critical state, falling in the same universality class. 

\section{Born-average Rényi entropies}
\label{sec:renyi}

\begin{figure}[h!] 
   \centering
   \includegraphics[width=\columnwidth]{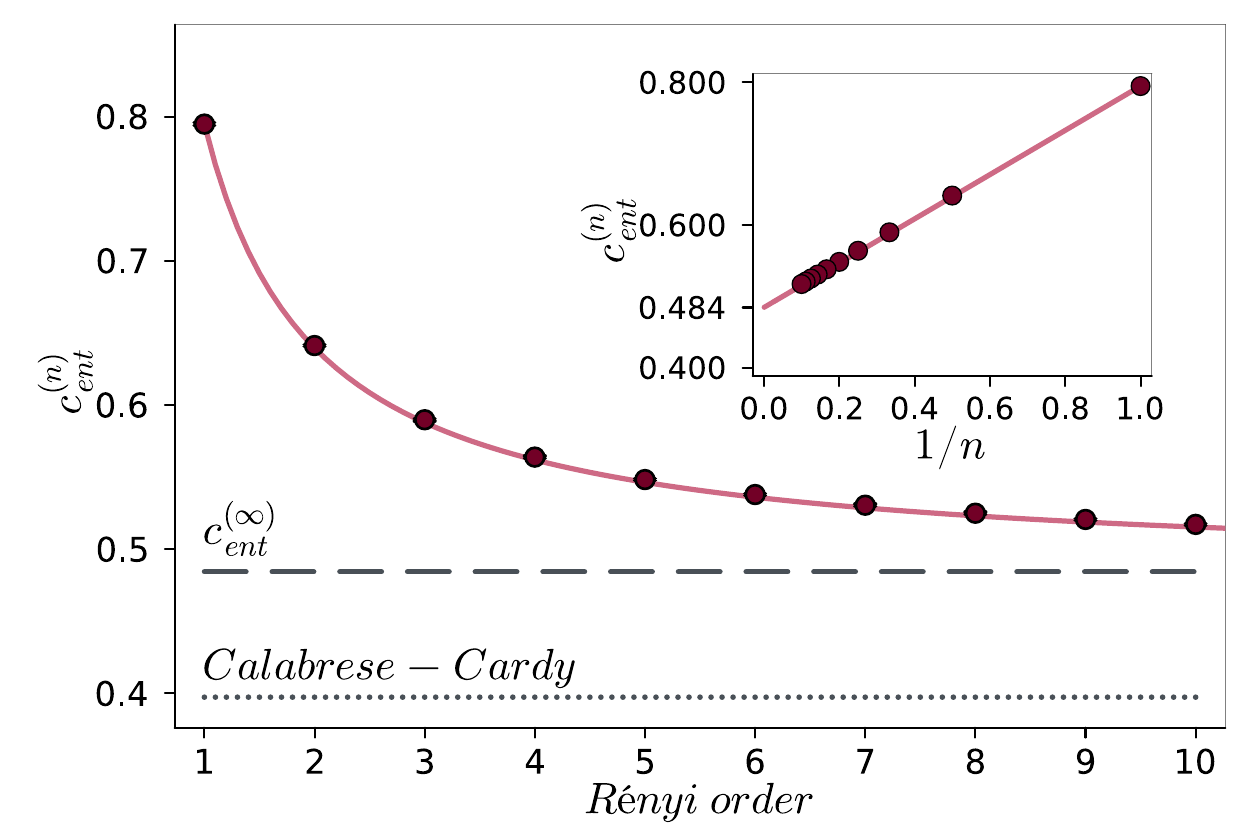}
   \caption{{\bf Born-average Rényi entropies of the self-dual critical point - their scaling dimensions}: $c_{n}$ versus Rényi order $n$. The inset shows that when $n\to\infty$, $c_{\rm ent}^{(n)}$ approaches a constant value $c_{\rm ent}^{(\infty)}=0.484(1)$. Here the solid line follows the scaling: $c_{\rm ent}^{(n)} = (c_{\rm ent}^{\rm vN}-c_{\rm ent}^{(\infty)})/n+c_{\rm ent}^{(\infty)}$, which is should be compared with the Calabrese Cardy formula for unitary CFT $c_{\rm ent}^{(n)} = c_{\rm ent}^{\rm vN}(1+n)/(2n)$. 
A similar scaling with Rényi indices was observed in the context of measurement-induced phase transitions~\cite{Pixley20mipt}.
	   The calculation uses parameters $L_x=512$ and $L_y=1024$ with periodic spatial boundary conditions, performing $2000$ Monte Carlo sweeps using the stochastic fermion evolution method. For $L_x$ ranging from 8 to 32, $1000$ Monte Carlo sweeps are performed for the $e$ configuration, conditional upon each of the 500 $m$ configurations. 
  }
   \label{fig:pureSrenyi}
\end{figure}

One can generalize Eq.~\eqref{eq:CalabreseCardy} to the Born average of the $n$-th order Rényi entropies:
\begin{equation}
\label{LabelEqFirstCumulantsOfnthRenyiEntropies}
\frac{1}{1-n}\sum_{em}P(em) \  \ln \bigl ({\rm tr} [\rho_A(em)]^{n}\bigr ) \ ,
\end{equation}
\\
Note that, in contrast, the Rényi entropies of the mixed state go beyond the Born average and correspond to the higher replica limits described by different critical theories at different critical locations~\cite{Fan24coherinfo}. 
As shown in Fig.~\ref{fig:pureSrenyi}, the Rényi entanglement scaling dimensions here clearly deviate from the Calabrese-Cardy formula for unitary CFT, where the central charge is the only fingerprint that governs all the Rényi entanglement entropy. In contrast, here our numerical results show that there are at least two independent exponents $c_{\rm ent}^{\rm vN}=0.795(1)$ and $c_{\rm ent}^{(\infty)}=0.484(1)$. 
From another perspective, given the limited knowledge about non-unitary CFTs, it is surprising to see that only two parameters are sufficient in describing all the Rényi entropies, which also appear in the case of entanglement transition~\cite{Pixley20mipt}.

\begin{figure*}[t!] 
   \centering
   \includegraphics[width=\textwidth]{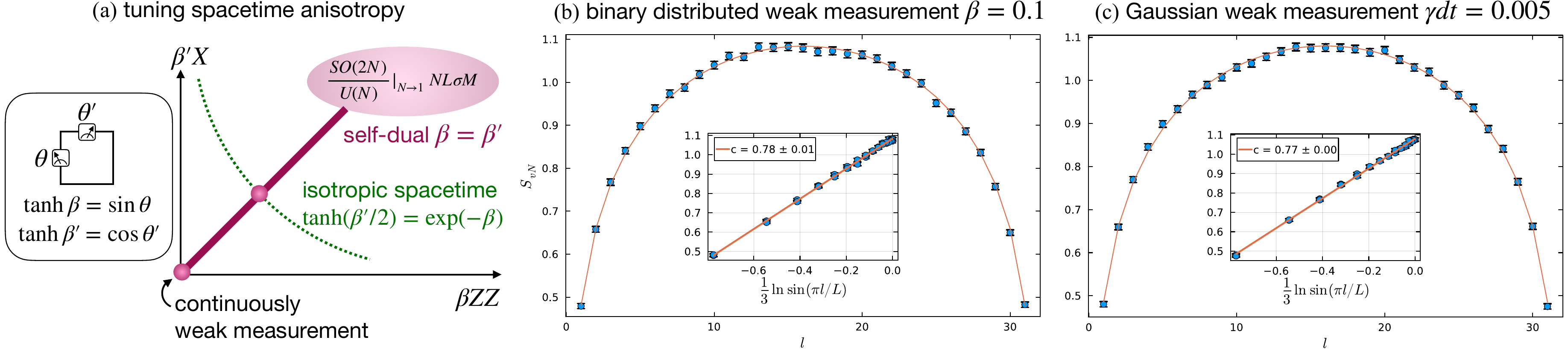} 
   \caption{
   {\bf Tuning the spacetime anisotropy by varying the measurement strength}.
   (a) Schematic: the green dashed line corresponds to the spacetime isotropic case we study in the main text, which locks $\beta'/2$ to the KW dual counterpart of $\beta/2$ because we assume a uniform measurement angle $\theta$ for the bond qubits in the bulk, which corresponds to an {\it isotropic} spacetime in the effective statistical model. As shown in the box, if we allow the measurement angle for the horizontal bond qubit $\theta'$ to be independently tuned, then we can tune $\beta$ and $\beta'$ independently, moving away from the spacetime isotropic line. Then a longer cylinder is needed to reach a steady (1+1)D quantum state. 
   The purple diagonal line $\beta=\beta'<\infty$ respects the self-duality and is expected to flow into the same universality class as described by $SO(2N)/U(N)\rvert_{N\to 1}$ $NL\sigma M$. 
   (b) Numerical computation for the binary weak measurement $\beta=\beta'=0.01$ for system size $L_x=32\ ,\ L_y=1000$.
   (c) Numerical computation for the Gaussian weak measurement for system size $L_x=32\ ,\ L_y=1000$. 
   Both (b) and (c) roughly agree with that of the finite strength binary measurement $\beta = \ln(1+\sqrt{2})\approx 0.8814$ shown in the main text: $c_{\rm ent}^{\rm vN}\approx 0.795(1)$. 
   }
   \label{fig:anisotropy}
\end{figure*}

%%%%%%%%%%%%%%%%%%%%%%%%%%%%%%%%%%%%%%%%%%%%%%%%%%%%%%%%%%%%%%%%%%%%%%%%%%%%%%%%%%%%%%%
\section{Comparison with continuous weak measurement}
\label{sec:continuous}
%%%%%%%%%%%%%%%%%%%%%%%%%%%%%%%%%%%%%%%%%%%%%%%%%%%%%%%%%%%%%%%%%%%%%%%%%%%%%%%%%%%%%%%

Here we verify the basic notion of universality - that the critical exponent does not depend on the microscopic details such as spacetime anisotropy, or whether the measurement is discrete or continuous. Namely, the whole self-dual line in Fig.~\ref{fig:anisotropy} with varying spacetime anisotropy exhibits the same entanglement entropy scaling. To be concrete, we show the entanglement entropy for the weak measurement limit when $\beta\ll 1$ while fixing $\beta'=\beta$, which can be realized from a space anisotropic 2D resource state. 
In the following we stick to the (1+1)D monitored quantum circuit representation of the problem. 

%%%%%%%%%%%%%%%%%%%%%%%%%%%%%%%%%%%%%%%%%%%%%%%%%%%%%%%%%%%%%%%%%%%%%%%
\subsubsection*{Bimodal discrete weak measurement}
%%%%%%%%%%%%%%%%%%%%%%%%%%%%%%%%%%%%%%%%%%%%%%%%%%%%%%%%%%%%%%%%%%%%%%%
Consider weakly measuring an $X$, the Kraus operator is 
\begin{equation}
M_s = \exp\left(\frac{\beta}{2} s X\right) / (2\cosh\beta) \ , \ s=\pm 1 \ ,
\end{equation}
with probability 
\begin{equation}
P(s) = \frac{1}{2\cosh\beta} \bra{\psi} e^{\beta s X} \ket{\psi} = \frac{1+s\tanh\beta\bra{\psi}X\ket{\psi}}{2} \ ,
\end{equation}
conditioned upon the state. Check the mean and the variance of this bimodal distribution of the coupling constant to $X$:
\begin{equation}
\mathbb{E}\left(\frac{\beta}{2}s\right) = \left(\frac{\beta}{2}\tanh\beta\right) \langle X\rangle \quad , 
\quad \text{Var}\left(\frac{\beta}{2}s\right) = \left(\frac{\beta}{2}\right)^2 \ ,
\end{equation}
from which we see that $\tanh\beta\in[0,1]$ expresses the fidelity of the measurement outcome $s$ with the true quantum expectation value $\langle X\rangle$ of the quantum state. Note that at the weak measurement limit $\beta\ll 1$, we have 
\begin{equation}
\mathbb{E}\left(\frac{\beta}{2}s\right) = 2 \text{Var}\left(\frac{\beta}{2}s\right) \langle X\rangle
\end{equation}
One can simply replace $X$ by any other Pauli observables for the equations above. 
We check that for a small size calculation with very weak measurement strength and highly spacetime anisotropy, the entanglement scaling (Fig.~\ref{fig:anisotropy}b) agrees with $c_{\rm ent}^{\rm vN}=0.795(1)$ as reported for the spacetime isotropic case in the main text, within numerical error bar. 

%%%%%%%%%%%%%%%%%%%%%%%%%%%%%%%%%%%%%%%%%%%%%%%%%%%%%%%%%%%%%%%%%%%%%%%
\subsubsection*{Gaussian continuous weak measurement}
%%%%%%%%%%%%%%%%%%%%%%%%%%%%%%%%%%%%%%%%%%%%%%%%%%%%%%%%%%%%%%%%%%%%%%%
Next we consider the continuous weak measurement~\cite{Graham23majorana}, which turns to the following Kraus operator
\begin{equation}
\label{LabelEqAppContinuousKrausOperator}
M(\alpha) = \left(\frac{\gamma dt}{\pi}\right)^{1/4} \exp\left(-\gamma dt\frac{(\alpha-X)^2}{2}\right) \ ,\ \alpha\in(-\infty, +\infty) \ ,
\end{equation}
with a continuous measurement outcome $\alpha$, satisfying the normalization condition $\int_{-\infty}^{\infty}  M(\alpha) d\alpha= 1$. The Born's rule dictates
\begin{equation}
P(\alpha) =  \sqrt{\frac{\gamma dt}{\pi}} \bra{\psi}\exp\left(-\gamma dt (\alpha-X)^2\right) \ket{\psi} \ .
\end{equation}
In the early derivation of the stochastic Schrödinger equation~\cite{kurt06continuous}, $X$ was referred to a continuous degree of freedom like the position in space:
\begin{equation*}
P(\alpha) =  \sqrt{\frac{\gamma dt}{\pi}} \int e^{-\gamma dt (\alpha-x)^2} |\bra{x}\ket{\psi}|^2 dx  \ .
\end{equation*}
and an approximation was made that the wave function distribution $|\bra{x}\ket{\psi}|^2$ is much narrower than the Gaussian distribution in such quantum state diffusion, resulting in:
\begin{equation*}
P(\alpha) \approx   \sqrt{\frac{\gamma dt}{\pi}}  \exp\left(-\gamma dt(\alpha-\bra{\psi}X\ket{\psi} )^2\right) \ ,
\end{equation*}
such that the mean and the variance of the coupling constant of $X$ is
\begin{equation}
\mathbb{E}(\alpha\gamma dt) \approx \gamma dt\langle X\rangle  \quad \ , \quad
\text{Var}(\alpha\gamma dt) \approx \frac{\gamma dt}{2} \ .
\end{equation}
which also satisfies 
\begin{equation}
\mathbb{E}(\alpha\gamma dt) = 2 \text{Var}(\alpha\gamma dt) \langle X\rangle \ ,
\end{equation}
as physically required for the consistency of the quantum measurement. 
Using this Gaussian distribution function instead of the bimodal distribution, our numerical computation for a small size (Fig.~\ref{fig:anisotropy}c) does not show significant deviation from $0.795(1)$ reported for the spacetime isotropic case for size $L_x=512$.

%%%%%%%%%%%%%%%%%%%%%%%%%%%%%%%%%%%%%%%%%%%%%%%%%%%%%%%%%%%%%%%%%%%%%%%%%%%%%%%%%%%%%%%
\section{Central charge of Ising+}
\label{sec:isingnorm}
%%%%%%%%%%%%%%%%%%%%%%%%%%%%%%%%%%%%%%%%%%%%%%%%%%%%%%%%%%%%%%%%%%%%%%%%%%%%%%%%%%%%%%%

Here we discuss the normalization issue of the Ising+ model. 
In our main text we normalize the probability function of disorder:
\begin{equation}
P(m)' = \frac{\mathcal{Z}(m)^2}{\mathcal{Z}_{\rm Ising}} \ , \quad \sum_m P(m)' = 1 \ ,
\end{equation}
where we denote $\mathcal{Z}_{\rm Ising} = \sum_m \mathcal{Z}(m)^2$ with Ising coupling constant $\tanh^{-1}\tan^2\theta$ as defined in the main text. $\mathcal{Z}(m)^2$ can be represented as a tensor network or a CC network, same as $\mathcal{Z}_{\rm Ising}$. 
However, it is {\it not} straightforward that $\mathcal{Z}(m)^2/\mathcal{Z}_{\rm Ising}$ can be represented as a network, because {\it inverting a network} - $1/\mathcal{Z}_{\rm Ising}$ is a highly nonlocal nonlinear operation. 
In our numerical CC network calculation (as a bilayer tensor network), the transfer matrix generates $\mathcal{Z}(m)^2$ only, without the normalization factor $\mathcal{Z}_{\rm Ising}$. Therefore the quenched disorder averaged free energy that the Majorana fermion picks up is contributed by two terms:
\begin{equation}
\begin{split}
F_{\rm unnorm}=&-\sum_m P(m)' \ln \mathcal{Z}(m)^2\\
=&
-\sum_m P(m)' \ln [P(m)'\mathcal{Z}_{\rm Ising}]\\
=&
-\sum_m P(m)' \ln P(m)' - \ln \mathcal{Z}_{\rm Ising} \\
=& 
F_{\rm Ising+} + F_{\rm Ising} \\
=&
 {\rm const}\times L_y - \left(c_{\rm Casimir} + \frac{1}{2}\right)\frac{\pi L_y}{6L_x} +\cdots\ ,
\end{split}
\end{equation}
where $F_{\rm Ising+}$ is the Shannon entropy of the disorder that scales with $c_{\rm Casimir}$, the effective central charge for Ising+ that we define in this paper. And $F_{\rm Ising}$ is the free energy of the clean Ising model with central charge $1/2$. Therefore, the Majorana fermion sees a combination of $c_{\rm Casimir}$ for Ising+ and $1/2$ from clean Ising. The normalization of the partition function subtracts the $1/2$ contribution. 

More generally, for $N$ replicas, 
\begin{equation}
\begin{split}
F_{N\ \rm unnorm} =& -\ln \sum_m P(m)^{'N}\mathcal{Z}_{\rm Ising}^N\\
=& -\ln \sum_m P(m)^{'N} -N\ln \mathcal{Z}_{\rm Ising}\\
=& -\ln \sum_m P(m)^{'N} +N F_{\rm Ising}\\
=& F_N +N F_{\rm Ising}\ ,
\end{split}
\end{equation}
the unnormalized version of the free energy picks up a background term with the Ising free energy {\it linearly} coupled to the replica index $N$. Thus taking the derivative with $N$ yields a replica-independent constant contribution originating from the critical Ising CFT. This is why in our definition the Ising+ has central charge 0 rather than $1/2$: $F_{N=1} = F_{N=1\ \rm unnorm} - F_{\rm Ising} \propto (1/2 - 1/2) =0$. 

Lyapunov spectrum. Since we do not have a {\it transfer matrix} for $\mathcal{Z}(m)^2/\mathcal{Z}_{\rm Ising}$, we cannot define its Lyapunov spectrum. Instead, the Lyapunov spectrum that we compute corresponds to the unnormalized model described by partition function $\mathcal{Z}(m)^2 = {\rm tr} (T^{L_y})$, where $T$ is the transfer matrix. 

%%%%%%%%%%%%%%%%%%%%%%%%%%%%%%%%%%%%%%%%%%%%%%%%%%%%%%%%%%%%%%%%%%%%%%%%%%%%%%%%%%%%%%%
\section{Supplementary data for finite-size data collapse}
\label{sec:fss}
\begin{figure*}[t] 
   \centering
   \includegraphics[width=.75\textwidth]{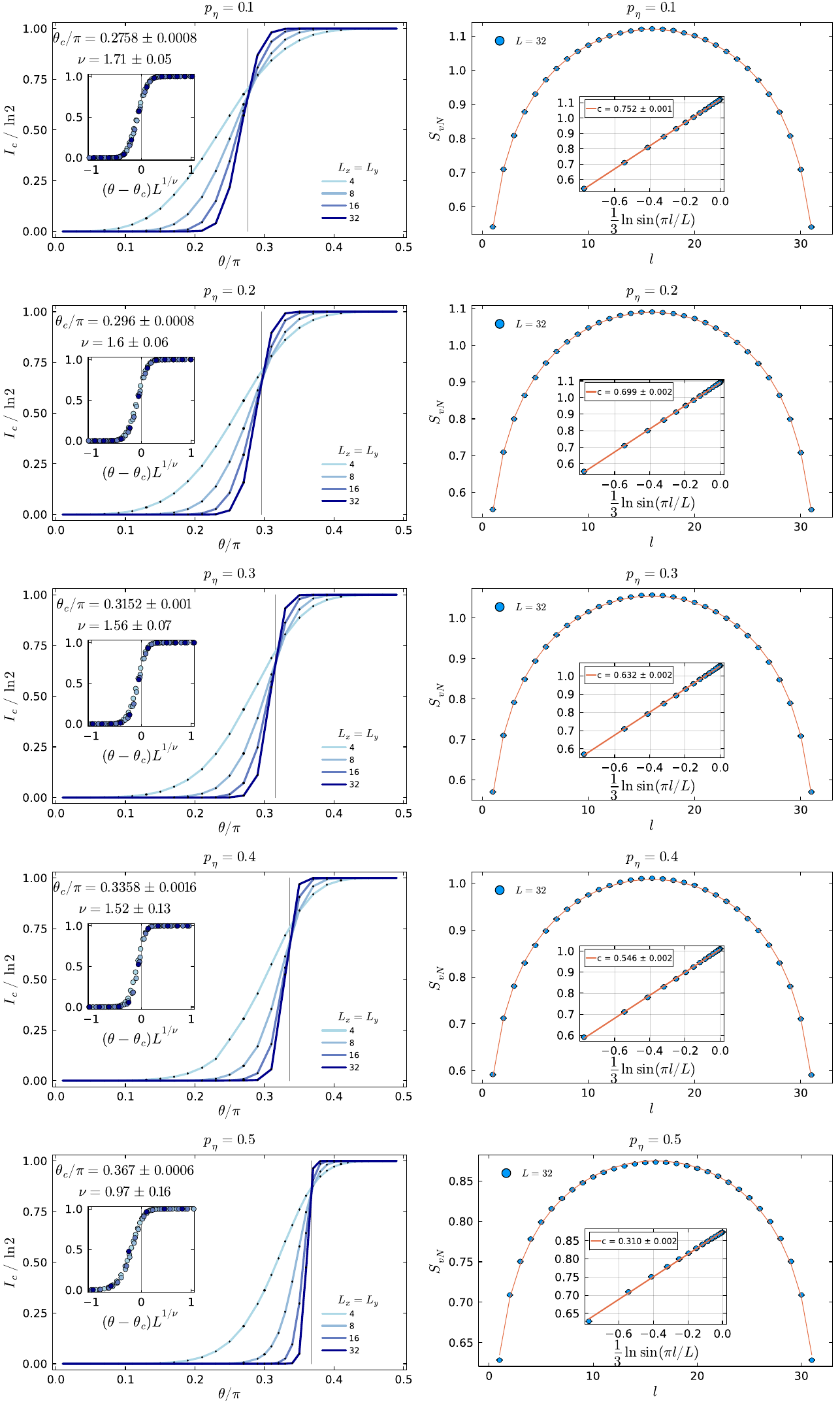}
   \caption{{\bf Phase transitions driven by $\theta$ with varying mass of $e$-vortex}: the first column shows the finite size scaling of coherent information signalling the phase transition; the second column shows the Born average bipartite von Neumann entanglement entropies at the critical point.}
   \label{fig:emass}
\end{figure*}

\begin{figure*}[h!] 
   \centering
   \includegraphics[width=.75\textwidth]{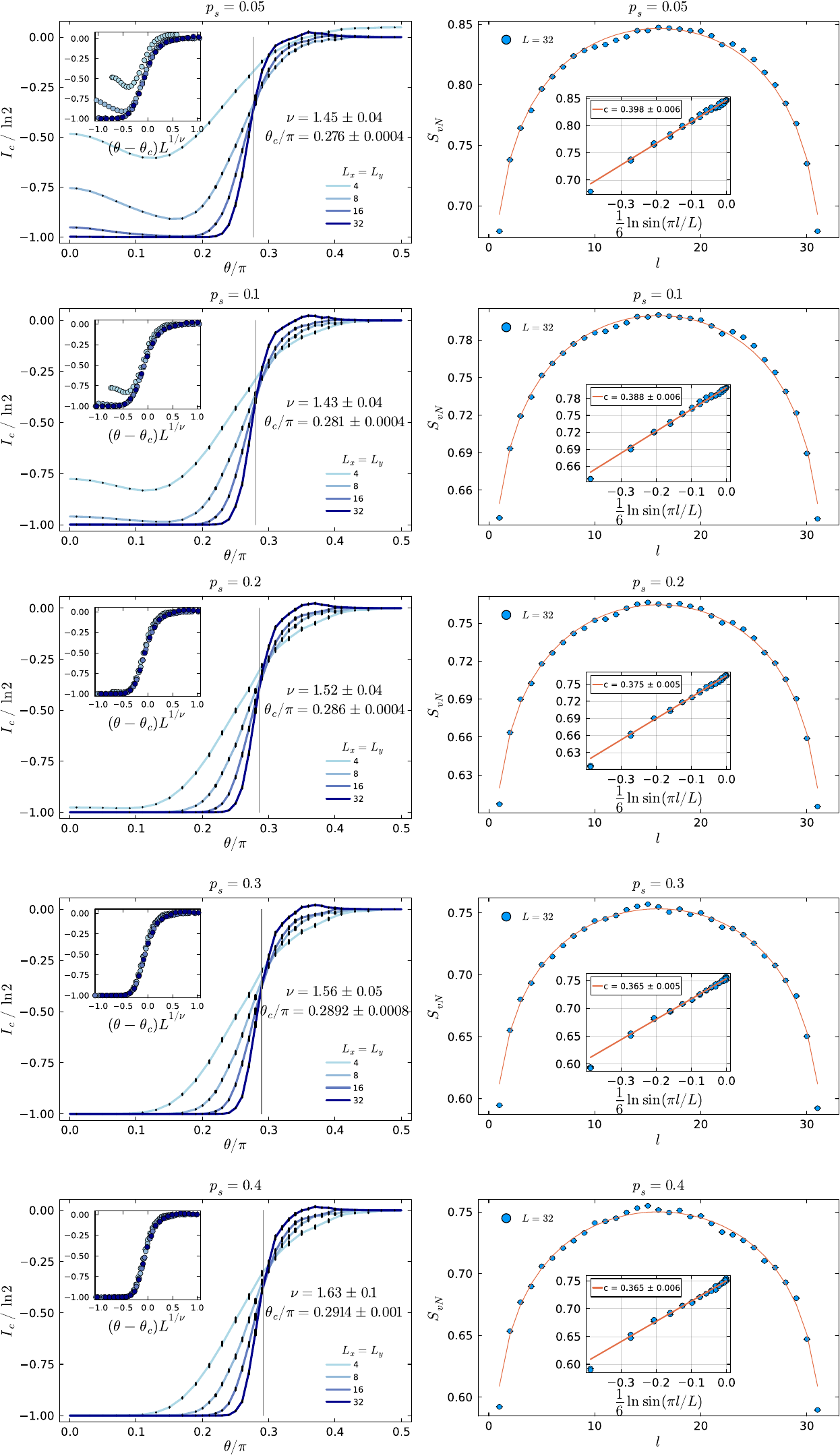}
   \caption{{\bf Phase transitions driven by $\theta$ with varying noise of $e$-vortex}: the first column shows the finite size scaling of the Born-averaged  second Rényi coherent information signalling the phase transition; the second column shows the Born average bipartite von Neumann entanglement entropies of the purified Choi state at the critical point.}
   \label{fig:enoise}
\end{figure*}

\begin{figure*}[h!] 
   \centering
   \includegraphics[width=\textwidth]{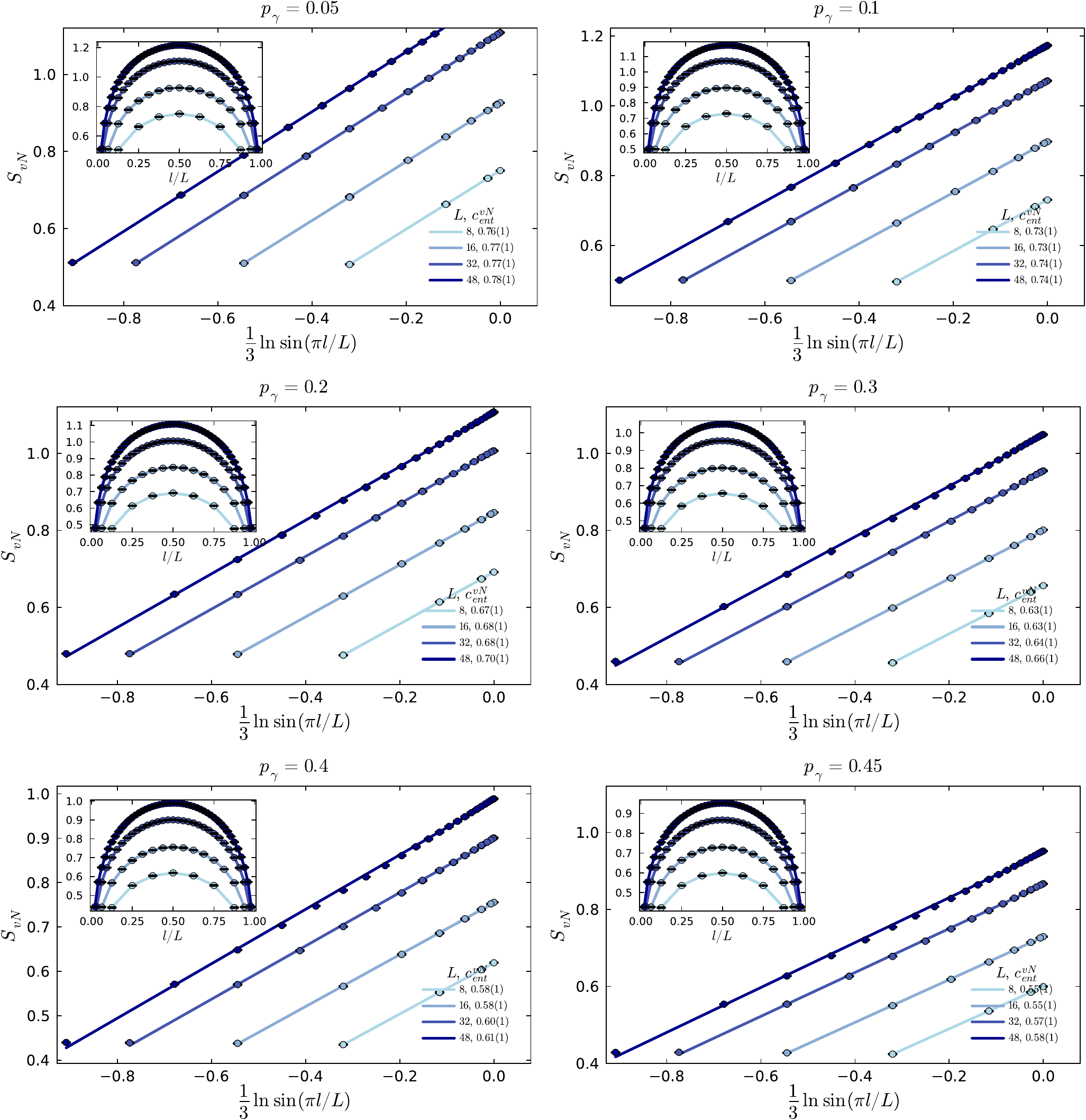}
   \caption{{\bf Von Neumann entanglement entropy scaling for the self-dual mixed state interpolating between weak self-dual and clean Ising by tuning the mass of $e$ and $m$ vortices}. The computation is performed at the exactly known self-dual location $\theta=\pi/4$, with Monte Carlo sampling combined with Gaussian fermion evolution. }
   \label{fig:emmass}
\end{figure*}

%%%%%%%%%%%%%%%%%%%%%%%%%%%%%%%%%%%%%%%%%%%%%%%%%%%%%%%%%%%%%%%%%%%%%%%%%%%%%%%%%%%%%%%
We close the manuscript by providing supplementary data for the critical states along the three critical line by tuning the $e$-mass (Fig.~\ref{fig:emass}), $e$-noise (Fig.~\ref{fig:enoise}), and $em$-mass (Fig.~\ref{fig:emmass}), respectively. 

\end{document}